\documentclass[aps,pra,superscriptaddress,reprint,a4paper,longbibliography]{revtex4-2}

\usepackage[utf8]{inputenc}
\usepackage[T1]{fontenc}
\usepackage{amsmath,amssymb,amsfonts,amsthm}
\usepackage{mathtools}
\usepackage{graphicx}% Include figure files
\usepackage{dcolumn}% Align table columns on decimal point
\usepackage{bm}% bold math
\usepackage[table]{xcolor}
\usepackage{tikz} %for ORCID
\usepackage{dsfont}
\usepackage{float}
\usepackage{braket}
\usepackage{xfrac}

\graphicspath{ {./figs/} }
\definecolor{myurlcolor}{rgb}{0,0,0.7}
\definecolor{myrefcolor}{rgb}{0.8,0,0}
\usepackage{hyperref}
\hypersetup{
    unicode=true, bookmarksnumbered=false, bookmarksopen=false, breaklinks=false, pdfborder={0 0 0},
    colorlinks=true,
    linkcolor=myrefcolor,citecolor=myurlcolor,urlcolor=myurlcolor
}

\theoremstyle{plain}% default

\makeatletter
\newcommand{\mrm}[1]{\mathrm{#1}}

\newcommand{\dd}{\mathrm{d}}
\newcommand{\ee}{\mathrm{e}}
\newcommand{\ii}{\mathrm{i}}
\newcommand{\TT}{\mathsf{T}}

\DeclareMathSymbol{\shortminus}{\mathbin}{AMSa}{"39}

\newcommand{\Tr}{\mathrm{Tr}}
\usepackage{xparse}
\NewDocumentCommand\trace{g}{
  \IfNoValueTF{#1}
    {\Tr}
    {\Tr\!\left\{#1\right\}}
}
\newcommand{\E}[2]{\mathbb{E}_{#2}\!\left[#1\right]}
\newcommand{\EE}[1]{\mathbb{E}\!\left[{#1}\right]}

\newcommand{\proj}[1]{\ket{#1}\!\!\bra{#1}}
\newcommand{\abs}[1]{\left|#1\right|}

\newcommand{\eref}[1]{(\ref{#1})}
\newcommand{\eqnref}[1]{Eq.~(\ref{#1})}
\newcommand{\eqnsref}[2]{Eqs.~(\ref{#1}) and (\ref{#2})}
\newcommand{\figref}[1]{Fig.~\ref{#1}}
\newcommand{\fref}[1]{Fig.~\ref{#1}}

\newcommand{\secref}[1]{Sec.~\ref{#1}}
\newcommand{\appref}[1]{App.~\ref{#1}}

\newcommand{\citeref}[1]{Ref.~\cite{#1}}
\newcommand{\refcite}[1]{Ref.~\cite{#1}}

\newcommand{\dt}{{\dd t}}
\newcommand{\Dt}{{\delta t}}

\renewcommand{\vec}[1]{\bm{#1}}

\newcommand{\Jqvec}[1]{\hat{\vec{#1}\hspace{0.1cm}}\hspace{-0.1cm}}

\newcommand{\dW}{{\dd W}}

\newcommand{\Ham}{\hat{H}}
\newcommand{\LinOp}{\hat{L\,}\!}
\newcommand{\UnitOp}{\hat{U}}

\newcommand{\Bhat}[1]{\hat{B}_{#1}}
\newcommand{\Banihil}{\Bhat{t}}
\newcommand{\Bcreat}{\Bhat{t}^\dagger}
\newcommand{\Banihilk}{\Bhat{k}}
\newcommand{\Bcreatk}{\Bhat{k}^\dagger}
\newcommand{\measE}{\,\hat{\!E}}
\newcommand{\bigO}{O}

\newcommand{\projector}[2]{\ket{#1} \!\! \bra{#2}}

\newcommand{\GenOp}{\hat{A}}
\newcommand{\Fop}{\hat{F}}

\newcommand{\Lin}{\mathcal{L}}
\newcommand{\Jx}{\hat{J}_x}
\newcommand{\Jy}{\hat{J}_y}
\newcommand{\Jz}{\hat{J}_z}
\newcommand{\estJx}{\est{\hat{J}}_x}
\newcommand{\estJy}{\est{\hat{J}}_y}

\newcommand{\jz}{\hat{s}_z}
\newcommand{\Smap}{\mathcal{S}}
\newcommand{\IjzI}{\hat{\jmath}_z}

\newcommand{\sz}{\hat{\sigma}_z}
\newcommand{\cc}{\scriptscriptstyle \mathrm{(c)}} %conditional evolution for 
\newcommand{\rhoc}{\rho_{\cc}}
\newcommand{\rhoct}{\rhoc(t)}

\newcommand{\rhoun}{\Tilde{\rho}}

\newcommand{\est}[1]{\tilde{#1}}
\newcommand{\brkt}[1]{\langle #1 \rangle}
\newcommand{\brktc}[1]{\big\langle #1 \big\rangle_{\!\cc}}
\newcommand{\kcoll}{\kappa_\mrm{coll}}
\newcommand{\kloc}{\kappa_\mrm{loc}}

\newcommand{\coll}{\mrm{coll}}
\newcommand{\loc}{\mrm{loc}}
\newcommand{\I}{\mathds{1}}
\newcommand{\D}{\mathcal{D}}
\renewcommand{\H}{\mathcal{H}}
\newcommand{\Vx}{\mrm{V}_x}
\newcommand{\Vy}{\mrm{V}_y}
\newcommand{\Vz}{\mrm{V}_z}
\newcommand{\estVx}{\est{\mrm{V}}_x}
\newcommand{\estVy}{\est{\mrm{V}}_y}
\newcommand{\estVz}{\est{\mrm{V}}_z}
\newcommand{\estCxy}{\est{\mathrm{C}}_{xy}}

\newcommand{\Cxy}{\mathrm{C}_{xy}}
\newcommand{\Czy}{\mathrm{C}_{zy}}
\newcommand{\Cxz}{\mathrm{C}_{xz}}

\newcommand\scalemath[2]{\scalebox{#1}{\mbox{\ensuremath{\displaystyle #2}}}}

\newcommand{\F}{\mathcal{F}}
\newcommand{\FF}{\mathfrak{F}}
\newcommand{\Unitary}{\mathcal{U}}
\newcommand{\Diag}[1]{\mathrm{diag}\!\left(#1\right)}
\newcommand{\CSS}{{\scriptscriptstyle \mathrm{CSS}}} %CSS 

\newcommand{\cov}{\mrm{cov}}

\newcommand{\Zeta}{\mathcal{Z}}

\definecolor{lime}{HTML}{A6CE39}
\DeclareRobustCommand{\orcidicon}{
	\begin{tikzpicture}
	\draw[lime, fill=lime] (0,0) 
	circle [radius=0.16] 
	node[white] {{\fontfamily{qag}\selectfont \tiny ID}};
	\draw[white, fill=white] (-0.0625,0.095) 
	circle [radius=0.007];
	\end{tikzpicture}
	\hspace{-2mm}
}
\foreach \x in {A, ..., Z}{\expandafter\xdef\csname orcid\x\endcsname{\noexpand\href{https://orcid.org/\csname orcidauthor\x\endcsname}
			{\noexpand\orcidicon}}
}
\makeatother
\begin{document}

\title{Noisy atomic magnetometry with Kalman filtering and measurement-based feedback}

\author{J\'{u}lia Amor\'{o}s-Binefa\orcidA{}}
\email{j.amoros-binefa@cent.uw.edu.pl}
\affiliation{Centre of New Technologies, University of Warsaw, Banacha 2c, 02-097 Warsaw, Poland.}

\author{Jan Ko\l{}ody\'{n}ski\orcidB{}}
\email{jankolo@ifpan.edu.pl}
\affiliation{Centre of New Technologies, University of Warsaw, Banacha 2c, 02-097 Warsaw, Poland.}
\affiliation{Institute of Physics, Polish Academy of Sciences, Aleja Lotnik\'{o}w 32/46, 02-668 Warsaw, Poland.}

\begin{abstract}
Sensing a magnetic field with an atomic magnetometer operated in real time presents significant challenges, primarily due to sensor nonlinearity, the presence of noise, and the need for one-shot estimation. To address these challenges, we propose a comprehensive approach that integrates measurement, estimation and control strategies. Specifically, this involves implementing a quantum non-demolition measurement based on continuous light-probing of the atomic ensemble. The resulting photocurrent is then directed into an Extended Kalman Filter to produce instantaneous estimates of the system's dynamical parameters. These estimates, in turn, are utilised by a Linear Quadratic Regulator, whose output is applied back to the system through a feedback loop. This procedure automatically steers the atomic ensemble into a spin-squeezed state, yielding a quantum enhancement in precision. Furthermore, thanks to the feedback proposed, the atoms exhibit entanglement even when the measurement data is discarded. To prove that our approach constitutes the optimal strategy in realistic scenarios, we derive ultimate bounds on the estimation error applicable in the presence of both local and collective decoherence, and show that these are indeed attained. Additionally, we demonstrate that for large ensembles 
the EKF not only reliably predicts its own estimation error in real time, but also accurately estimates spin squeezing at short timescales.
\end{abstract}

\maketitle

%%%%%%%%%%%%%%%%%%%%%%%%%%%%%%%%%%%%%%%%%%%%%%%%%%%%%%%%%%%%%%%%%%%%%%%%%%%%%%%%%%%%%%%%%%%
%%%%%%%%%%%%%%%%%%%%%%%%%%%%%%%%%%%%%%%%%%%%%%%%%%%%%%%%%%%%%%%%%%%%%%%%%%%%%%%%%%%%%%%%%%%
\section{Introduction}
\label{sec:introduction}
Optical magnetometers that rely on atomic ensembles pumped and probed with laser light~\cite{budker_optical_2007} constitute ultraprecise magnetic field sensors, achieving sensitivities comparable to state-of-the-art superconducting quantum interference devices~\cite{clarke2004squid}. Not only do they not require cryogenic cooling, but they have also recently been miniaturised to chip scales~\cite{kitching_chip-scale_2018}. As a result, they promise breakthroughs, e.g., when used to sense magnetic fields in medical applications~\cite{jensen_magnetocardiography_2018,boto_moving_2018,limes_portable_2020,zhang_recording_2020}, as well as in the search for new exotic physics~\cite{Pospelov2013,pustelny_global_2013}. These tasks, in particular, fall into the category of real-time sensing problems in which the sensor is employed to track a time-varying signal (magnetic field) while continuously acquiring measurement data. Such a scenario may be considered the most demanding, as it requires the sensing procedure to be performed only once, with the sensor being controlled ``on the fly.'' Although entanglement---a crucial resource in quantum metrology~\cite{DAriano2001,Giovannetti2004}---has been shown to significantly enhance the sensitivity in experiments involving sufficiently many (independent and identical) repetitions~\cite{Sewell2012,Koschorreck2010,Wasilewski2010}, there is yet to be an experimental demonstration that this can be done in real time with the entanglement being created by measurement backaction~\cite{Kuzmich1998}, despite some prominent achievements~\cite{Kuzmich2000,Shah2010,Kong2020}.

Apart from substantial experimental challenges, an important hurdle is the proposal and verification of an accurate dynamic model of the atomic noise, which would then allow for the tools of control and statistical inference theory to be used in the design of a future device. This contrasts with the setting of optomechanical sensors operating at cryogenic temperatures~\cite{Aspelmeyer2014}, in which case quantum Gaussian stochastic models have been proposed and verified~\cite{Iwasawa2013,wieczorek_optimal_2015}, allowing for spectacular demonstrations of cooling and controlling such devices in real time~\cite{rossi_observing_2019}, while incorporating both measurement-based~\cite{Wilson2015,Rossi2018,Sudhir2017} (also with the use of levitated nanoparticles~\cite{Setter2018,Magrini2021}) and coherent~\cite{Ernzer2023} feedback methods~\cite{nurdin2017linear,hamerly2012advantages,Yamamoto2014}.

In our work, we close this gap by proposing a quantum dynamical model of a continuously monitored atomic (spin-$1/2$) ensemble that incorporates measurement-based feedback, and---in contrast to previous works~\cite{Geremia2003,Stockton2004,Molmer2004,Madsen2004,deutsch_quantum_2010}---accounts for decoherence without relying on any approximations. It allows us to simulate the exact evolution of moderate-sized ensembles by resorting to novel numerical tools~\cite{Rossi2021}. As a result, we can identify how to improve previous approaches~\cite{Geremia2003,Stockton2004,Molmer2004,Madsen2004}, in order to build an effective model that correctly describes typical experiments involving a large number of atoms, e.g.~$N\approx10^{5}\!-\!10^{13}$~\cite{Kuzmich2000,Wasilewski2010,Shah2010,Koschorreck2010,Sewell2012,Kong2020}. Such a reliable model allows us to then construct for the first time an efficient \emph{estimation+control scheme} that leverages atomic entanglement, i.e.~spin squeezing~\cite{Ma2011,Pezze2018RMP}, to sense a magnetic field in real time at the ultimate quantum limit~%
\footnote{For clarity, we consider here the task of sensing a magnetic field that is unknown but of a constant value. For generalisation and application of our results to tracking time-varying and fluctuating magnetic fields, see~\citeref{Amoros-Binefa2025}}.

In particular, the effective dynamical model goes beyond linearity while its noise remains Gaussian, as it is defined within a frame that rotates with the ensemble's overall spin at the Larmor frequency. The model is then used to construct an \emph{Extended Kalman Filter} (EKF)~\cite{crassidis2011optimal,simon2006}, which estimates the field along with the first and second moments of the angular momentum. These estimates are in turn fed back into a closed-loop system, where a \emph{Linear Quadratic Regulator} (LQR)~\cite{Stockton2004,Magrini2021} stabilises the state of the atoms.

The optimality of the procedure is verified by deriving and saturating bounds on precision that apply to \emph{any} sensing scheme involving measurement-based feedback. These are a consequence of collective (cf.~\refcite{Amoros-Binefa2021}) and/or local atomic decoherence incorporated in the dynamics. Moreover, as such noise-induced quantum limits scale at best linearly with the number of atoms $N$ (and the sensing time), they exclude the possibility of surpassing the so-called \emph{standard quantum limit} (SQL)---disproving previous conjectures based on numerical evidence~\cite{rossi_noisy_2020}.

Finally, by resorting to the exact dynamical model we demonstrate that the proposed sensing scheme steers the atomic ensemble into a state whose entanglement is amplified by the information acquired from past measurements, i.e.~a \emph{conditionally} spin-squeezed state~\cite{Pezze2018RMP}. However, as our scheme interprets and feeds back measurement outcomes ``on the fly,'' the final atomic state also exhibits \emph{unconditional} spin squeezing~\cite{Pezze2018RMP}, i.e.~its entanglement is assured without the need to store or reinterpret past measurement data. Furthermore, in the experimental regime~\cite{Kuzmich2000,Shah2010,Kong2020} in which our effective model applies, we show that the EKF also reliably predicts the conditional spin squeezing, so that this form of entanglement can be used for other tasks.

The manuscript is organised as follows:~in \secref{sec:setup} we present the setup of the atomic magnetometer we choose to consider. \secref{sec:simulation_system} discusses the numerical simulation of the exact sensor model and how to approximate it by introducing the comoving Gaussian picture. Then, in \secref{sec:bounds}, we build upon the work in \refcite{Amoros-Binefa2021} by extending it to include local decoherence, and establish ultimate bounds on the achievable precision in magnetic field estimation. Subsequently, in \secref{sec:estimation_and_control}, we detail our chosen estimation and control strategies, namely the EKF and LQR. The final two sections, \secref{sec:performance} and \secref{sec:squeezing}, present our results. Specifically, section \ref{sec:performance} demonstrates that in the large-$N$ regime, our proposed EKF+LQR strategy can attain the noise-induced ultimate bounds on precision. Moreover, \secref{sec:squeezing} reveals that the introduction of LQR feedback prepares the atomic state in a multipartite entangled state, as indicated by the emergence of unconditional spin squeezing. Finally, \secref{sec:conclusions} summarises our results and discusses their implications.

%%%%%%%%%%%%%%%%%%%%%%%%%%%%%%%%%%%%%%%%%%%%%%%%%%%%%%%%%%%%%%%%%%%%%%%%%%%%%%%%%%%%%%%%%%%
%%%%%%%%%%%%%%%%%%%%%%%%%%%%%%%%%%%%%%%%%%%%%%%%%%%%%%%%%%%%%%%%%%%%%%%%%%%%%%%%%%%%%%%%%%%
\section{Setup}
\label{sec:setup}
\begin{figure}[t!]
    \includegraphics[width=\columnwidth]{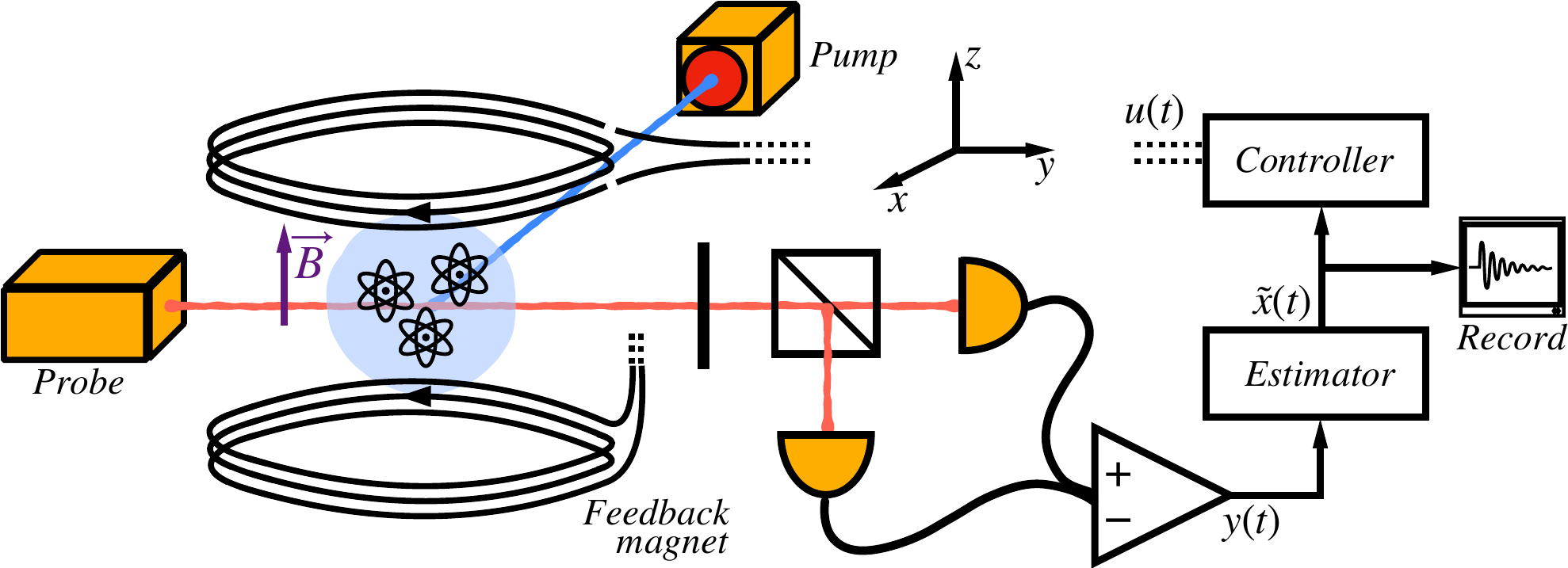}
    \caption{\textbf{Atomic magnetometry with estimation and control}. The magnetometer consists of an ensemble of spin-1/2 atoms, initially pumped along $x$ into a coherent spin state, which then Larmor precess around the $z$ axis due to the constant magnetic field $B$ to be determined. The total atomic spin is probed along the $y$ direction by the second beam that weakly interacts with the ensemble, so that the corresponding spin component can be continuously monitored via balanced photodetection. The output photocurrent $y(t)$ is then used to construct estimates of the temporary dynamical parameters of the atomic ensemble, $\Tilde{\vec{x}}(t)$ (also containing the $B$-field estimate), and feed them into the controller that instantaneously decides on the value, $u(t)$, of the extra magnetic field to be induced through the feedback coil, so that arbitrary bias can be applied in real time to the estimated magnetic field $B$.}
    \label{fig:setup}
\end{figure}

The main goal of the magnetometry experiment depicted in \fref{fig:setup} is to estimate a constant~\cite{Note1} magnetic field $B$ aligned with the $z$ axis. An ensemble of $N$ atoms is used to indirectly probe the magnetic field $B$, while being pumped with circularly polarised light along the $x$ direction, see \fref{fig:setup}, such that only two energy levels of each atom effectively contribute to the light-probing process~\cite{Budker2013}. As a consequence, we may treat the atomic ensemble as a collection of $N$ spin-$1/2$ particles, whose spin precesses around the $z$ axis at a Larmor frequency $\omega = \gamma B$ induced by the magnetic field $B$, where $\gamma$ is the gyromagnetic ratio. Moreover, the evolution of the total spin is then described through the use of collective angular momentum operators, $\hat{J}_\alpha = \sum_{i = 1}^N \hat{\sigma}_\alpha^{(i)}/2$ with $\alpha = x, y, z$, that form an (orientation) vector $\Jqvec{J}=(\hat{J}_x,\hat{J}_y,\hat{J}_z)^\TT$. 

Furthermore, assuming that the atoms are initially pumped (along $x$;~see \figref{fig:setup}) into a coherent spin state (CSS), the mean and the variance for each component of $\Jqvec{J}(t)$ are given at time $t=0$ by $\brkt{\Jqvec{J}}_\CSS = (N/2,0,0)^\TT$ and $\vec{\mrm{V}}^\CSS = (0,N/4,N/4)^\TT$~\cite{Ma2011}, respectively, where $\mrm{V}_\alpha^\CSS \coloneqq \brkt{\hat{J}_\alpha^2}_\CSS-\brkt{\hat{J}_\alpha}_\CSS^2$. However, as explicitly shown in \fref{fig:CSS_bloch}, it is useful to visualise such a state as a quasiprobability distribution on a (generalised) Bloch sphere---formally the Wigner function projected onto a sphere (see \appref{ap:Wigner}~\cite{Pezze2018RMP})---which is centred at $\brkt{\Jqvec{J}}_\CSS$ with the width in the $x$, $y$, $z$ directions specified by the elements of $\vec{\mrm{V}}^\CSS$.

Once pumped, the atoms are continuously monitored by a linearly polarised probe beam directed along the $y$ axis, as shown in \figref{fig:setup}. The probe light is sufficiently detuned from the relevant atomic transition, so that its interaction with the atoms can be considered linear while still inducing backaction due to the quantum non-demolition (QND) character of the measurement~\cite{Geremia2006,deutsch_quantum_2010}. In particular, upon interaction with the atoms the probe-beam polarisation gets (Faraday) rotated by an angle proportional to the total angular momentum component along the probe propagation, i.e.~$\hat{J}_y$. As a result, the output photocurrent of a differential photodetection measurement, which registers small polarisation-angle deviations, is given by~\cite{Belavkin1989,handel_modelling_2005}
\begin{align} 
	\label{eq:photocurrent}
    y(t) dt = 2\eta \sqrt{M} \brktc{\Jy(t)}dt + \sqrt{\eta} \, dW,
\end{align}
where $\eta$ is the detection efficiency and $dW$ denotes the stochastic Wiener increment, satisfying $\E{dW^2}{}=dt$ according to It\^{o} calculus~\cite{Gardiner1985}. The white-noise fluctuations in \eqnref{eq:photocurrent} arise due to the shot noise of the photodetection process. However, by fixing their strength to unity, we effectively renormalise the photocurrent $y(t)$. As a result, it is the \emph{measurement strength} parameter $M$ that---while incorporating all the relevant experimental electronic, light-matter couplings etc.~\cite{Geremia2006,Amoros-Binefa2025}---parametrises the ratio between the atomic contribution to the detected signal (first term in \eqnref{eq:photocurrent}) and the magnitude of white noise (second term in \eqnref{eq:photocurrent}).

An essential feature of the above formalism is the incorporation of measurement backaction exerted onto the atoms~\cite{Belavkin1989,handel_modelling_2005}. In particular, within \eqnref{eq:photocurrent}, the mean $\brktc{\Jy (t)} = \trace{\rhoc (t) \Jy }$ is evaluated with respect to the \emph{conditional} atomic state, $\rhoc (t) \equiv \rho(t|\vec{y}_{\le t})$, i.e.~the one most consistent (minimising the  mean-square distance~\cite{handel_modelling_2005}) with the particular measurement trajectory observed, $\vec{y}_{\le t} = \{y(\tau)\}_{0\leq \tau \leq t}$. Additionally, to explicitly write the dynamics of $\rhoc (t)$, we must also account for the measurement-based control strategy introduced in \figref{fig:setup}. This strategy assumes that based on the photocurrent record (potentially the whole history), $\vec{y}_{\le t}$, estimates of some dynamical parameters for the atomic system are made---denoted by a vector $\Tilde{\vec{x}}(t)$ in \figref{fig:setup}, which includes the estimate of the Larmor frequency, $\est{\omega} = \gamma \Tilde{B}$, and hence the $B$ field. The estimates, $\Tilde{\vec{x}}(t)$, are then used to set the control (scalar) field to a specific value $u(\vec{y}_{\le t})$, altering the additional magnetic field applied instantaneously along the estimated $B$ and thus modifying the Larmor frequency at time $t$:~$\omega\to\omega+u(\vec{y}_{\le t})$. As the control field $u(\vec{y}_{\le t})$ depends on the whole measurement record $\vec{y}_{\le t}$, the dynamics of the atomic ensemble at each time step is not only affected by the backaction of the current measurement but also dependent on the entire measurement record through the addition of control.

Bearing this in mind and simplifying notation by denoting the control field as $u(t)\equiv u(\vec{y}_{\le t})$, we write the stochastic master equation (SME) that governs the dynamics of the conditional atomic state $\rhoc (t)$ as~\cite{Belavkin1989,handel_modelling_2005}
\begin{align}
    d\rhoc = &-i (\omega + u(t)) [\Jz,\rhoc] dt \nonumber \\
    &+ \frac{\kloc}{2} \sum_{j=1}^N \D[\sz^{(j)}] \rhoc dt \, + \kcoll \D[\Jz]\rhoc dt  \nonumber \\
    &+ M\D[\Jy]\rhoc dt + \sqrt{\eta M} \H[\Jy] \rhoc dW, 
    \label{eq:SME}
\end{align} 
where the superoperators $\D$ and $\H$ are defined for any operator $\GenOp$ and state $\rho$ as \mbox{$\D[\GenOp]\rho \coloneqq  \GenOp \rho  \GenOp^\dagger - \frac{1}{2}\{\GenOp^\dagger\GenOp,\rho\}$} and \mbox{$\H[ \GenOp] \rho \coloneqq  \GenOp \rho + \rho  \GenOp^\dagger - \trace{(\GenOp+\GenOp^\dagger)\rho}\rho$}. 

The last two terms in \eqnref{eq:SME} arise due to the backaction of the continuous quantum measurement:~the first term represents the measurement-induced decoherence in the basis of the observable being probed, $\hat{J}_y$; the second term accounts for the stochastic jump dictated by the photocurrent recorded during a particular time step, $dW = [y(t)/\sqrt{\eta} - 2 \sqrt{\eta M} \brktc{\Jy}]dt$ according to \eqnref{eq:photocurrent}. This last term, crucially nonlinear in $\rhoc$, opens doors for conditional squeezing of the atomic state~\cite{Geremia2003}. However, to account for the impact of noise and verify the robustness of our estimation strategies, we also incorporate in \eqnref{eq:SME} local and global decoherence terms. These terms effectively dephase the atomic state along the $z$ direction of the estimated $B$ field at rates $\kloc$ and $\kcoll$ for local and collective dephasing, respectively. Local dephasing acts independently on each individual atom $j$ in the basis of $\sz^{(j)}/2$, while the collective term acts globally within the basis of the collective atomic spin operator $\hat{J}_z$.

%%%%%%%%%%%%%%%%%%%%%%%%%%%%%%%%%%%%%%%%%%%%%%%%%%%%%%%%%%%%%%%%%%%%%%%%%%%%%%%%%%%%%%%%%%%
%%%%%%%%%%%%%%%%%%%%%%%%%%%%%%%%%%%%%%%%%%%%%%%%%%%%%%%%%%%%%%%%%%%%%%%%%%%%%%%%%%%%%%%%%%%
\section{Simulating the system dynamics} 
\label{sec:simulation_system}
In this work, we explicitly simulate for the first time the SME \eref{eq:SME}. However, even though we adapt the most recent numerical methods~\cite{Rossi2021} to incorporate decoherence and feedback, as discussed below, such a `brute-force' approach is possible only for atomic ensembles of moderate size. That is why, our goal is still to propose an effective model of dynamics that applies to relevant experiments~\cite{Kuzmich2000,Shah2010,Kong2020}, and can thus be used to design the building blocks of an estimation+control scheme---here, EKF+LQR introduced later in \secref{sec:estimation_and_control}. Importantly, the numerical solution allows us to validate our approach, so that our results can then be extrapolated to relevant ensemble sizes, beyond the reach of `brute-force' numerics.

\begin{figure}[t]
    \centering
    \includegraphics[width=\columnwidth]{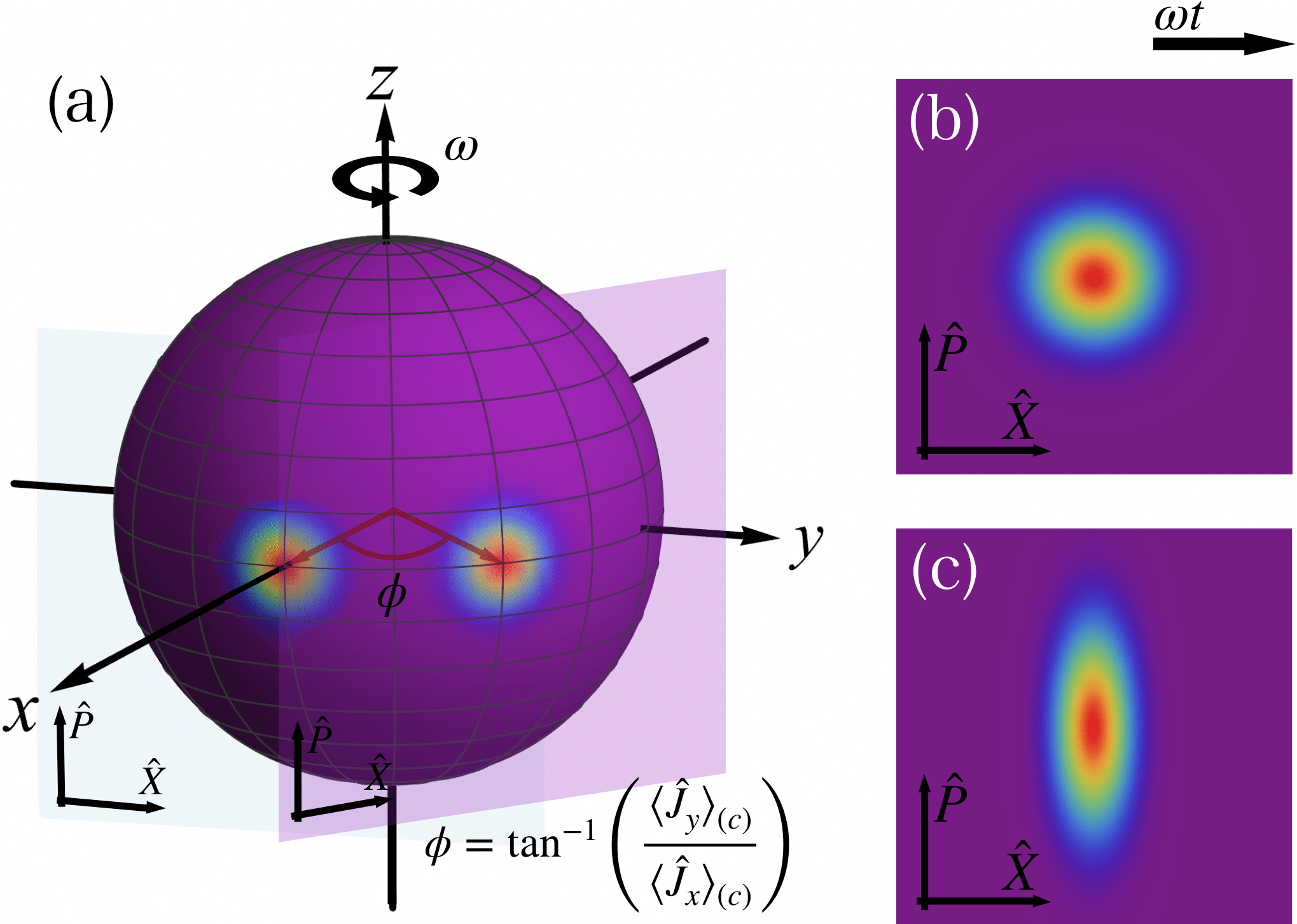}
    \caption{\textbf{Bloch sphere representation of the atomic state:~linear-Gaussian (LG) and comoving Gaussian (CoG) approximations.} The 3D-plot in (a) depicts the Wigner function of a CSS pointing along $x$ in a generalised Bloch sphere for $N = 100$. For large enough atomic ensembles ($N\gg1$), the curvature of the sphere at the maximum of the CSS Wigner function can be approximated by the LG plane, where the effective phase-space quadratures, $\hat{X}$ and $\hat{P}$, are defined as in \eqnref{eq:canonical_X&P}. The continuous measurement of the $y$ spin component, which induces spin squeezing of the atomic state, can then be interpreted as just squeezing (antisqueezing) the $\hat{X}$ ($\hat{P}$) quadrature at short times at which the LG approximation still holds. Subfigure (c) visualizes the effect of continuous spin squeezing on the Wigner function of a CSS in the LG plane [shown in (b)]. In order for the Gaussian picture to remain valid at longer times, we let the LG plane rotate at the Larmor frequency $\omega$ together with the atomic spin, and refer to this description as the \emph{comoving Gaussian} (CoG) approximation.}
    \label{fig:CSS_bloch}
\end{figure}

First, however, let us provide the physical motivation and reason of why our effective model succeeds in describing dynamics of large atomic ensembles. As shown in \figref{fig:CSS_bloch}, for a highly polarised ensemble, it is common to approximate the Wigner distribution of the total spin on a Bloch sphere with the distribution of a single bosonic mode in a plane (phase space)~\cite{Pezze2018RMP}, by using the Holstein-Primakoff transformation~\cite{Molmer2004,Madsen2004}. However, such an approximation employed in previous works~\cite{Geremia2003,Stockton2004,Molmer2004,Madsen2004,deutsch_quantum_2010} is insufficient for our purposes, as it fails to correctly capture the Larmor precession of the spin. Thus, in order not to lose necessary information about the curvature of the Bloch sphere, we move to a frame rotating at the Larmor frequency (see \figref{fig:CSS_bloch}) in which the effective bosonic modes coprecess with the spin. A similar approach has been pursued when studying the emergence of a quantum chaotic behaviour in such a system~\cite{Munoz-Arias2020}. Formally, as done below, this corresponds to tracking the evolution of the ensemble spin operators up to their second-order moments, with higher-order contributions being ignored.

%%%%%%%%%%%%%%%%%%%%%%%%%%%%%%%%%%%%%%%%%%%%%%%%%%%%%%%%%%%%%%%%%%%%%%%%%%%%%%%%%%%%%%%%%%%
\subsection{Exact model:~numerical solution}
\label{sec:exact_dyn}
Optically pumped magnetometers operate with atomic numbers in the range $N\approx10^{5}\!-\!10^{13}$~\cite{Kuzmich2000,Wasilewski2010,Shah2010,Koschorreck2010,Sewell2012,Kong2020}. This precludes any naive numerical simulations of the ensemble dynamics, since the dimension of the underlying Hilbert space scaling exponentially with $N$, i.e.~as $2^N$ for spin-$1/2$ atoms. However, assuming that the system preserves permutational invariance over the entire duration of its evolution---meaning that any pair of atoms within the ensemble is interchangeable---the dimension of the density matrix can be reduced to scale polynomially with $N$. In particular, for a collection of spin-$1/2$ atoms, as the density matrix then possesses a direct-sum structure with each block being associated with a spin number $j$ ranging from $0$ $(\tfrac{1}{2})$ to $N/2$ for even (odd) $N$, its complexity scales as $\bigO(N^3)$~\cite{Chase2008,Baragiola2010,Shammah2018}. Moreover, if the evolution is induced by collective processes---i.e.~generated by collective operators that are themselves permutationally invariant---any initial state supported by the totally symmetric subspace (with $j=N/2$), e.g.~CSS, must evolve within it, further reducing the complexity to $\bigO(N)$~\cite{Chase2008,Baragiola2010,Shammah2018}.

In this work, we use the numerical solution of SME \eref{eq:SME} as a benchmark, which preserves the permutational symmetry (or even the totally symmetric subspace in the $\kloc=0$ case). Specifically, for moderate $N$, we employ the code of~\citet{Rossi2021} that exploits the symmetries of the system as described above. It resorts to numerical integration of an SME by constructing the Kraus operators of the weak measurement at each time step, while also guaranteeing the positivity of the density matrix~\cite{Rouchon2015,Albarelli2018}.

%%%%%%%%%%%%%%%%%%%%%%%%%%%%%%%%%%%%%%%%%%%%%%%%%%%%%%%%%%%%%%%%%%%%%%%%%%%%%%%%%%%%%%%%
\subsection{Approximate model:~comoving Gaussian picture}
\label{sec:CoG_dyn}
%

%%%%%%%%%%%%%%%%%%%%%%%%%%%%%%%%%%%%%%%%%%%%
\subsubsection{Linear-Gaussian regime}
The exploitation of permutational symmetry is still not sufficient to reach experimentally relevant values of $N$. One approach is to further assume that the $B$ field is small and that the impact of local decoherence is negligible. As a result, by considering small enough timescales, $t \lesssim 1/(M + \kcoll)$, we can approximate $\brktc{\Jx(t)}$ with its unconditional average value $\brkt{\Jx(t)}=\tfrac{N}{2} \ee^{-(M + \kcoll)t/2}$~\cite{Amoros-Binefa2021}. Geometrically, as depicted in \fref{fig:CSS_bloch}, this corresponds to effectively approximating the surface of the generalised Bloch sphere by a plane perpendicular to the collective angular momentum vector pointing in the $x$ direction~\cite{Geremia2003,Albarelli2017,Amoros-Binefa2021}. This plane then defines an effective phase space with position and momentum operators given by
\begin{equation} 
	\label{eq:canonical_X&P}
 	\hat{X} \coloneqq \Jy/\sqrt{|\brkt{\Jx(t)}|} \; \;  \text{and} \; \; \hat{P} \coloneqq \Jz/\sqrt{|\brkt{\Jx(t)}|},
\end{equation}
which satisfy the canonical commutation relation $[\hat{X},\hat{P}] \approx i$, as long as $\Jx \approx |\brkt{\Jx(t)}|\openone$ for sufficiently large $N$~\cite{Madsen2004,Molmer2004}. As SME \eref{eq:SME} then becomes equivalent to a set of differential equations for the first and second moments of the quadratures in \eref{eq:canonical_X&P} that are linear in $\langle\hat{X}\rangle$ and $\langle\hat{P}\rangle$, as well as in the magnetic field $B$~\cite{Madsen2004,Molmer2004}, we refer to such a regime as being \emph{linear Gaussian} (LG)~\cite{Amoros-Binefa2021}.

%%%%%%%%%%%%%%%%%%%%%%%%%%%%%%%%%%%%%%%%%%%%
\subsubsection{Beyond the linear-Gaussian regime}
In real-life magnetometers~\cite{Kuzmich2000,Wasilewski2010,Shah2010,Koschorreck2010,Sewell2012,Kong2020}, the atomic spin must precess multiple times over the course of the detection process to collect a sufficient signal. This precludes the LG approximation from  actually being useful. Therefore, to describe the system as approximately Gaussian at all times, we allow the LG plane (see \fref{fig:CSS_bloch}) to Larmor precess with the mean angular-momentum vector $\brkt{\Jqvec{J}(t)}$ at frequency $\omega$~\cite{Munoz-Arias2020}. We refer to this as the \emph{comoving Gaussian} (CoG) approximation. We expect this approach to be valid under the following conditions:~the ensemble is large enough, i.e., $N \gg 1$;~the squeezing due to the continuous measurement is not too strong to wrap the Wigner function around the Bloch sphere; and the local decoherence is moderate, allowing the dynamics to be well described by only the first and second moments.

In particular, by considering the conditional evolution within the Heisenberg picture of the mean angular momenta $\brktc{\hat{J}_\alpha(t)}$, as well as their covariance matrix with elements $\mathrm{C}^{\cc}_{\alpha \beta}(t) \coloneqq \frac{1}{2} \left(\brktc{\{\hat{J}_\alpha(t),\hat{J}_\beta(t)\}}-2\brktc{\hat{J}_\alpha(t)}\brktc{\hat{J}_\beta(t)} \right)$ such that $\mathrm{V}^{\cc}_{\alpha}(t)\coloneqq \mathrm{C}^{\cc}_{\alpha\alpha}(t)$ ($\alpha,\beta = x, y,z$), we derive in \appref{ap:Ito} based on SME \eref{eq:SME} the following set of coupled stochastic differential equations (dropping the explicit $t$ dependence of all the quantities for convenience):
%the $(\textrm{c})$-sub/superscript
\begin{widetext}
\begin{subequations}
\label{eq:dynamical_model}
\begin{align}  
    &  d\brktc{\Jx} \!=\!  - (\omega \!+\! u(t))  \brktc{\Jy}  dt \!- \frac{1}{2}(\kcoll \!+\! 2 \kloc \!+\! M)  \brktc{\Jx} dt \!+\! 2\sqrt{\eta M}  \Cxy^{\cc} \, dW \\ 
    &  d\brktc{\Jy} \!=\! (\omega \!+\! u(t))  \brktc{\Jx} dt \!- \frac{1}{2}(\kcoll \!+\! 2\kloc)  \brktc{\Jy} dt \!+\! 2\sqrt{\eta M}  \Vy^{\cc} \, dW \label{eq:dJy}\\
    & d \Vx^{\cc} \! = \! - 2 (\omega \! + \! u(t))  \Cxy^{\cc} dt \! + \! \kcoll \! \left( \! \Vy^{\cc} \!+\!  \brktc{\Jy}^2 \!-\! \Vx^{\cc} \!\right)\!dt \!+\! \kloc \!\left( \!\frac{N}{2} \!-\!2 \Vx^{\cc} \!\right)\!dt \!+\!  M\! \left( \! \Vz^{\cc} \!-\! \Vx^{\cc} \!-\! 4 \eta {\Cxy^{\cc}}^2 \!\right) \!dt \label{eq:dVx}\\
    &   d\Vy^{\cc} \!=\! 2 (\omega \!+\! u(t))  \Cxy^{\cc} dt \!+\! \kcoll \!\left(\!  \Vx^{\cc} \!+\!  \brktc{\Jx}^2 \!-\!  \Vy^{\cc}\!\right) \!dt \!+\! \kloc \!\left(\! \frac{N}{2} \!-\!2 \Vy^{\cc} \!\right) \!dt \!-\! 4 \eta M  {\Vy^{\cc}}^2 dt \label{eq:dVy}\\
    &   d\Vz^{\cc} \!=\!  M \!\left( \!\Vx^{\cc} \!+\!  \brktc{\Jx}^2 \!-\! \Vz^{\cc}\! \right) \!dt \label{eq:dVz}\\
    &   d\Cxy^{\cc} \!=\! (\omega \!+\! u(t)) \!\left(\!\Vx^{\cc} \!-\!  \Vy^{\cc}\!\right) \!dt \!-\! \kcoll \!\left(\!2\Cxy^{\cc} \!+\! \brktc{\Jx}\brktc{\Jy}\!\right)\!dt\!-\! 2\kloc \Cxy^{\cc} dt \!-\! \frac{1}{2} M \Cxy^{\cc} \left( 1 + 8\eta \Vy^{\cc}\right)\!dt \label{eq:dCxy}\\
   &d\omega \!=\! 0,
\end{align}
\end{subequations}
\end{widetext}
where in Eqs.~(\ref{eq:dVx}-\ref{eq:dCxy}) we importantly ignore all the (stochastic) contributions that involve the third-order moments, which can be found in \appref{ap:Ito}. 

In what follows, we simulate the exact dynamics \eref{eq:SME} of the density matrix for low values of $N$ to verify that Eqs.~\eref{eq:dynamical_model} correctly describe the evolution of the lowest moments for modest values of decoherence and measurement-strength parameters. As we observe the agreement to improve with the atomic number at short timescales (see \appref{ap:verify_CoG} for further details) for the experimentally relevant regimes of large $N\approx10^{5}\!-\!10^{13}$~\cite{Kuzmich2000,Wasilewski2010,Shah2010,Koschorreck2010,Sewell2012,Kong2020}, we subsequently use Eqs.~\eref{eq:dynamical_model} to simulate the dynamics of the atomic sensor with sufficient accuracy. 

Crucially, regardless of the size of the ensemble, we construct the EKF based on the nonlinear dynamical model \eref{eq:dynamical_model}. The output of the filter provides us with real-time estimates of dynamical parameters, i.e.~of $\vec{x}(t) = ( \brktc{\Jx},\brktc{\Jy},\Vx^{\cc},\Vy^{\cc},\Vz^{\cc},\Cxy^{\cc},\omega )^\TT$. In turn, we use these estimates to devise the control strategy determining $u(t)$ employed in either \eqnref{eq:SME} or Eqs.~\eref{eq:dynamical_model}.

%%%%%%%%%%%%%%%%%%%%%%%%%%%%%%%%%%%%%%%%%%%%%%%%%%%%%%%%%%%%%%%%%%%%%%%%%%%%%%%%%%%%%%%%
%%%%%%%%%%%%%%%%%%%%%%%%%%%%%%%%%%%%%%%%%%%%%%%%%%%%%%%%%%%%%%%%%%%%%%%%%%%%%%%%%%%%%%%%
\section{Ultimate Limits on Precision} 
\label{sec:bounds}
With an established scalable method for simulating the system, our attention now turns to one of the fundamental questions in atomic magnetometry: how to most accurately infer the true value of the Larmor frequency $\omega$ for a particular measurement record $\vec{y}_{\leq t}$. With the photocurrent being continuously acquired, employing a Bayesian approach to estimation is apt in this scenario, offering a systematic way of continually updating our knowledge of the parameter as new data become available. 

Typically, in Bayesian estimation theory we seek an optimal estimator $\est{\omega}_t (\vec{y}_{\leq t})$ of $\omega$ that minimises the average mean squared error (AMSE)~\cite{Van-Trees,Teklu2009},
\begin{align} 
		\label{eq:AMSE}   
		\EE{\Delta^2 \est{\omega}_t}
    & \coloneqq \E{\left(\est{\omega}_t(\vec{y}_{\leq t})-\omega\right)^2}{p(\vec{y}_{\leq t},\omega)}\\
    & =  \int\!\! d\omega \, p(\omega) \! \int\!\!\D\vec{y}_{\leq t} \; p(\vec{y}_{\leq t}|\omega) \left(\est{\omega}_t (\vec{y}_{\leq t})-\omega\right)^2 \nonumber,
\end{align}
where the averaging $\EE{\cdot}$ is performed over all measurement trajectories up to time $t$, $\int\!\!\D\vec{y}_{\leq t}$, and also over all possible values of the estimated parameter, $\int\!\! d\omega$. The \emph{prior distribution} $p(\omega)$ in \eqnref{eq:AMSE} represents our knowledge of $\omega$ before collecting any measurement data, while the likelihood $p(\vec{y}_{\leq t}|\omega)$ is the probability of observing a measurement record $\vec{y}_{\le t}$ given the parameter value $\omega$. The optimal estimator minimising the AMSE is generally given by the mean of the posterior distribution~\cite{Van-Trees,Teklu2009}: 
\begin{equation}
	\label{eq:mMSEest}
  \est{\omega}_t^\mrm{opt}(\vec{y}_{\le t}) = \E{\omega}{p(\omega|\vec{y}_{\leq t})}= \int \!d\omega\; \omega \, p(\omega|\vec{y}_{\le t}).
\end{equation}

Constructing the posterior distribution $p(\omega|\vec{y}_{\le t})$ is a hard task. However, in the case of systems with linear dynamics and additive Gaussian noise, the posterior does not have to be explicitly reconstructed since the optimal estimator \eref{eq:mMSEest} is given by the \emph{Kalman filter} (KF)~\cite{kalman_new_1960,kalman_new_1961}. For nonlinear systems, other methods exist, such as the EKF~\cite{simon2006} or the unscented Kalman filter~\cite{julier1995,julier2000}, that allow one to efficiently tackle the problem, but do not guarantee optimality.

%%%%%%%%%%%%%%%%%%%%%%%%%%%%%%%%%%%%%%%%%%%%%%%%%%%%%%%%%%%%%%%%%%%%%%%%%%%%%%%%%%%%%%%%
\subsection{Noiseless performance at small timescales}
In the case of our system, the LG regime is exactly the scenario in which the KF provides the optimal estimation strategy. For instance, in the absence of noise, i.e.~for $\kcoll = 0$ and $\kloc = 0$ in \eqnref{eq:SME}, the AMSE of the KF (estimator) then follows the Heisenberg scaling in $N$ and a superclassical scaling in time~\cite{Geremia2003,Amoros-Binefa2021}:
\begin{equation}
    \EE{\Delta^2 \est{\omega}_t} = \frac{1}{N^2} \frac{3}{\eta M t^3} \; \;  \textrm{for} \; \; t \ll (N M)^{-1}.
    \label{eq:Heisenberg}
\end{equation}
However, both of the aforementioned scalings should be taken with a pinch of salt, as they rely on the assumption of $t$ being small enough for the LG regime to be valid. In particular, the scaling in time must eventually become at most quadratic due to the Hamiltonian being bounded in \eqnref{eq:SME}~\cite{Kurdzialek2023}. Furthermore, by ignoring quantum fluctuations in the $\hat{J}_x$ direction, we disregard the fact that the optimal time $t$ actually decreases with $N$ in \eqnref{eq:Heisenberg}, which puts the Heisenberg scaling into question~\cite{Caprotti2023}.

Nonetheless, the emergence of superclassical scalings $N^2$ and $t^3$ in \eqnref{eq:Heisenberg} is a manifestation of generating conditional spin squeezing~\cite{Ma2011} at short timescales and, hence, the ensemble then exhibits interatomic entanglement~\cite{Pezze2018RMP}. In particular, as depicted in \fref{fig:CSS_bloch}, the continuous measurement of $\Jy$ ($\hat{X}$ quadrature in the LG plane) squeezes its variance in detriment of the variance of $\Jz$ ($\hat{P}$ quadrature in the LG plane) as the time evolves. Since our interest is in preparing a state highly sensitive to small variations of $\omega$, we wish for it to have a maximal polarisation along $x$ and maximal squeezing along $y$. How closely our state aligns with this particular geometry is quantified by the (Wineland) squeezing parameter $\xi_{y}^2$~\cite{Kitagaba_Ueda_SSS,Wineland1994}, which effectively compares any state with a CSS pointing along $x$. Hence, we define its inverse as the relevant \emph{spin-squeezing parameter}~\cite{Ma2011}, i.e.,
\begin{equation}
    \xi^{-2}_{y}(t) \coloneqq
    \left(
    \!\left.\frac{\Vy^{\cc}(t)}{\brktc{\Jx(t)}^2}
    \right/
    \frac{\Vy^\CSS}{\brkt{\Jx}_\CSS^2}
    \right)^{-1} = \frac{\brktc{\Jx(t)}^2}{N\;\Vy^{\cc}(t)},
    \label{eq:spin_squeez_par}
\end{equation}
which guarantees spin squeezing at time $t$ when $\xi^{-2}_{y}(t) > 1$, whereas for $\xi^{-2}_{y}(t) \le 1$ any spin squeezing, and hence any multiparticle entanglement~\cite{Pezze2018RMP}, cannot be certified. 

In experiments, $\xi^{-2}\approx10$(10dB) and $\xi^{-2}\approx100$(20dB) have been achieved with magnetically sensitive~\cite{Bohnet2014} and atomic-clock~\cite{hosten_measurement_2016} states, respectively, for an ensemble of $N\approx10^5$ rubidium atoms by conducting cavity-enhanced premeasurements. Recently, $\xi^{-2}\approx2.8$(4.5dB) was demonstrated for $N\approx10^{11}$ by also taking into account measurements conducted after the sensing phase of the protocol (retrodiction)~\cite{bao_spin_2020}.

%%%%%%%%%%%%%%%%%%%%%%%%%%%%%%%%%%%%%%%%%%%%%%%%%%%%%%%%%%%%%%%%%%%%%%%%%%%%%%%%%%%%%%%%
\subsection{Noisy bounds}
When moving away from the LG regime, the optimality of the estimation method, e.g.~of the EKF, cannot be assured. However, the AMSE \eref{eq:AMSE} can \emph{always} be lower bounded by the Bayesian Cram\'{e}r-Rao bound~\cite{Van-Trees,Fritsche2014}:
\begin{equation} \label{eq:BCRB}
    \EE{\Delta^2 \est{\omega}_t } \geq \frac{1}{\F[p(\omega)] + \int\! d\omega\, p(\omega)\, \F[p(\vec{y}_{\le t}|\omega)]}
\end{equation}
where $\F[ \; \cdot \;]$ is the Fisher information computed with respect to the estimated parameter $\omega$, see \appref{ap:CSbound}. The bound is dictated by two distinct contributions:~one coming from our prior knowledge about the parameter, and the other associated with the information about the parameter contained within the measured data. Importantly, as bound \eref{eq:BCRB} always applies for a given measurement scheme determining $p(\vec{y}_{\le t}|\omega)$, it proves the optimality of the estimation strategy considered when saturated. 

Nonetheless, both \eqnsref{eq:AMSE}{eq:BCRB} still depend on a particular choice of the measurement scheme. Hence, in order to construct a benchmark applicable in any scenario, we determine a further lower bound on AMSE \eref{eq:AMSE} that is \emph{independent} of both the estimation method and the measurement strategy. In particular, the presence of decoherence allows us to derive such a bound in \appref{ap:CSbound}, which for a Gaussian prior distribution, $p(\omega)=\mathcal{N}(\mu_0,\sigma_0)$, reads:
\begin{equation}
    \EE{\Delta^2 \est{\omega}_t }  
    \geq
    \frac{1}{\frac{1}{\sigma_0^2} + \left(\dfrac{\kcoll}{t} + \dfrac{2\kloc}{t N}\right)^{-1}}
    \underset{\sigma_0\to\infty}{\geq} \frac{\kcoll}{t} + \frac{2\kloc}{N t}.  
    \label{eq:CSlimit}
\end{equation} 
Bound \eref{eq:CSlimit}---that we refer to as the \emph{classical simulation} (CS) \emph{limit} following Refs.~\cite{Amoros-Binefa2021,matsumoto_metric_2010,Demkowicz2012}---applies at \emph{any timescale}, consistently vanishing when $\kcoll=\kloc=0$, i.e.~in absence of noise. Otherwise, it holds for \emph{any measurement-based feedback strategy}, independently of the initial state of the system, or the form of the measurements (also adaptive) involved;~see \appref{ap:CSbound} and \citeref{Amoros-Binefa2021}.

As a consequence, the CS limit \eref{eq:CSlimit} directly disproves the possibility of attaining the superclassical scalings of $N^2$ and $t^3$ in the presence of decoherence. In particular, the first term in \eqnref{eq:CSlimit} sets an $N$-independent bound dictated by the collective decoherence~\cite{Amoros-Binefa2021}, while the second term arising from the local noise follows the SQL of $1/Nt$---leaving room only for a constant-factor quantum enhancement~\cite{Demkowicz2012}. The latter observation unfortunately disproves the conjecture about breaching the SQL-like scaling in $N$ despite local dephasing, formulated in \refcite{rossi_noisy_2020} based on numerical evidence.

%%%%%%%%%%%%%%%%%%%%%%%%%%%%%%%%%%%%%%%%%%%%%%%%%%%%%%%%%%%%%%%%%%%%%%%%%%%%%%%%%%%%%%%%
%%%%%%%%%%%%%%%%%%%%%%%%%%%%%%%%%%%%%%%%%%%%%%%%%%%%%%%%%%%%%%%%%%%%%%%%%%%%%%%%%%%%%%%%
\section{Estimation and control} 
\label{sec:estimation_and_control}
With a universal lower bound established for the AMSE, let us propose the estimation and control strategies that we anticipate to yield the lowest possible estimation error, while remaining feasible for implementation.

A natural choice of an estimator tailored to the nonlinear Gaussian dynamical model derived in Eqs.~\eref{eq:dynamical_model} is the EKF~\cite{crassidis2011optimal,simon2006}. However, even though the CoG approximation accounts for the coprecession of the LG plane with the mean angular-momentum vector $\brkt{\Jqvec{J}(t)}$, the measurement direction is physically fixed to $y$ and cannot be varied, so that, e.g., the stochastic term in \eqnref{eq:dJy} is always determined by $\Vy$. That is why, the principal aim of the measurement-based feedback that we introduce is to keep $\brkt{\Jqvec{J}(t)}$ pointing along its initial $x$ direction, so that the measurement may induce squeezing perpendicularly to $\brkt{\Jqvec{J}(t)}$ at all times, prolonging the LG regime of \figref{fig:CSS_bloch}. 

For this purpose, we use the LQR to find the control law, which we expect to be optimal in the LG regime~\cite{Stockton2004}. Within our scheme, the control field $u(t)$ provided by the LQR is built from the estimates of the EKF, unlike other measurement-based control strategies that rely on feeding back directly photocurrent \eref{eq:photocurrent}~\cite{wiseman1994_Feedback,Thomsen2002}.

%%%%%%%%%%%%%%%%%%%%%%%%%%%%%%%%%%%%%%%%%%%%%%%%%%%%%%%%%%%%%%%%%%%%%%%%%%%%%%%%%%%%%%%%
\subsection{Estimator:~Extended Kalman Filter}
\label{sec:EKF}
Within the CoG approximation, the ensemble dynamics is completely described by a vector of dynamical parameters, $\vec{x}(t) = ( \brktc{\Jx},\brktc{\Jy},\Vx^{\cc},\Vy^{\cc},\Vz^{\cc},\Cxy^{\cc},\omega )^\TT$ appearing in Eqs.~\eref{eq:dynamical_model}, referred to as the \emph{state} in estimation theory~\cite{crassidis2011optimal}, which evolves according to a system of coupled nonlinear stochastic equations of the form
\begin{align}
		\label{eq:state_dyn}
    \dot{\vec{x}}(t) &= \vec{f}[\vec{x}(t),u(t),\vec{\xi},t],
\end{align}
with function $\vec{f}$ determined by the dynamical model \eref{eq:dynamical_model}, and $\vec{\xi}$ denoting a vector of independent Langevin-noise terms---here, $\vec{\xi} = (\xi, 0)^\TT$ with the Wiener increment in Eqs.~\eref{eq:dynamical_model} then corresponding to $dW=\xi dt$~\cite{Gardiner1985}.

Additionally, the \emph{observation} of the true state $\vec{x}$ is performed according to the measurement model \eref{eq:photocurrent}, which can be conveniently written as
\begin{align}
    y(t) &= h[\vec{x}(t),\zeta,t] = H \, \vec{x}(t) + \zeta,
    \label{eq:observation}
\end{align}
where a general $h$ function is linear in $\vec{x}$ for the case of \eqnref{eq:photocurrent}, with $H=2\eta\sqrt{M}(0,1,0,0,0,0,0)$. Moreover, we must now impose the condition that $\zeta = \sqrt{\eta}\,\xi$ in the quantum setting, as the observation noise $\zeta$ is correlated with the state noise $\xi$ due to the quantum backaction~\cite{Belavkin1989}.

Let us denote by $\est{\vec{x}}(t)$ the \emph{EKF estimator} of the state $\vec{x}(t)$ at time $t$, and its corresponding error matrix by $\EE{\Delta^2 \est{\vec{x}}(t)} \coloneqq \E{(\est{\vec{x}}(t) - \vec{x}(t))(\est{\vec{x}}(t) - \vec{x}(t))^\TT}{p(\vec{y}_{\leq t},\vec{x}(0))}$. Although the latter can in principle be computed only when having access to the true state dynamics, the EKF also provides its estimate for the error matrix, which we refer to as the \emph{EKF covariance} $\Sigma(t)$. Initially setting at $t=0$---prior to taking any measurements---$\est{\vec{x}}(0)$ and $\Sigma(0)=\EE{\Delta^2 \est{\vec{x}}(0)}$ to be the mean and covariance of the prior distribution for the state, respectively, the EKF estimator is found by simultaneously integrating the following differential equations along a particular photocurrent record $\vec{y}_{\le t} = \{y(\tau)\}_{0\leq \tau \leq t}$, i.e.~\cite{simon2006}:
\begin{subequations}
\label{eq:EKF_alg}
	\begin{align}
	    \dot{\est{\vec{x}}} &= \vec{f}[\est{\vec{x}},u,0,t] + K (y(t)-h[\est{\vec{x}},0,t]) 
	    \label{eq:EKF_dyn}\\
	    \dot{\Sigma} &= (F - G S R^{-1} H)\Sigma + \Sigma(F - G S R^{-1} H)^\TT + \nonumber\\
	    & \; \; \; \, + G(Q - S R^{-1} S^\TT) G^\TT - \Sigma H^\TT R^{-1} H \Sigma,
	    \label{eq:Sigma_dyn}
	\end{align}
\end{subequations}
which are coupled via the Kalman gain $K \coloneqq (\Sigma H^\TT - G S) R^{-1}$, whose explicit $t$ dependence we drop above, similarly to the dynamical matrices $F(t)$ and $G(t)$.

The matrices $Q \coloneqq \E{\vec{\xi} \, \vec{\xi}^\TT}{} =(1, 0 \, ; 0, 0)$, $R \coloneqq \E{\zeta^2}{} = \eta$ and $S \coloneqq \E{\vec{\xi} \zeta}{}= (\sqrt{\eta},\, 0)^\TT$ that appear in the Riccati equation \eref{eq:Sigma_dyn} (and in the Kalman gain $K$) correspond to the covariance and correlation matrices of the noise vectors and, importantly, are predetermined. Moreover, the dynamical matrices $F(t) \coloneqq \nabla_{\vec{x}} \vec{f}|_{(\est{\vec{x}},u,0)}$ and $G(t) \coloneqq \nabla_\xi \vec{f} |_{\est{\vec{x}}}$ (and $H \coloneqq \nabla_{\vec{x}} h$~%
\footnote{Within our observation model \eref{eq:observation}, the function $h$ is linear, so that $H$ is time-invariant.}) 
are defined as the Jacobian matrices of function $\vec{f}$ (and $h$), whose symbolic form can normally be precomputed---as done in \appref{ap:EKF} for the dynamical model \eref{eq:dynamical_model}. However, as these are evaluated at, and hence depend on, the current value of the EKF estimator $\est{\vec{x}}(t)$, their exact (numerical) forms must be reevaluated at each step of the EKF algorithm \eref{eq:EKF_alg}. 

This stands in stark contrast to the special case of a linear model, i.e.~when both $\vec{f}$ and $h$ are linear in $\vec{x}(t)$, so that all $F$, $G$ and $H$ become independent of $\est{\vec{x}}(t)$. As a result, the Ricatti equation \eref{eq:Sigma_dyn} can be solved independently of \eqnref{eq:EKF_dyn}, i.e.~prior to taking any measurements. This is the special scenario in which the EKF estimator consistently simplifies to the KF, with its covariance guaranteed to coincide with the true error at all times, i.e.~$\Sigma(t)=\EE{\Delta^2 \est{\vec{x}}(t)}$~\cite{simon2006,crassidis2011optimal}.

% As this does not hold for nonlinear dynamical models such as model \eref{eq:dynamical_model}, we have to simulate the dynamics of the atomic sensor in order to have access to the true state $\vec{x}(t)$ at all times. As a result, we can then explicitly compute the error matrix $\EE{\Delta^2 \est{\vec{x}}(t)}$ by averaging over sufficiently many measurement records. By then inspecting its diagonal entries, $\EE{\Delta^2 \est{\vec{x}}(t)}_{\ell\ell}$, we obtain AMSEs for estimating the conditional means, (co)variances, and Larmor frequency, as appearing in Eqs.~\eref{eq:dynamical_model}, i.e.,
\begin{equation}
\EE{\Delta^2 \est{O}}\coloneqq \E{\est{O}_{\cc}(t)-O_{\cc}(t)}{p(\vec{y}_{\leq t},\hat{O}(0))},
\label{eq:true_errors}
\end{equation}
where $\est{O} \in \{\brkt{\estJx},\!\brkt{\estJy},\!\estVx,\!\estVy, \!\estVz, \!\est{C}_{xy},\est{\omega}\}$.

%%%%%%%%%%%%%%%%%%%%%%%%%%%%%%%%%%%%%%%%%%%%%%%%%%%%%%%%%%%%%%%%%%%%%%%%%%%%%%%%%%%%%%%%
\subsection{Controller:~Linear Quadratic Regulator}
\label{sec:LQR}
As motivated at the beginning of this section, a naive control strategy---that we refer to as \emph{field compensation}---would be to just feed back the EKF estimate of the Larmor frequency, i.e.~set $u(t) = - \est{\omega}(t)$ in Eqs.~\eref{eq:dynamical_model}, which should simply cancel the Larmor precession.

However, as will become clear below (see e.g.~\figref{fig:sq} in \secref{sec:squeezing}), such a solution is unstable due to the estimate of the Larmor frequency, $\est{\omega}(t)\approx\omega$, being only approximate. This leads to an error in compensating for the precession that accumulates over time. That is why, we resort to LQR-theory, allowing us to construct a stable control law that is further guaranteed to be optimal in the LG regime.

In particular, we focus on the LG regime in which it is sufficient to describe the system by only two (rather than seven in $\vec{x}(t)$) dynamical parameters, i.e.~by state $\vec{z}(t) = (\brkt{\Jy}, \omega)^\TT$, which evolves under a linearised version of \eqnref{eq:state_dyn} obtained by further approximating the dynamical model \eref{eq:dynamical_model} at short timescales~\cite{Geremia2003,Amoros-Binefa2021}:
\begin{align} 
\label{eq:l&g_system}
    \dot{\vec{z}}(t) = A \, \vec{z}(t) + B \, u(t) + \sigma(t) \, \vec{\xi},
\end{align} 
where now $A \coloneqq (0 , J; 0, 0)$, $B \coloneqq (J, 0)^\TT$ and $\sigma(t) \coloneqq (2\sqrt{\eta M} \Vy^{\cc} ,0;0,1)$, with $\vec{\xi} = (\xi, 0)^\TT$ being the same stochastic term as in \eqnref{eq:state_dyn}, such that $dW=\xi dt$. Note that for this LG system, the variance of $\Jy$, $\Vy^{\cc}$, is a deterministic function with an analytical form~\cite{Stockton2004,Amoros-Binefa2021}.

Then, the LQR corresponds to the form of $u(t)$ that \emph{linearly} depends on the state vector, here $\vec{z}(t)$, while minimising a given \emph{quadratic} cost function~\cite{crassidis2011optimal}:
\begin{align}
    I(u) &= \int_0^\infty\!\!dt\;
    \left[\vec{z}^\TT(t) P \vec{z}(t) + u(t) V u(t) \right]
    \label{eq:quad_cost} \\
    &=  \int_0^\infty \!\!dt \; 
    \left[p_J \brktc{\Jy}^2 + p_\omega \, \omega^2 + \nu \, u^2(t) \right].
    \label{eq:quad_cost_ex} 
\end{align}
Here, following \refcite{Stockton2004}, we have already chosen the (time-independent) cost matrices $P\ge0$ and $V>0$ to take a diagonal form, $P = (p_J, 0 ; 0 , p_\omega)$ and $V = \nu$ with $p_J,p_\omega\ge0$ and $\nu>0$. For such a choice, it becomes clear from \eqnref{eq:quad_cost_ex} that the LQR, which minimises $I(u)$, not only counteracts the Larmor precession by compensating for $\omega>0$, but also importantly aims at zeroing the angular-momentum component $\brktc{\Jy}$ at any time.

\begin{figure*}[t!]
    \centering
    \includegraphics[width=\textwidth]{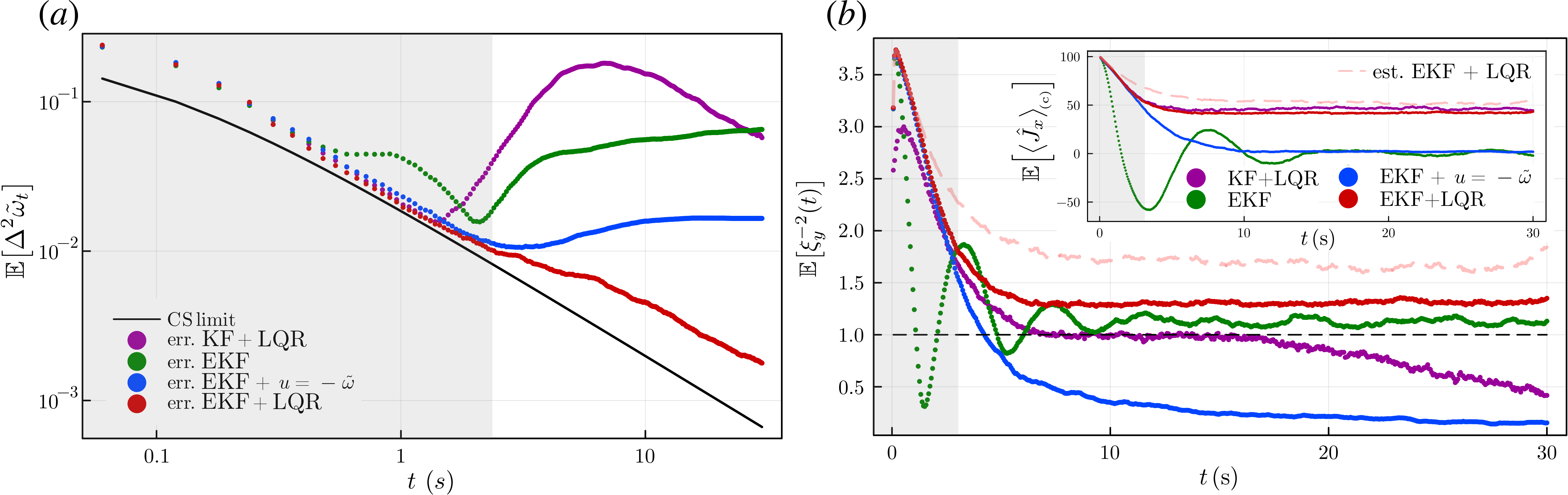}
    \caption{\textbf{Performance of different estimation and control strategies.} \emph{Subplot (a)} presents the evolution of the estimation error (AMSE) for different estimation+control strategies that involve either KF (purple) or EKF (all other) as estimators, and either no feedback (green), field compensation (blue) or the LQR (red) as feedback methods. These strategies are compared against the CS limit (black), which sets the ultimate bound on the attainable error. The EKF+LQR strategy outperforms all other schemes and maintains a decreasing error trend even beyond the LG regime (shaded grey area). \emph{Subplot (b)} shows the dynamics of the spin-squeezing parameter \eref{eq:spin_squeez_par} for all estimation+control strategies, while its \emph{inset} depicts the corresponding evolutions of the spin polarisation $\brkt{\Jx}$ (all with consistent colouring). For the EKF+LQR strategy, both also include the values predicted by the EKF (pink dashed line), which are overoptimistic. The parameters used in SME \eref{eq:SME} for simulations are $N = 200$, $\kcoll = 0.02$, $\kloc = 0$, $M = 0.3$, $\omega = 1$ and $\eta = 1$. The KF and EKF estimators are initialised with the mean $\est{\vec{x}}(0) = (N/2,0,0,N/4,N/4,0,\mu_0)^\TT$ and covariance $\Sigma(0) = \Diag{0,0,0,0,0,0,\sigma_{0}^2}$ dictated by the initial CSS state of the atoms, and the Gaussian prior distribution for $\omega\sim\mathcal{N}(\mu_0,\sigma_0^2)$. All results are obtained after averaging over $\nu = 1000$ measurement trajectories, whereas $\omega$-averaging is avoided by choosing its true value $\omega=1$ for a prior with $\mu_0 = \omega + \sigma_0=1.5$ and $\sigma_0=0.5$.}
    \label{fig:err}
\end{figure*}

Crucially, thanks to the dynamics \eref{eq:l&g_system} being linear, the LQR minimising \eqnref{eq:quad_cost} can be generally found by ignoring the stochastic part $\vec{\xi}$ in \eqnref{eq:l&g_system}, which only increases the attainable minimal cost \eref{eq:quad_cost} (on average)~\cite{crassidis2011optimal}. Moreover, given also a linear observation model, e.g.~\eqnref{eq:observation}, the optimal control problem can be solved independently to the state estimation task~\cite{crassidis2011optimal}. In particular, the LQR solution is then given by:
\begin{align}
    u(t) &= -K_C \, \est{\vec{z}}(t), \label{eq:LQR_control_law}\\
    K_C &\coloneqq V^{-1} B^\TT \Sigma_C , \label{eq:gain_matrix}\\
    0 &= A^\TT \Sigma_C + \Sigma_C A + P - \Sigma_C B V^{-1} B^\TT \Sigma_C \label{eq:control_Riccati}
\end{align}
where the optimal control field $u(t)$ is linearly related at any time to the state estimator, $\est{\vec{z}}(t)$ (i.e.~the KF for state \eref{eq:l&g_system} and observation \eref{eq:observation} dynamics), by the gain (matrix) $K_C$. The gain is defined in \eqnref{eq:gain_matrix} and involves the solution of the algebraic Riccati equation \eref{eq:control_Riccati} for the matrix $\Sigma_C$. Now, as the matrices $A$ and $B$ in \eqnref{eq:l&g_system} are time-independent, all $\Sigma_C$, $K_C$, as well as the LQR, can be determined analytically~\cite{Stockton2004}. In particular, the LQR in our case reads
\begin{equation} 
	\label{eq:LQR}
    u(t) = - \est{\omega}(t) - \lambda \;\brktc{\estJy(t)},
\end{equation}
where $\lambda \coloneqq \sqrt{p_J/\nu}$ is a constant parameter that should be appropriately chosen. Note that by letting $\lambda=0$ we recover the (naive) field compensation strategy. 

Furthermore, as in what follows we will use the LQR \eref{eq:LQR} also beyond the LG regime---in particular, in the CoG regime in which the EKF is used to estimate the full state $\vec{x}(t) = ( \brktc{\Jx},\brktc{\Jy},\Vx^{\cc},\Vy^{\cc},\Vz^{\cc},\Cxy^{\cc},\omega )^\TT$---we generalise the control law \eref{eq:LQR_control_law} to read $u(t) = -K_C \,\Xi\, \est{\vec{x}}(t)$, where $\Xi\coloneqq(0,1,0,\dots,0,0;0,0,0\dots,0,1)$ just selects the relevant state components of $\vec{x}(t)$ that appear in $\vec{z}(t)$, while $\est{\vec{x}}(t)$ is now the EKF estimator in \eqnref{eq:EKF_dyn}.

%%%%%%%%%%%%%%%%%%%%%%%%%%%%%%%%%%%%%%%%%%%%%%%%%%%%%%%%%%%%%%%%%%%%%%%%%%%%%%%%%%%%%%%%
%%%%%%%%%%%%%%%%%%%%%%%%%%%%%%%%%%%%%%%%%%%%%%%%%%%%%%%%%%%%%%%%%%%%%%%%%%%%%%%%%%%%%%%%
\section{Real-time sensing performance} 
\label{sec:performance}
We benchmark the performance of the EKF+LQR strategy by demonstrating its superiority over other estimation+control strategies that involve less sophisticated inference (KF rather than EKF) and feedback (field compensation rather than LQR) methods. Furthermore, we verify whether and at what timescales the CS limit \eref{eq:CSlimit} (induced by the global and/or local decoherence) can be attained---then proving the complete sensing scheme to be optimal, i.e., being optimised over not only the estimation+control strategy, but also the initial atomic state and any measurements involved.

In order to do so, we focus on identifying the evolution in time of the AMSE \eref{eq:AMSE}, $\EE{\Delta^2 \est{\omega}_t}$;~see, e.g., \figref{fig:err}(a). However, in order to simultaneously monitor the dynamics of quantum correlations and coherence exhibited by the atomic ensemble, we also investigate the evolution of the spin-squeezing parameter \eref{eq:spin_squeez_par}, $\EE{\xi^{-2}_{y}(t)}$, as well as the ensemble polarisation, $\EE{\brktc{\Jx}}$;~see the main and inset plots in \figref{fig:err}(b), respectively.

Importantly, as the performance must be quantified on average, all the three quantities have to be averaged, $\EE{\cdot}$, over sufficiently many measurement trajectories obtained when simulating the dynamics~\footnote{For polarisation, despite $\EE{\brktc{\Jx}}=\brkt{\Jx}$, the unconditional dynamics of $\rho(t)$ cannot be simply determined from \eqnref{eq:SME} due to the presence of the control $u(t)$.}. Furthermore, AMSE \eref{eq:AMSE} must be averaged over the prior distribution $p(\omega)$, which represents our \emph{a priori} knowledge about the Larmor frequency. However, in order to reduce the number of trajectories computed and improve the clarity of the presented plots, we avoid averaging over $p(\omega)$, but rather present measurement-trajectory averages for a fixed parameter value that is representative of the assumed prior, i.e.~for $\omega = \mu \pm \sigma_0$ given a Gaussian prior $p(\omega)=\mathcal{N}(\mu_0,\sigma_0^2)$. For such an educated choice, the AMSE is consistently always greater than the CS limit \eref{eq:CSlimit} evaluated for the given $\sigma_0>0$ [see e.g.~\figref{fig:err}(a)] that, however, is always valid on average~%
\footnote{Choosing, e.g., $\omega=\mu_0$ would yield zero error at $t=0$ and naively suggest the CS limit \eref{eq:CSlimit} to be then ``breached'', what cannot happen when explicitly averaging over $p(\omega)$.}.

In what follows, we firstly focus on relatively low atomic numbers, $N$, for which we can explicitly simulate the true dynamics of the atomic ensemble along a particular measurement trajectory, as described in \secref{sec:exact_dyn}. This allows us to demonstrate the superiority of the EKF+LQR strategy without making any approximations. However, as we observe and confirm (see \appref{ap:verify_CoG}) that, with an increase of $N$, the evolution of atoms can be well described at short timescales by the CoG approximation of \secref{sec:CoG_dyn}, we then use it not only to construct the EKF, but also for simulations. As a result, we are able to consider an experimentally relevant number of atoms, e.g.~$N=10^5$~\cite{Bohnet2014,hosten_measurement_2016} used in \figref{fig:attaining_bounds} below, for which we may in detail demonstrate the optimality of the EKF+LQR strategy by saturating the CS limit \eref{eq:CSlimit}. What is more, we show large spin squeezing to then be generated and long maintained despite decoherence---also for the \emph{unconditional dynamics} thanks to the feedback (LQR), i.e., when averaging over the measurement trajectories.

%%%%%%%%%%%%%%%%%%%%%%%%%%%%%%%%%%%%%%%%%%%%%%%%%%%%%%%%%%%%%%%%%%%%%%%%%%%%%%%%%%%%%%%%
\subsection{Identifying the best estimation and control strategy}
\label{sec:numerical_results}
In \fref{fig:err}(a) we first compare the AMSE of the frequency estimate for four different estimation+control strategies when only the collective decoherence is present ($\kcoll>0$, $\kloc=0$). This allows us to also consider the simplest estimation strategy based on the KF and linearised dynamics \eref{eq:l&g_system}~\cite{Geremia2003,Amoros-Binefa2021}, which is not applicable as soon as $\kloc>0$ in Eqs.~\eref{eq:dynamical_model}. In particular, we compare the EKF with no control (in green), the EKF with field compensation (denoted by $u(t)= -\est{\omega}(t)$, in blue), EKF+LQR ($u(t)$ as in \eqnref{eq:LQR} with $\lambda = 1$, in red), and the (linearised) KF combined with the LQR (in purple). As evident from \figref{fig:err}(a), the EKF+LQR approach consistently outperforms all other strategies. Moreover, the results highlight the importance of using an estimator (EKF) that can handle nonlinearities in the system, rather than a linear one (KF). Additionally, they stress the necessity of devising an appropriate feedback strategy following the principles of the LQR optimal-control theory.

Additional figures are presented in \figref{fig:err}(b), in order to show that the EKF+LQR strategy (red) is the only one that keeps the ensemble both spin squeezed and polarised along $x$ (main and inset, respectively) significantly beyond the LG regime ($t \!\lesssim\!(M+\kcoll)^{-1}$~\cite{Amoros-Binefa2021}, shaded in grey). When no control is applied (green), the atomic state is still squeezed on average $\EE{\xi_y^{-2}(t)}>1$, but quickly depolarises with the precession. When attempting to cancel the precession with just the estimate of the frequency, $u= -\est{\omega}$ (blue), both the polarisation and spin squeezing \eref{eq:spin_squeez_par} are rapidly lost. Controlling the sensor with an EKF+LQR instead of using $u = -\est{\omega}$ achieves an ideal outcome, preserving both spin squeezing and polarisation well past the coherence time $1/\kcoll$ at about $\EE{\xi_y^{-2}}\approx1.25$(0.97dB), and extending all the way to $t=30$, as shown in \figref{fig:err}(b). Correct estimation is also crucial;~employing a KF (purple) instead of an EKF, even combined with the best feedback strategy (LQR), results in a worse performance in terms of spin squeezing, although polarisation is maintained.

\subsubsection{Using the EKF to estimate spin squeezing}
A key advantage of using Kalman filtering techniques, or more generally Bayesian inference~\cite{Van-Trees}, over, e.g., model-free machine-learning methods~\cite{Carleo2019,Dawid2023,Duan_2025}, is the fact that the former provide errors for their estimates, which are accurate as long as the dynamical model can be trusted. As noted in \secref{sec:EKF}, this is the case for the KF when LG stochastic models are considered, for which the KF's covariance is assured to represent the true average error, i.e.~$\Sigma(t)=\EE{\Delta^2 \est{\vec{x}}(t)}$~\cite{simon2006,crassidis2011optimal}. For quantum LG models without feedback, the KF can thus be directly used to reconstruct conditional dynamics of any (Gaussian) quantum observable $\hat{X}(t)$~\cite{wieczorek_optimal_2015}, even when only the \emph{unconditional} dynamics is available due to, e.g., the model of measurement backaction being unknown~\cite{Kong2020}. 

In our case, to only have access to the unconditional dynamics of the system would mean having an unconditional evolution dictated by \eqnref{eq:SME} without any feedback (since $u(t)$ is trajectory dependent, i.e.~$u(t) \equiv u(t|\vec{y}_{\leq t})$) and with the last $dW$-dependent term dropped. Then, the term $\brktc{\hat{J}_y(t)}$ in the detection model \eref{eq:photocurrent} should be reinterpreted as $J_y(t)\sim\mathcal{N}(\brkt{\Jy(t)},\Delta^2\Jy(t))$, so that it represents the particular value of $\hat{J}_y(t)$ occurring at time $t$---being drawn from the corresponding unconditional (Gaussian) Wigner function. 

For any such unconditional dynamics, as long as it is LG, after initialising the KF to $\est{x}(0)=\brkt{\hat{X}(0)}$ and $\Sigma_{xx}(0)=\Delta^2 \hat{X}(0)$, the KF directly provides us in real time with $\est{x}(t)=\brktc{\hat{X}(t)}$ and $\Sigma_{xx}(t)=\Delta^2_{\cc}\hat{X}$ based on the measurement data being recorded~%
\footnote{
	However, in case of KF and (quantum) LG dynamics, its covariance $\Sigma(t)$ does not depend on a particular measurement trajectory and, hence, can be computed without inspecting it.
}, i.e.~the mean and variance, respectively, of the conditional (Gaussian) Wigner function correctly describing $\hat{X}(t)$ given the measurement record~\cite{Tsang2009,Zhang2017}. Hence, as covariances of the KF then represent conditional variances of quantum observables, these can be directly used to, e.g., certify entanglement in QND-based experiments~\cite{Kong2020}.

In contrast, as we possess an explicit model of the \emph{conditional} dynamics \eref{eq:SME}, its solution (in the Heisenberg picture) for any moment of a quantum observable already correctly accounts for the measurement record observed and, in our case, a given strategy of the measurement-based feedback)~\cite{handel_modelling_2005}. Such moments, in particular $\brktc{\hat{J}_\alpha(t)}$ and $\Delta^2_{\cc}\hat{J}_\alpha(t)\equiv\mrm{V}_\alpha^{\cc}$, then constitute (classical) dynamical parameters that can be tracked in real time, in the same way as the Larmor frequency $\omega$. Thus, one should view the CoG model \eref{eq:dynamical_model} as a nonlinear approximation of SME \eref{eq:SME}, which captures the conditional evolution of classical parameters (means and variances for the observables of interest), whereas the EKF is rather just a tool to infer these in an efficient way, with $\omega$ being estimated in parallel. As a result, in contrast to the KF discussed above, the EKF is initialised with the estimates of $\brktc{\Jqvec{J}_\alpha(0)}=\brkt{\Jqvec{J}}_\CSS$ and $\Delta^2_{\cc}\Jqvec{J}_\alpha(0)\equiv\vec{\mrm{V}}_\alpha^{\cc}(0)=\vec{\mrm{V}}^\CSS$ being determined by those of the initial CSS state and known exactly with no errors, i.e., with the corresponding covariance elements of the EKF initially set to zero, i.e., $\Sigma_{OO}(0)=0$ for all $O\ne\omega$ in \eqnref{eq:true_errors}.

Nonetheless, the EKF estimates of relevant quantum means and variances, if accurate, can be directly used to, e.g., predict the conditional spin squeezing of the ensemble. Still, one should be careful with such a procedure, as the EKF may overestimate on average both the spin-squeezing parameter \eref{eq:spin_squeez_par} and the ensemble polarisation, $\EE{\xi^{-2}_{y}(t)}$ and $\EE{\brktc{\Jx}}$, as shown with the pink dashed lines in the main and inset plots of \figref{fig:err}(b), respectively. This is a result of operating at low $N=200$ in \figref{fig:err}, for which the CoG model \eref{eq:dynamical_model} used to construct the EKF approximates well dynamics \eref{eq:SME} only at short timescales ($t \ll (M+\kcoll)^{-1}$). However, we show in what follows that, for relevant sizes of atomic ensembles, e.g., $N=10^5$ in \figref{fig:attaining_bounds} below, as long as the CoG approximation \eref{eq:dynamical_model} is valid, the estimates provided by the EKF correctly predict the spin-squeezing parameter \eref{eq:spin_squeez_par} on average.

\subsubsection{Benchmarking against a classical strategy with a strong measurement}
In order to complete the discussion about the role of continuous spin squeezing and the necessity to generate entanglement in achieving the AMSEs shown in \figref{fig:err}, we decide to further benchmark the real-time estimation+control strategies against a \emph{classical scenario} in which no entanglement is generated. As an alternative we consider the scheme in which the experimenter, rather than continuously probing the atomic ensemble until a given time $t$, performs any possible \emph{strong measurement} at $t$. In such a case, rather than following the conditional dynamics \eref{eq:dynamical_model}, the ensemble evolves undisturbed (following Eqs.~\eref{eq:dynamical_model} with $M\to0$) until it is destructively measured. As elaborated on in \appref{app:classical_strategy}, the AMSE within such a scheme can still be constrained by the Bayesian Cram\'{e}r-Rao Bound \eref{eq:BCRB} but with $\F[\cdot]$ being replaced by the quantum Fisher information (QFI)~\cite{Helstrom1976}. 

We demonstrate in \appref{app:classical_strategy}, by resorting to exact numerical simulations for the real-time scenarios as above and explicitly computing the relevant QFI for the classical scenario, that the classical limit is indeed surpassed by the EKF+LQR strategy despite the presence of relatively weak collective decoherence. Moreover, as within the classical strategy there is no mechanism to counteract the decoherence, its usage becomes pointless at longer times $t$, at which the atoms reach a steady state that ceases to be sensitive to any variations of the estimated Larmor frequency $\omega$. We show this effect explicitly in \appref{app:classical_strategy} by choosing either the collective or local decoherence to be relatively strong, in order to stress that, for the EKF+LQR strategy---because information about the estimated $\omega$ keeps growing over time as the ensemble stabilises in a metrologically useful state---the AMSE keeps decreasing over long timescales, while the classical strategy quickly becomes useless in a single-shot scenario.

\begin{figure*}[t!]
    \centering
    \includegraphics[width=\textwidth]{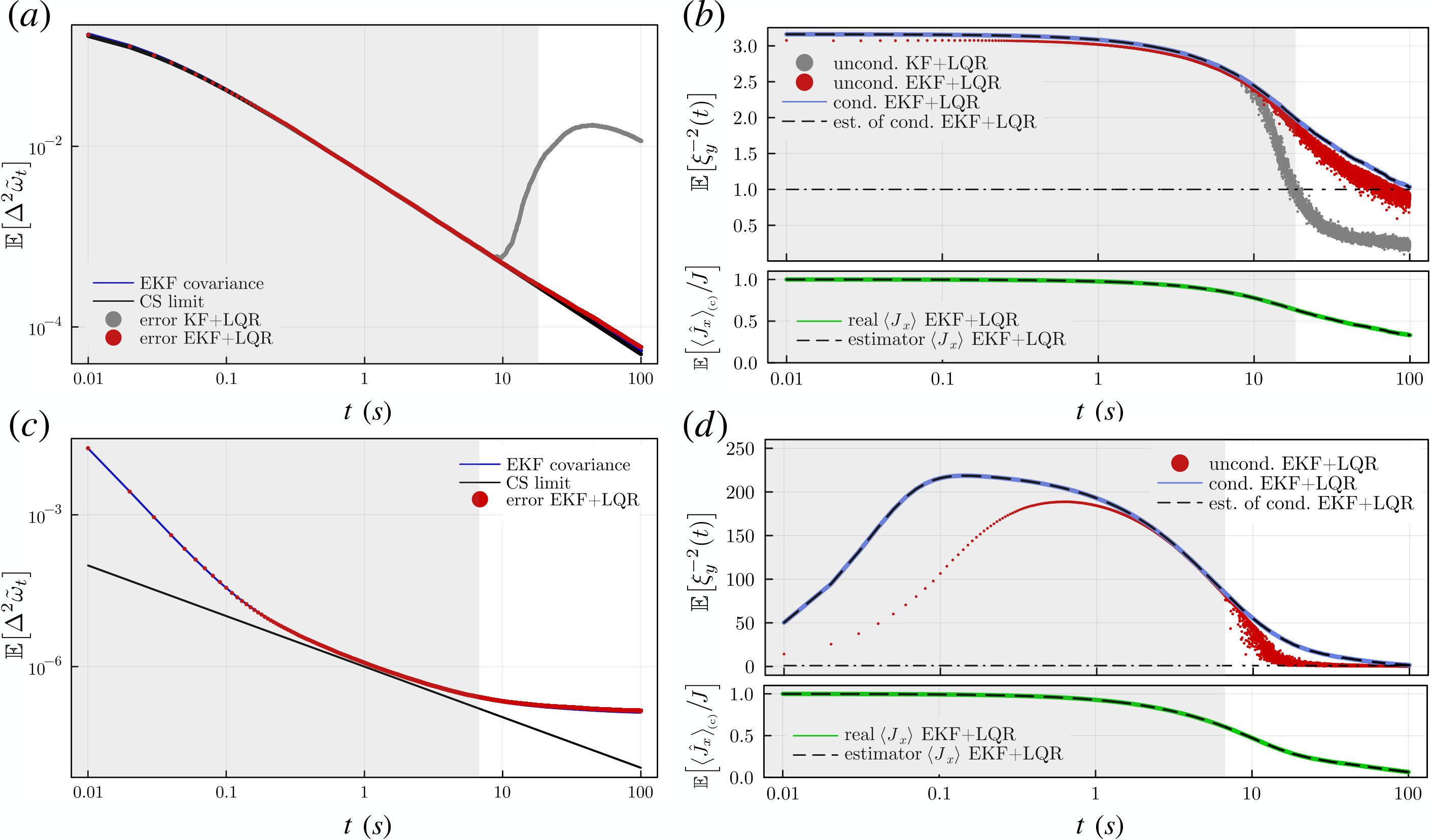}
    \caption{\textbf{Performance in estimation and spin squeezing extrapolated to large atomic ensembles}, here $N=10^5$. Subplots (\emph{a}) and (\emph{b}) (\emph{upper row}) depict the case of pure collective decoherence $\kcoll = 0.005$, whereas subplots (\emph{c}) and (\emph{d}) (\emph{lower row}) deal with pure local decoherence $\kloc = 0.05$. \emph{Left column}:~(\emph{a}) and (\emph{c}) show the error (AMSE) attained by the EKF+LQR strategy (red circles) when estimating $\omega$ and its average prediction by the EKF (blue line), $\EE{\Sigma_{\omega\omega}}$, both being lower bounded by the CS limit \eref{eq:CSlimit} (black line). For collective decoherence, the performance of KF+LQR strategy is also included (grey circles) to emphasise its failure beyond the LG regime (pink shading in all subplots). \emph{Right column}:~(\emph{b}) and (\emph{d}) illustrate the evolution of the spin-squeezing parameter \eref{eq:spin_squeez_par} for conditional (blue line) and unconditional (red circles) dynamics, as compared with its classical threshold (horizontal dash-dot line). The evolution of the ensemble polarisation $\brkt{\Jx}=\mathbb{E}[\brktc{\Jx}]$ (green line) is also shown in both cases in extra lower plots. Both the conditional spin squeezing and the polarisation in (\emph{b}) and (\emph{d}) are estimated very accurately by the EKF on average (superimposed dashed black lines). The above data are simulated employing the CoG model \eref{eq:dynamical_model} with other parameters set to $M = 0.05$, $\omega = 1$, and $\eta = 1$. As in \figref{fig:err}, the KF and EKF estimators are initialised with mean $\Tilde{\vec{x}}(0) = (N/2,0,0,N/4,N/4,0,\mu_0)^\TT$ and covariance $\Sigma(0) = \Diag{0,0,0,0,0,0,\sigma_{0}^2}$ dictated by the initial CSS state and the Gaussian prior distribution for $\omega\sim\mathcal{N}(\mu_0,\sigma_0^2)$. All results are obtained after averaging over $\nu = 20000$ measurement trajectories, while $\omega$ averaging is avoided by choosing the prior with $\sigma_0=0.5$ and $\mu_0 = \omega + \sigma_0=1.5$.
    }
    \label{fig:attaining_bounds}
\end{figure*}

\begin{figure*}[t!]
    \centering
    \includegraphics[width=\textwidth]{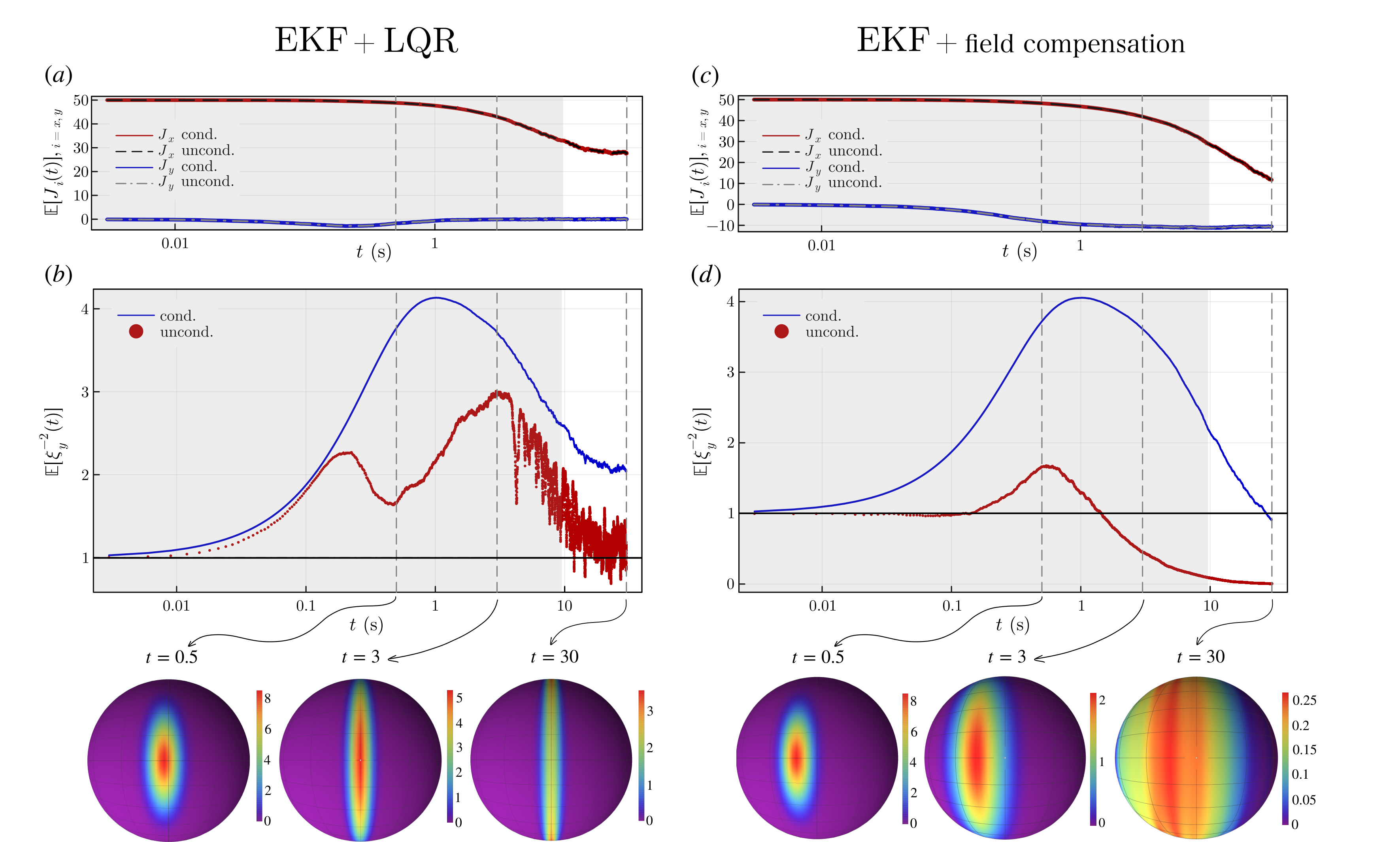}
    \caption{\textbf{Conditional versus unconditional spin squeezing.} The exact ($N=100$) spin-squeezing dynamics with collective decoherence is shown depending on the control strategy:~LQR (\emph{left column}) versus (naive) field compensation (\emph{right column}). \emph{Top row:} Subplots (\emph{a}) and (\emph{c}) depict evolution of angular momentum components in $\Jx$ (red) and $\Jy$ (blue) directions, in particular their conditional and unconditional means that consistently match. \emph{Middle row:} Subplots (\emph{b}) and (\emph{d}) compare the dynamics of the average unconditional (in red) and conditional (in blue) spin-squeezing parameters \eref{eq:spin_squeez_par}, also verifying whether they surpass the classical value (horizontal black line). Vertical dashed grey lines mark the relevant times for which we explicitly plot the spherical Wigner functions (\emph{bottom row}) representing the instantaneous unconditional state. Note that for the LQR control (\emph{left}), even though the width along $y$ of the distribution progressively narrows with time, the amplitude of the Wigner function also decays. The other parameters used in the SME \eref{eq:SME} for simulations read:~$\kcoll = 0.005$, $\kloc = 0$, $M = 0.1$, $\omega = 1$, $\eta = 1$. The EKF is initialised with mean $\Tilde{\vec{x}}(0) = (N/2,0,0,N/4,N/4,0,\mu_0)^\TT$ and covariance $\Sigma(0) = \Diag{0,0,0,0,0,0,\sigma_{0}^2}$ dictated by the initial CSS state, and the Gaussian prior distribution for $\omega\sim\mathcal{N}(\mu_0,\sigma_0^2)$. All results are obtained after averaging over $\nu = 500$ measurement trajectories, while $\omega$ averaging is avoided by choosing the prior with $\sigma_0=0.5$ and $\mu_0 = \omega + \sigma_0=1.5$.
    }
    \label{fig:sq}
\end{figure*}

%%%%%%%%%%%%%%%%%%%%%%%%%%%%%%%%%%%%%%%%%%%%%%%%%%%%%%%%%%%%%%%%%%%%%%%%%%%%%%%%%%%%%%%%
\subsection{Extending the results to high $N$}
As brute-force numerics become impossible, in order to extend the simulations of dynamics \eref{eq:SME} to high atomic numbers, $N$, we postulate that the CoG model \eref{eq:dynamical_model} can be used not only within the EKF construction \eref{eq:EKF_alg} but also to replace \eqnref{eq:SME} when simulating the (conditional) dynamics of the first and second moments of the angular-momentum operators, while incorporating feedback. In \appref{ap:verify_CoG}, by direct comparison with the exact solution of the SME \eref{eq:SME}, we show that with an increase in $N$ the CoG model predicts increasingly better polarisation $\brktc{\Jx}$ and variance $\Vy^{\cc}$ of the atomic ensemble that specify the spin-squeezing parameter \eref{eq:spin_squeez_par}, as long as the LG regime ($t\lesssim(M\!+\!2\kloc\!+\!\kcoll)^{-1}$) is considered. Moreover, if in particular the task of Larmor frequency estimation for the EKF+LQR scheme is of interest, the CoG model can be used for simulations far beyond the coherence time ($t\gtrsim(M\!+\!2\kloc\!+\!\kcoll)^{-1}$) unless significant collective decoherence ($\kcoll>0$) is present, which makes the CoG model mildly but persistently inaccurate (below 1\% of rel.~error) despite the increasing $N$;~see \appref{ap:verify_CoG}.

In \figref{fig:attaining_bounds}, we present the so-extrapolated results for $N=10^5$ (cf.~Refs.~\cite{Bohnet2014,hosten_measurement_2016}) to explicitly show that, for such a sufficiently large atomic number, the EKF+LQR strategy can be considered optimal within the LG regime, as its corresponding AMSE (in red) attains the CS limit \eref{eq:CSlimit} (in black) for both collective and local decoherence;~see Figs.~\ref{fig:attaining_bounds}(a) and \ref{fig:attaining_bounds}(c), respectively. Furthermore, it provides estimates of $\omega$ that also improve with time for timescales beyond the LG regime, $t > (M\!+\!2\kloc\!+\!\kcoll)^{-1}$, at which the KF+LQR strategy would fail (see the grey line in \figref{fig:attaining_bounds}(a)), or would not even be applicable in \figref{fig:attaining_bounds}(c).

Strikingly, in both Figs.~\ref{fig:attaining_bounds}(a) and \ref{fig:attaining_bounds}(c), i.e.~both for collective and local decoherence, the average covariance of the EKF (in blue), $\EE{\Sigma_{\omega\omega}}$, follows the true AMSE (in red). This confirms that, despite the nonlinearity of the CoG model \eref{eq:dynamical_model}, the (trajectory-dependent) error provided by the EKF can be trusted. In particular, the covariance provided by the EKF along \emph{any} measurement trajectory  correctly predicts the AMSE in the LG regime. Moreover, at longer timescales, it does not fluctuate significantly and concentrates on the AMSE upon averaging over only a small number of repetitions.

Not only is the CS limit \eref{eq:CSlimit} not guaranteed to be generally tight, but, for local decoherence, it also diminishes as $\propto1/N$, making its attainability even less likely. Still, as shown in \figref{fig:attaining_bounds}(c), the AMSE of the EKF+LQR strategy (in red), superimposed on the EKF covariance (in blue), attains the CS limit (in black) for a short time window, so that its optimality can then be guaranteed---answering positively the open question posed in \refcite{rossi_noisy_2020}.

%%%%%%%%%%%%%%%%%%%%%%%%%%%%%%%%%%%%%%%%%%%%%%%%%%%%%%%%%%%%%%%%%%%%%%%%%%%%%%%%%%%%%%%%
%%%%%%%%%%%%%%%%%%%%%%%%%%%%%%%%%%%%%%%%%%%%%%%%%%%%%%%%%%%%%%%%%%%%%%%%%%%%%%%%%%%%%%%%
\section{Conditional versus unconditional spin squeezing} 
\label{sec:squeezing}
When discussing \emph{spin-squeezing} results, we can either focus on the spin-squeezing parameter \eref{eq:spin_squeez_par} evaluated along a specific measurement trajectory (i.e.~\emph{conditional}), relevant for real-time magnetometry, or examine the entire feedback loop system as a mechanism for generating an entangled state independent of our observations (i.e.~\emph{unconditional}). While the \emph{conditional} state of the system $\rhoc(t) \equiv \rho(t|\vec{y}_{\leq t})$ is understood as that most closely describing the state given a particular measurement record $\vec{y}_{\leq t}$, an \emph{unconditional} state $\rho(t)$ describes the system when we discard, or do not have access to, the measurement outcomes, which formally corresponds to averaging the conditional state over all the possible past measurement trajectories, i.e.~$\rho(t) = \E{\rhoc(t)}{p(\vec{y}_{\leq t})}$.

In the absence of feedback, the impact that continuously measuring the system has on the unconditional dynamics of $\rho(t)$ is simply to introduce extra collective decoherence---e.g., in the case of SME \eref{eq:SME} with $u(t)=0$ after taking $\E{\cdot}{p(\vec{y}_{\leq t})}$ of both sides, the measurement induces only the extra \mbox{$M\D[\Jy]$ term}. On the contrary, in the presence of feedback, the effective master equation describing the unconditional evolution cannot be easily deduced from the conditional dynamics, e.g.~from \eqnref{eq:SME}, unless restrictive assumptions (e.g.~Markovianity) are made~\cite{nurdin2017linear}, which are not fulfilled for the LQR-based control strategy described in \secref{sec:LQR}. However, as such restrictive feedback scenarios are known to \emph{unconditionally} drive the system into a spin-squeezed state~\cite{wiseman1994_Feedback,Thomsen2002}, we confirm that this is also the case here.

In Figs.~\ref{fig:attaining_bounds}(b) and \ref{fig:attaining_bounds}(d), we demonstrate that the EKF+LQR strategy is not only capable of generating conditional spin squeezing (blue lines), as already shown in \figref{fig:err}(b) for $N=200$, but also yields significant unconditional spin squeezing (red circles) above the classical limit (horizontal solid black line). Furthermore, as \figref{fig:attaining_bounds} presents extrapolated results using the CoG model \eref{eq:dynamical_model} for $N=10^5$, thanks to dealing with a large atomic ensemble, the EKF very accurately estimates both the average conditional spin-squeezing parameter \eref{eq:spin_squeez_par} and the polarization, i.e.~in \figref{fig:attaining_bounds}(b\&d) the dashed black lines coincide with the blue and green lines, respectively---in contrast to the previously considered \figref{fig:err}(b) and its inset, with pink dashed lines largely overestimating both the spin-squeezing parameter and the polarisation. While \figref{fig:attaining_bounds}(b) highlights the advantages of using the EKF over the KF (grey dots) for maintaining the multiparticle entangled state beyond the LG regime---the unconditional spin squeezing is lost at timescales about $10$ times shorter for the KF---the reliability of these conclusions is compromised by the limitations of the CoG approximation at long times;~see \appref{ap:verify_CoG}. 

That is why we return in \figref{fig:sq} to simulating the exact SME \eref{eq:SME} in the presence of only the collective decoherence ($N=100$, $\kcoll=0.005$), where we study in detail the phenomenon of conditional versus unconditional spin squeezing by further comparing the EKF+LQR strategy (left column) with the naive field compensation (right column). Similarly to Figs.~\ref{fig:attaining_bounds}(b) and \ref{fig:attaining_bounds}(d), in Figs.~\ref{fig:sq}(b) and \ref{fig:sq}(d) we compare the average conditional spin-squeezing parameter (blue) and the unconditional spin squeezing of the average state (red), of which the latter also breaches the classical value (solid black horizontal line) beyond the LG regime for the EKF+LQR strategy employed in \figref{fig:sq}(b). This is confirmed by the (spherical) Wigner distribution plots (snapshots in time at $t=0.5,3,30$), which for the EKF+LQR strategy are clearly steadily ``squeezed'' in the $y$ direction even at $t=30\,>\,1/(M+\kcoll)\approx10$, in contrast to the field-compensation strategy, in which case the Wigner distribution begins to lose its shape already at $t=3$ within the LG regime. 

As the $\omega$ estimate of the EKF is initially set in \figref{fig:sq} to $\est{\omega}(0)>\omega$, the control operation initially overcompensates for the Larmor precession and rotates the spin in the counterclockwise direction when viewed along $z$---this is manifested by the spin components $\brkt{\Jy}=\mathbb{E}[\brktc{\Jy}]$ acquiring negative values in both (a) and (c), as well as by the corresponding Wigner function being shifted to the left, e.g.~at $t=0.5$ for both control strategies. An analogous behaviour would occur when choosing $\est{\omega}(0)<\omega$, in which case the control operation would initially undercompensate the Larmor precession, so that the spin rotates clockwise around $z$ (Wigner function shifts to the right), before either the LQR control increases the counter-rotation and stabilises the spin along $x$ (left column) or the stability is lost when the naive field-compensation strategy is pursued (right column).

%%%%%%%%%%%%%%%%%%%%%%%%%%%%%%%%%%%%%%%%%%%%%%%%%%%%%%%%%%%%%%%%%%%%%%%%%%%%%%%%%%%%%%%%
%%%%%%%%%%%%%%%%%%%%%%%%%%%%%%%%%%%%%%%%%%%%%%%%%%%%%%%%%%%%%%%%%%%%%%%%%%%%%%%%%%%%%%%%
\section{Conclusions} 
\label{sec:conclusions}
The quantum dynamics of a (spin-$1/2$) atomic magnetometer is studied;~first by exact numerical simulation, and later by introducing and verifying a comoving Gaussian approximation suitable for large atomic ensembles, while incorporating a measurement-based feedback loop, as well as local and collective decoherence in the form of dephasing along the magnetic field direction.

Based on the effective Gaussian description, an efficient estimation+control scheme is proposed that comprises an EKF and an LQR. In parallel, ultimate limits on real-time sensitivity are derived that apply to \emph{all} schemes involving measurement-based feedback and continuous measurements, given the decoherence considered. By saturating these, the EKF+LQR strategy is proven to be optimal at short timescales of experimental relevance. This is possible thanks to the enhancement of sensitivity by the continuous measurement, which generates interatomic entanglement (conditional spin squeezing) in real time. The entanglement emerges even if an experimenter disregards the recorded measurements---the atomic ensemble is driven by the LQR control into a state that is also entangled ``on average.'' Additionally, since, for large atomic numbers ($N\gtrsim10^5$), the EKF correctly predicts from measurements not only the magnetic field, but also the means and variances of ensemble spin operators, its use is sufficient for an experimenter to infer the conditional spin squeezing and, hence, benefit fully from it in real time.

Our work demonstrates for the first time that practical measurement-based feedback schemes can significantly enhance the operation of atomic sensors by exploiting quantum entanglement despite the noise. As a result, we believe that it paves the way for such schemes to be employed in state-of-the-art atomic magnetometers.
\begin{acknowledgments}
We thank Piotr Sza\'{n}kowski, Francesco Albarelli, Matteo A.~C.~Rossi, Marco G.~Genoni, Klaus M\o{}lmer, M\u{a}d\u{a}lin Gu\c{t}\u{a}, and Morgan W.~Mitchell for many helpful comments. This work was supported by the Quantum Optical Technologies project that was carried out within the International Research Agendas programme of the Foundation for Polish Science cofinanced by the European Union under the European Regional Development Fund, as well as the Project C’MON-QSENS! that was supported by the National Science Centre (2019/32/Z/ST2/00026), Poland under QuantERA, which has received funding from the European Union's Horizon 2020 research and innovation programme under Grant Agreement No.~731473. J.A.-B. also acknowledges "Excellence Initiative - Research University (2020-2026)" of the Ministry of Science and Higher Education (Grant No.~BOB-IDUB-622-361/2023). This research was funded in whole or in part by the National Science Centre, Poland under Grant No.~2023/50/E/ST2/00457.
\end{acknowledgments}

%%%%%%%%%%%%%%%%%%%%%%%%%%%%%%%%%%%%%%%%%%%%%%%%%%%%%%%%%%%%%%%%%%%%%%%%%%%%%%%%%%%%%%%%
%%%%%%%%%%%%%%%%%%%%%%%%%%%%%%%%%%%%%%%%%%%%%%%%%%%%%%%%%%%%%%%%%%%%%%%%%%%%%%%%%%%%%%%%
\section*{Data availability} 
The data that support the findings of this article are openly available~\cite{data_ref}.

%%%%%%%%%%%%%%%%%%%%%%%%%%%%%%%%%%%%%%%%%%%%%%%%%%%%%%%%%%%%%%%%%%%%%%%%%%%%%%%%%%%%%%%%%%%%%
%%%%%%%%%%%%%%%%%%%%%%%%%%%%%%%%%%%%%%%%%%%%%%%%%%%%%%%%%%%%%%%%%%%%%%%%%%%%%%%%%%%%%%%%%%%%%
%Appendices
%%%%%%%%%%%%%%%%%%%%%%%%%%%%%%%%%%%%%%%%%%%%%%%%%%%%%%%%%%%%%%%%%%%%%%%%%%%%%%%%%%%%%%%%%%%%%
%%%%%%%%%%%%%%%%%%%%%%%%%%%%%%%%%%%%%%%%%%%%%%%%%%%%%%%%%%%%%%%%%%%%%%%%%%%%%%%%%%%%%%%%%%%%%
\appendix

%%%%%%%%%%%%%%%%%%%%%%%%%%%%%%%%%%%%%%%%%%%%%%%%%%%%%%%%%%%%%%%%%%%%%%%%%%%%%%%%%%%%%%%%%%%%%%%%%%%%%%%%%%%%%%%%%%%%
%%%%%%%%%%%%%%%%%%%%%%%%%%%%%%%%%%%%%%%%%%%%%%%%%%%%%%%%%%%%%%%%%%%%%%%%%%%%%%%%%%%%%%%%%%%%%%%%%%%%%%%%%%%%%%%%%%%%
%%%%%%%%%%%%%%%%%%%%%%%%%%%%%%%%%%%%%%%%%%%%%%%%%%%%%%%%%%%%%%%%%%%%%%%%%%%%%%%%%%%%%%%%%%%%%%%%%%%%%%%%%%%%%%%%%%%%
\section{Wigner function on a sphere} 
\label{ap:Wigner}

State representations on the Bloch sphere are useful to visualise spin squeezing for an ensemble of (spin-$1/2$) atoms, and to gain intuition about properties of the overall quantum state. For that reason, we choose to compute the Wigner quasiprobability distribution and map it into the Bloch sphere as described in \refcite{Pezze2018RMP}, i.e.,
\begin{align}
    W(\theta,\phi) = \sqrt{\frac{N+1}{4\pi}} \sum_{k = 0}^N \sum_{q=-k}^{k} \rho_{kq} Y_{kq}(\theta,\phi)
    \label{eq:Wigner_quasi}
\end{align}
where the $Y_{kq}(\theta,\phi)$ are complex spherical harmonics, for which the coefficients $\rho_{kq} = \sum_{m_1,m_2 = -J}^J \rho_{m_1,m_2} t_{kq}^{m_1 m_2}$
are determined by the part of the density matrix supported by the \emph{totally symmetric subspace}, in particular its elements $\rho_{m_1,m_2}$ written in the angular momentum basis for the maximal total spin $J=N/2$, with the Clebsch-Gordan coefficients being $t_{kq}^{m_1 m_2} \coloneqq (-1)^{J-m_1-q} \brkt{J, m_1 ; J, -m_2 | k,q}$~\cite{Agarwal_wigner,Schmied2011}. Note that the exact density matrix is needed to generate the Wigner quasiprobability distribution \eref{eq:Wigner_quasi};~therefore, only in the case where we can solve SME \eref{eq:SME} exactly ($N\lesssim300$) can we compute $W(\theta,\phi)$. 

Throughout the manuscript, we present Wigner functions mapped onto Bloch spheres to illustrate the distribution of coherent spin states (e.g.~in \figref{fig:CSS_bloch}), or to show how the EKF+LQR (estimation+control) strategy unconditionally squeezes the atomic ensemble (in \figref{fig:sq}) despite collective decoherence. Note that in the presence of local decoherence the state is no longer supported only by the totally symmetric subspace, so that $W(\theta,\phi)$ no longer captures all its properties.

%\vspace{1cm}
%%%%%%%%%%%%%%%%%%%%%%%%%%%%%%%%%%%%%%%%%%%%%%%%%%%%%%%%%%%%%%%%%%%%%%%%%%%%%%%%%%%%%%%%%%%%%%%%%%%%%%%%%%%%%%%%%%%%
%%%%%%%%%%%%%%%%%%%%%%%%%%%%%%%%%%%%%%%%%%%%%%%%%%%%%%%%%%%%%%%%%%%%%%%%%%%%%%%%%%%%%%%%%%%%%%%%%%%%%%%%%%%%%%%%%%%%
%%%%%%%%%%%%%%%%%%%%%%%%%%%%%%%%%%%%%%%%%%%%%%%%%%%%%%%%%%%%%%%%%%%%%%%%%%%%%%%%%%%%%%%%%%%%%%%%%%%%%%%%%%%%%%%%%%%%
\section{Derivation of the CoG dynamical model \eqref{eq:dynamical_model}} 
\label{ap:Ito}
The set of stochastic differential equations \eref{eq:dynamical_model} can be derived by carefully applying the rules of It\^{o} calculus, e.g., by noting that the differential of any two functions of time and a stochastic process, $f$ and $g$, reads $d (f g) = f dg + g df + df dg$. In our case, these functions are the means, variances and covariances of some quantum observable $\hat{\bigO}$, whose dynamical evolution can then be computed by substituting the conditional dynamics \eref{eq:SME} of $d\rhoc$ into $d\brkt{\hat{\bigO}} = \Tr\{\hat{\bigO} \, d\rhoc\}$. In particular, considering $\hat{J}_\alpha$, $\mathrm{V}^{\cc}_{\alpha}$ and $\mathrm{C}^{\cc}_{\alpha\beta}$ with $\alpha,\beta = x,y,z$ appearing in Eqs.~\eref{eq:dynamical_model}, which satisfy 
\begin{subequations}
\label{eq:deriv_model}
\begin{align}  
    &d\brktc{\hat{J}_\alpha} = \trace{\hat{J}_\alpha \, d\rho_{\cc}}\\ 
    &d \mrm{V}^{\cc}_\alpha = d\brktc{\hat{J}_\alpha^2} -  d (\brktc{\hat{J}_\alpha}^2) \\
    &\qquad\, = d\brktc{\hat{J}_\alpha^2} - 2\brktc{\hat{J}_\alpha} d(\brktc{\hat{J}_\alpha}) - d\brktc{\hat{J}_\alpha} d\brktc{\hat{J}_\alpha} \nonumber \\
    &d\mrm{C}^{\cc}_{\alpha\beta} = \frac{1}{2}d\brktc{\hat{J}_\alpha\hat{J}_\beta} + \frac{1}{2}d\brktc{\hat{J}_\beta\hat{J}_\alpha} - d(\brktc{\hat{J}_\alpha}\brktc{\hat{J}_\beta})  \nonumber \\
    &\qquad\; = \frac{1}{2}d\brktc{\hat{J}_\alpha\hat{J}_\beta} + \frac{1}{2}d\brktc{\hat{J}_\beta\hat{J}_\alpha} - \brktc{\hat{J}_\beta}d\brktc{\hat{J}_\alpha} \nonumber \\
    &\qquad\qquad - \brktc{\hat{J}_\alpha}d\brktc{\hat{J}_\beta} - d\brktc{\hat{J}_\alpha}d\brktc{\hat{J}_\beta},
\end{align}
\end{subequations}
we derive, by working to the relevant order $O(dt^{3/2})$,
\begin{widetext}
\begin{subequations}
\label{eq:full_dynamical_model}
\begin{align}  
    &  d\brktc{\Jx} =  - (\omega \!+\! u(t))  \brktc{\Jy}  dt \!- \frac{1}{2}(\kcoll \!+\! 2 \kloc \!+\! M)  \brktc{\Jx} dt \!+\! 2\sqrt{\eta M}  \Cxy^{\cc} \, dW \\ 
    &  d\brktc{\Jy} = (\omega \!+\! u(t))  \brktc{\Jx} dt \!- \frac{1}{2}(\kcoll \!+\! 2\kloc)  \brktc{\Jy} dt \!+\! 2\sqrt{\eta M}  \Vy^{\cc} \, dW \\
    &d\brktc{\Jz}  = -\frac{1}{2} M \brktc{\Jz} dt \!+\! 2\sqrt{\eta M} \Czy^{\cc} \, dW \\
    & d \Vx^{\cc}  =  - 2 (\omega \! + \! u(t))  \Cxy^{\cc} dt \! + \! \kcoll \! \left( \! \Vy^{\cc} \!+\!  \brktc{\Jy}^2 \!-\! \Vx^{\cc} \!\right)\!dt \!+\! \kloc \!\left( \!\frac{N}{2} \!-\!2 \Vx^{\cc} \!\right)\!dt \!+\!  M\! \left( \! \Vz^{\cc} \!-\! \Vx^{\cc} \!-\! 4 \eta {\Cxy^{\cc}}^2 \!\right) \!dt \nonumber\\
    & \;\;\;\;\;\;\;\;\;\;\;\; + 2\sqrt{\eta M} \left( \frac{1}{2}\cov_{\cc}(\Jx,\Jx,\Jy) + \frac{1}{2}\cov_{\cc}(\Jy,\Jx,\Jx) \right) dW \\
    &   d\Vy^{\cc} = 2 (\omega \!+\! u(t))  \Cxy^{\cc} dt \!+\! \kcoll \!\left(\!  \Vx^{\cc} \!+\!  \brktc{\Jx}^2 \!-\!  \Vy^{\cc}\!\right) \!dt \!+\! \kloc \!\left(\! \frac{N}{2} \!-\!2 \Vy^{\cc} \!\right) \!dt \!-\! 4 \eta M  {\Vy^{\cc}}^2 dt \nonumber \\
    & \;\;\;\;\;\;\;\;\;\;\;\, + 2 \sqrt{\eta M} \, \cov_{\cc}(\Jy,\Jy,\Jy) \, dW  \\
    &   d\Vz^{\cc} =  M \!\left( \!\Vx^{\cc} \!+\!  \brktc{\Jx}^2 \!-\! \Vz^{\cc}\! \right) \!dt + 2\sqrt{\eta M} \left(\frac{1}{2} \cov_{\cc}(\Jz,\Jz,\Jy) + \frac{1}{2} \cov_{\cc}(\Jy,\Jz,\Jz) \right) \, dW\\
    &   d\Cxy^{\cc} = (\omega \!+\! u(t)) \!\left(\!\Vx^{\cc} \!-\!  \Vy^{\cc}\!\right) \!dt \!-\! \kcoll \!\left(\!2\Cxy^{\cc} \!+\! \brktc{\Jx}\brktc{\Jy}\!\right)\!dt\!-\! 2\kloc \Cxy^{\cc} dt \!-\! \frac{1}{2} M \Cxy^{\cc} \left( 1 + 8\eta \Vy^{\cc}\right)\!dt \nonumber \\
    &\;\;\;\;\;\;\;\;\;\;\;\;\;\;  + 2\sqrt{\eta M} \left(\frac{1}{4}\cov_{\cc}(\Jx \Jy^2) + \frac{1}{2}\cov_{\cc}(\Jy \Jx \Jy) + \frac{1}{4} \cov_{\cc}(\Jy^2 \Jx) \right) dW \\
    &d\Czy^{\cc}  = (\omega \! + \! u(t)) \Cxz^{\cc} dt  - \frac{1}{2}\left(\kcoll  + 2\kloc + M\left(1 + 8\eta\Vy^{\cc}\right)\right) \Czy^{\cc} dt \\
    &\;\;\;\;\;\;\;\;\;\;\;\;\;\;  + 2\sqrt{\eta M} \bigg( \frac{1}{4}\cov_{\cc}(\Jz,\Jy,\Jy) +  \frac{1}{2}\cov_{\cc}(\Jy,\Jz,\Jy) + \frac{1}{4} \cov_{\cc}(\Jy,\Jy,\Jz) \bigg) dW \\
    &d\Cxz^{\cc}  = - (\omega \! + \! u(t))\Czy^{\cc} dt  - \frac{1}{2} \left(\kcoll + 2\kloc + 4M \right) \Cxz^{\cc} dt - M\brktc{\Jz}\brktc{\Jx} dt - 4 \eta M \Cxy^{\cc}\Czy^{\cc} dt \\ 
    &\;\;\;\;\;\;\;\;\;\;\;\;\;\;  + 2\sqrt{\eta M} \bigg(\frac{1}{4} \cov_{\cc}(\Jx,\Jz,\Jy) + \frac{1}{4} \cov_{\cc}(\Jy,\Jx,\Jz)  + \frac{1}{4} \cov_{\cc}(\Jz,\Jx,\Jy) + \frac{1}{4} \cov_{\cc}(\Jy,\Jz,\Jx) \bigg) dW
\end{align}
\end{subequations}
where, for any three operators $\hat{A}$, $\hat{B}$, and $\hat{C}$, we define 
\begin{equation}
	\cov_{\cc}(\hat{A}, \hat{B},  \hat{C}) \coloneqq \brktc{\hat{A}  \hat{B}  \hat{C}} - \brktc{\hat{A}}\brktc{\hat{B}  \hat{C}} - \brktc{\hat{B}} \brktc{\hat{A}  \hat{C}} - \brktc{\hat{C}}\brktc{\hat{A}  \hat{B}} + 2 \brktc{\hat{A}}\brktc{\hat{B}}\brktc{\hat{C}}.
\end{equation}
\end{widetext}

In order to be able to construct an EKF from the equations above, as well as to have a manageable system of stochastic differential equations that approximately describe our sensor, we drop the third-order moments from Eqs.~\eref{eq:full_dynamical_model}, i.e., we perform a cut-off approximation. Crucially, third-order moments appear only in the stochastic terms of the dynamical equations for the second order moments, reducing the impact of the approximation in the accuracy of the CoG model. 

Additionally, in the main text we also omit the differential equations for $\brktc{\Jz}$, $\Cxz^{\cc}$, or $\Czy^{\cc}$, since these quantities are consistently zero throughout the time evolution. This follows from their initial (CSS-state) conditions, i.e.  $\brkt{\Jz(0)} = \Cxz^{\cc}(0) = \Czy^{\cc}(0) = 0$, and their exclusively decaying evolution when the stochastic kicks due to the third-order moments are disregarded.

%%%%%%%%%%%%%%%%%%%%%%%%%%%%%%%%%%%%%%%%%%%%%%%%%%%%%%%%%%%%%%%%%%%%%%%%%%%%%%%%%%%%%%%%%%%%%%%%%%%%%%%%%%%%%%%%%%%%
%%%%%%%%%%%%%%%%%%%%%%%%%%%%%%%%%%%%%%%%%%%%%%%%%%%%%%%%%%%%%%%%%%%%%%%%%%%%%%%%%%%%%%%%%%%%%%%%%%%%%%%%%%%%%%%%%%%%
%%%%%%%%%%%%%%%%%%%%%%%%%%%%%%%%%%%%%%%%%%%%%%%%%%%%%%%%%%%%%%%%%%%%%%%%%%%%%%%%%%%%%%%%%%%%%%%%%%%%%%%%%%%%%%%%%%%%
\section{Discrete-time picture of a measurement-based feedback scheme}
\label{ap:discr_pict}
Any conditional dynamics of a quantum system that undergoes a continuous measurement and measurement-based feedback, such as SME \eref{eq:SME}, corresponds to the continuous-time limit ($\Dt\to0$) of a discretised evolution consisting of a sequence of completely positive and trace-preserving (CPTP) maps, $\Phi_\Dt$ of duration $\Dt$, that are interspersed by weak sequential measurements~\cite{Belavkin1989}.

In particular, the system state at time $t=k\,\delta t$, i.e.~after $k$ steps of the discretised sequence, generally reads
\begin{align}
	\label{eq:cm_discr}
  \rho[k|\vec{y}_k] 
  \coloneqq
  \frac{\Phi_k\!\!\left[\!\hat{E}_{y_{k}}  \dots \Phi_1\!\!\left[\!\hat{E}_{y_{1}}\rho_{0}\hat{E}_{y_{1}}^{\dagger}\!\right]\! \dots \hat{E}_{y_{k}}^{\dagger} \! \right]}{p(\vec{y}_k|\omega)},
\end{align}
where the map acting on the state in between measurements for a time $\delta t$,
\begin{align} \label{eq:map_phi}
    \Phi_k \coloneqq \Phi_{\Dt}^{(k)} (\omega,\vec{y}_k),
\end{align}
is determined by the system intermediate evolution, which incorporates feedback and thus generally depends on the whole past measurement record, $\vec{y}_k = \{y_1,\dots,y_k\}$, apart from the estimated parameter $\omega$.

The denominator in \eqnref{eq:cm_discr} is the \emph{likelihood}, i.e.~the probability of observing the measurement trajectory $\vec{y}_k$ given a particular value of the estimated parameter $\omega$:
\begin{align}
	\label{eq:likelihood_discr}
	p(\vec{y}_k|\omega) 
  =  \trace{\Phi_k\!\!\left[\!\hat{E}_{y_{k}}  \dots \Phi_1\!\!\left[\!\hat{E}_{y_{1}}\rho_{0}\hat{E}_{y_{1}}^{\dagger}\!\right]\! \dots \hat{E}_{y_{k}}^{\dagger} \! \right]}.
\end{align}
The measurement operators $\hat{E}_{y_{\ell}}$ in \eqnsref{eq:cm_discr}{eq:likelihood_discr} are determined by the continuous measurement being performed and yield a positive operator-valued measure (POVM), $\{\hat{E}_{y_k}^\dagger \hat{E}_{y_k}\}_{y_k}$ with $\sum_{y_k} \hat{E}_{y_k}^\dagger \hat{E}_{y_k}=\I$. For simplicity, we have set the measurements to be the same within a sequence;~however, such an assumption is unnecessary in the derivation of the precision bounds that follows. In particular, our analysis also applies to schemes in which the type of measurement is adaptively changed over the course of a single measurement trajectory.

Let us also note that state $\rho[k|\vec{y}_k]$ specified in \eqnref{eq:cm_discr} is actually the discretised version of the conditional density matrix introduced in the main text above \eqnref{eq:SME}, $\rhoct \equiv \rho(t|\vec{y}_{\leq t})$, which is then formally defined in the continuous-time limit as
\begin{equation}
	\rho(t|\vec{y}_{\leq t}) \coloneqq \lim_{\Dt\to0} \left\{ \rho[k|\vec{y}_k]\quad\text{s.t.}\quad k=\frac{t}{\Dt}\right\},
\end{equation}
now being conditioned on a continuous measurement record, $\vec{y}_{\leq t}\coloneqq\lim_{\Dt\to0}\{\vec{y}_{
k=t/\Dt}\}$.

\subsection{Separability of feedback maps in a sequence} \label{subsec:separability}
As the impact of decoherence and feedback is considered to be time invariant (Markovian) in between measurements, the dynamical map \eref{eq:map_phi} forms a semigroup with a fixed dynamical generator. The form of the generator, however, may vary stochastically from one step to another, as the feedback incorporated by the map is built on measurement outcomes that fluctuate. Nevertheless, because the increments of the measurement stochastic process are independent and stationary, map \eref{eq:map_phi} generally takes the form
\begin{equation}
    \Phi_k = \exp{\left\{\Dt \Lin_{\mrm{det}}^{(k)} + \Delta\! \mrm{Y}_{\!k} \Lin_{\mrm{Y}}^{(k)}\right\}},
    \label{eq:Phik_Levy_form}
\end{equation}
with the dynamical generator being split into its deterministic and stochastic parts and $\mrm{Y}_{\!k}$ denoting a general (discrete) L\'{e}vy process~\cite{Applebaum_2009}.

Importantly, the finite increment of the L\'{e}vy process $\mrm{Y}_{\!k}$ can always be expressed using the L\'{e}vy-It\^{o} decomposition as~\cite{Applebaum_2009}: 
\begin{equation} \label{eq:Levy_ito_decomp}
    \Delta \!\mrm{Y}_{\!k} = \mu_k \Dt + \sigma_k \Delta \! \mrm{W}_k +\!\! \int_{|x|<1} \!\!\!\!\!\!\!\!\!\! x \, \tilde{\mrm{N}}(\Dt,\dd x) +\!\! \int_{|x|\geq 1}\!\!\!\!\!\!\!\!\!\! x \, \mrm{N}(\Dt,\dd x),
\end{equation}
which includes a deterministic drift, a Gaussian diffusion with $\Delta\!\mrm{W}_k \sim \mathcal{N}(0,\Dt)$, and two jump terms, respectively.

The last term in \eqnref{eq:Levy_ito_decomp} represents large jumps of magnitude $|x|\geq 1$, and the first integral term is a compensated Poisson process that accounts for small jumps of magnitude $|x| < 1$. Here, $x$ is the size of a jump, $\dd x$ is a small interval of jump sizes, and $\mrm{N}(\Dt,\dd x)$ is the number of jumps in the time interval $\Dt$ whose size falls in $\dd x$. However, for small jumps, we actually integrate over the \emph{compensated} Poisson measure, $\tilde{\mrm{N}}(\Dt,\dd x) \coloneqq \mrm{N}(\Dt,\dd x) - \nu(\dd x) \, \Dt$, where $\nu(\dd x)$ is the L\'{e}vy measure, telling us the expected number of jumps of size $\dd x$ per unit time (i.e.~$\EE{\mrm{N}(\Dt,\dd x)} = \nu(\dd x) \Dt$). In the special case where this L\'{e}vy measure is finite, i.e. $\int x \nu(\dd x) < \infty$, we can further simplify the L\'{e}vy-It\^{o} decomposition as
\begin{align}
    \Delta \!\mrm{Y}_{\!k} = \mu_k^\prime \Dt + \sigma_k \Delta \! \mrm{W}_k +\!\! \int_{|x|\in \mathbb{R} \setminus \{0\}} \!\!\!\!\!\!\!\!\!\! x \, \mrm{N}(\Dt,\dd x) 
\end{align}
where $\mu_k^\prime \coloneqq \mu_k - \int_{|x|<1} x \, \nu(\dd x)$. Crucially, the variance of the Poisson jumps  is proportional to $\Dt$, i.e.
\begin{equation}
    \mrm{Var}\!\left[\int_{|x|\in \mathbb{R}\setminus \{0\}}  \!\!\!\!\!\!\!\!\!\! x \, \mrm{N}(\Dt,\dd x) \right] = \Dt \int_{|x|\in \mathbb{R}\setminus \{0\}} \!\!\!\!\!\!\!\!\!\! x^2 \, \nu(\dd x).
\end{equation}
Therefore, it is always valid to expand the dynamical map \eqref{eq:Phik_Levy_form} to the lowest order in $\Dt$ as
\begin{align}
    \!\!\Phi_{\!k} \!=\! \I \!+\! \Dt \Lin_{\!\mrm{det}}^{(k)} \!+\! \Delta \!\mrm{Y}_{\!k} \Lin_{\mrm{Y}}^{(k)} \!\!+\! \frac{1}{2} (\!\Delta\!\mrm{Y}_{\!k}\!)^2 (\!\Lin_{\mrm{Y}}^{(k)})^2 \!\!+\! \bigO(\Dt^{\frac{3}{2}}), 
    \label{eq:lowest_order}
\end{align}
because all higher-order terms---denoted schematically as $\bigO(\Dt^{\sfrac{3}{2}})$---even when containing stochastic increments cannot contribute at the lowest $\Dt$ order. By the same arguments, we can rewrite \eqnref{eq:lowest_order} as
\begin{align} 
    \Phi_{\!k} &\!=\! \left(\!\I \!+\! \Dt \Lin_{\!\mrm{det}}^{(k)} \!+\! \bigO(\Dt^2)\! \right) \!\big(\!\I \!+\! \Delta \! \mrm{Y}_{\!k}\Lin_{\mrm{Y}}^{(k)} \!+\! \frac{1}{2}\! (\!\Delta\!\mrm{Y}_{\!k})^{2} (\!\Lin_{\mrm{Y}}^{(k)})^{2} \nonumber \\
    &+ \bigO(\Dt^{\sfrac{3}{2}}) \big) 
    = \Xi_{\mrm{det}} \circ \FF_{\mrm{Y}},
    \label{eq:phik_det_and_stoch}
\end{align}
which can be interpreted as the lowest order in the Suzuki-Trotter expansion, consisting of deterministic and stochastic components. Specifically, the deterministic part is given by
\begin{equation}
    \Xi_\mrm{det} \coloneqq \exp{\{\Dt \Lin_{\mrm{det}}^{(k)}\}}
\end{equation}
whereas the stochastic part reads
\begin{equation}
    \FF_{\mrm{Y}} \coloneqq \exp{\{ \Delta \!\mrm{Y}_{\!k} \Lin_{\mrm{Y}}^{(k)} \}}.
\end{equation}

This treatment accounts for the Markovian feedback of \citet{wiseman1994_Feedback}, as well as any Bayesian feedback~\cite{MarcoFrancescoNotes}. In particular, for Markovian feedback, by identifying the stochastic term as $\Delta\! \mrm{Y}_{\!k} \Lin_{\mrm{Y}}^{(k)}\cdot\equiv - \ii \Delta\! \mrm{Y}_{\!k} [\Fop, \, \cdot \,]$, one recovers the results of \refcite{wiseman1994_Feedback}, as shown explicitly in \appref{subsec:Markov_feedback}.  Considering instead a general Bayesian feedback that may utilise the whole measurement record, we identify the deterministic part of the generator in \eqnref{eq:Phik_Levy_form} as that encoding the parameter, i.e., $\Dt\Lin_{\mrm{det}}^{(k)}\equiv\Dt\Lin_\omega$, and the stochastic part as that depending on the measurement trajectory $\vec{y}_k$ via feedback, i.e., $\Delta\! \mrm{Y}_{\!k} \Lin_{\mrm{Y}}^{(k)}\equiv\Dt\Lin_{\vec{y}_k}$. Therefore, we obtain
\begin{align}
    \Phi_k %&= \I + \Dt \Lin_{\omega_k} + \Dt \Lin_{\vec{y}_k} + \bigO(\Dt^2) \nonumber \\
    &= \left(\I + \Dt \Lin_{\omega_k}+ \bigO(\Dt^2) \right) \left(\I + \Dt \Lin_{\vec{y}_k} + \bigO(\Dt^{\sfrac{3}{2}}) \right) \nonumber \\
    &= \Xi_{\omega} \circ \FF_{\vec{y}_k} + \bigO(\Dt^{\sfrac{3}{2}}),
    \label{eq:sep_feedback}
\end{align}
where we now analogously define $\Xi_{\omega}\coloneqq\exp\{\Dt \Lin_{\omega_k}\}$ and $\FF_{\vec{y}_k}\coloneqq\exp\{\Dt\Lin_{\vec{y}_k}\}$ as the maps responsible for noisy parameter encoding and the application of the feedback, respectively. The above analysis applies to any measurement-based feedback, but we consistently recover the feedback of our scheme in \eqnref{eq:SME} for the special case of $\Dt\Lin_{\vec{y}_k}\cdot \equiv - \ii \Dt \, u(\vec{y}_k) [\Ham, \, \cdot \,]$ and its continuous-time limit, as shown explicitly below in \appref{subsec:recovering_the_sme}.

Having established in \eqnref{eq:sep_feedback} that any measurement-based feedback and noisy parameter encoding can always be separated in the $\Dt\to0$ limit at the level of maps, we may decompose the sequential structure of the conditional state in \eqnref{eq:cm_discr} further. In particular, for small enough $\Dt$, we can always write
\begin{align}
    \!\!&\;
    \rho[k|\vec{y}_k] = \label{eq:cm_discr_wfeed} \\
    \!\!&\
    =\frac{\Xi_\omega\!\!\left[\FF_{\vec{y}_{k}}\!\!\left[\!\hat{E}_{y_{k}}  \dots \Xi_\omega\!\!\left[\FF_{\vec{y}_{1}}\!\!\left[\!\hat{E}_{y_{1}}\rho_{0}\hat{E}_{y_{1}}^\dagger\!\right]\!\right] \dots \hat{E}_{y_{k}}^{\dagger} \! \right]\! \right]}{p(\vec{y}_k|\omega)}, \nonumber 
\end{align}
with the likelihood defined now as
\begin{align}
    \!\!&\; 
    p(\vec{y}_k|\omega) = \label{eq:likelihood_discr_wfeed} \\
    \!\!&
    = \trace{\Xi_\omega\!\!\left[\FF_{\vec{y}_{k}}\!\!\left[\!\hat{E}_{y_{k}}  \dots \Xi_\omega\!\!\left[\FF_{\vec{y}_{1}}\!\!\left[\!\hat{E}_{y_{1}}\rho_{0}\hat{E}_{y_{1}}^\dagger\!\right]\!\right] \dots \hat{E}_{y_{k}}^{\dagger} \! \right]\! \right]}. \nonumber
\end{align}

\section{Classical Simulation bound for any protocol with local and global decoherence} 
\label{ap:CSbound}
Although the Larmor frequency $\omega$ in $\Xi_\omega$ may itself be allowed to follow a stochastic process~\cite{Amoros-Binefa2025}, here we focus on estimating a constant value by employing a Bayesian strategy. Within the Bayesian approach to estimation theory, the AMSE can be lower bounded by different classes of Bayesian bounds~\cite{Fritsche2014}. For our purposes, we choose the (marginal unconditional) Bayesian Cram\'{e}r-Rao bound (BCRB)~\cite{bobrovsky1987,Fritsche2014}:
\begin{align} \label{eq:ap_BCRB}
    \EE{\Delta^2 \est{\omega}_t} \geq (J_B)^{-1},
\end{align}
where $J_B$ is the Bayesian information (BI)~\cite{Van-Trees},
\begin{align} \label{eq:ap_BI}
    J_B \coloneqq \E{\left(\partial_{\omega} \ln p(\omega,\vec{y}_{\leq t}) \right)^2}{p(\omega,\vec{y}_{\leq t})}.
\end{align}
The BI can be split into two terms, $J_B = J_P + J_M$. The first term, $J_P$, represents the contribution of our prior knowledge about $\omega$, 
\begin{align} \label{eq:ap_J_P}
    J_P = \F[p(\omega)] \coloneqq \E{\left(\partial_{\omega} \ln p(\omega) \right)^2}{p(\omega)},
\end{align}
corresponding to the Fisher information (FI) of the prior distribution $p(\omega)$ evaluated with respect to the estimated parameter $\omega$. As in this work we assume that the \textit{a priori} knowledge about $\omega$ is described by a Gaussian distribution of mean $\mu$ and variance $\sigma^2$, 
\begin{align} \label{eq:ap_B0_dist}
    \omega \sim p(\omega) = \frac{1}{\sqrt{2\pi \sigma^2}} \ee^{-\frac{(\omega - \mu)^2}{2\sigma^2}},
\end{align}
its FI corresponds to just the inverse of the variance, i.e.,
\begin{align}
    J_P = \frac{1}{\sigma^2}.
\end{align}

The second term, namely, the contribution of the measurement record, or $J_M$, can be understood as averaging the FI of the likelihood over the prior distribution, i.e.,
\begin{align}
    J_M &\coloneqq \E{\left( \partial_{\omega} \ln p(\vec{y}_{\leq t}|\omega) \right)^2}{p(\omega,\vec{y}_{\leq t})} \nonumber \\
    &= \int d\omega \; p(\omega) \, \F[p(\vec{y}_{\leq t}|\omega)]
    \label{eq:ap_J_M}
\end{align}
with
\begin{equation} \label{eq:ap_Fisher_pyB}
    \F[p(\vec{y}_{\leq t}|\omega)] \coloneqq \E{\left( \partial_{\omega} \ln p(\vec{y}_{\leq t}|\omega) \right)^2}{p(\vec{y}_{\leq t}|\omega)}
\end{equation}
being the FI of $p(\vec{y}_{\leq t}|\omega)$ with respect to $\omega$, i.e.~the likelihood of observing a measurement trajectory $\vec{y}_{\leq t}$ given that the true value of the Larmor frequency is $\omega$.

Now, as $J_M$, or, equivalently, $\F[p(\vec{y}_{\leq t}|\omega)]$, depends on a particular measurement strategy assumed, in what follows we focus on establishing a universal upper bound on $J_M$ that applies no matter the measurement sequence, including measurement-based feedback, and which stems from our previous work~\cite{Amoros-Binefa2021}.

The proof that follows is fully based on the form of map $\Xi_\omega$ responsible for noisy $\omega$ encoding. Moreover, in \eqnref{eq:sep_feedback} we have shown that $\Xi_\omega$ emerges from $\Phi_k$ by separating it---up to order $\Dt^{\sfrac{3}{2}}$---into two terms, one including the $\omega$ encoding, i.e., $\Xi_\omega$, and the other one the feedback, $\FF_{\vec{y}_k}$. Therefore, the bounds we now derive apply to any form of measurement-based feedback, but are specific to map $\Xi_\omega$. In particular, let us further split the internal dynamics given by map $\Xi_\omega$ into two additional maps,
\begin{equation} \label{eq:splitting_of_dynamics}
	\Xi_\omega=\Omega\circ\Lambda_\omega,
\end{equation}
where $\Omega$ denotes the nonunitary evolution arising in between measurements due to the collective decoherence (of strength $\kcoll$), and channel $\Lambda_\omega$ encompasses both the unitary frequency-encoding and the nonunitary local decoherence (of strength $\kloc$). Note that in SME \eref{eq:SME}  both generators of the collective and local decoherence commute with one another and the $\omega$ encoding, and so must their resulting CPTP maps upon integration, i.e.~$[\Omega,\Lambda_\omega]=0$. However, even if the two maps were not to commute, map $\Xi_\omega$ could still be split to order $\Dt^2$, using the Trotter-Suzuki approximation. 
As a result, the conditional state $\rhoc[k]\equiv\rho[k|\vec{y}_k]$ given by \eqnref{eq:cm_discr_wfeed}, by substituting \eqnref{eq:splitting_of_dynamics}, can then be written as
\begin{align}
\label{eq:ap_ev_discr}
    \!\!&\;
    \rho[k|\vec{y}_k] = \\
    \!\!&\
    =\frac{\!\Omega \! \left[ \Lambda_{\omega} \! \! \left[ \FF_{\vec{y}_{k}}\!\!\left[  \!\hat{E}_{y_{k}} \!\dots \Omega \! \left[\Lambda_{\omega} \! \left[\FF_{\vec{y}_1}\!\!\left[\hat{E}_{y_{1}} \rho_{0}\hat{E}_{y_{1}}^{\dagger}\right]\right]\right]  \!\dots \hat{E}_{y_{k}}^{\dagger}\right]\right] \right]}{p(\vec{y}_k|\omega)}. \nonumber 
\end{align}
Although the above decomposition is valid for any type of measurement-based feedback in the $\Dt\ll1$ limit, in the case of the feedback considered in SME \eref{eq:SME}, it applies for any $\Dt>0$, as the feedback in \eqnref{eq:SME} effectively changes the Larmor frequency and, hence, commutes with both the $\omega$ encoding and the decoherence. 

Analogously, the denominator of \eqnref{eq:ap_ev_discr} is now the special case of the general likelihood \eref{eq:likelihood_discr}, and reads
\begin{align}
    \!\!&\; 
    p(\vec{y}_k|\omega) = \label{eq:likelihood_discr_2} \\
    \!\!&
    = \!\trace{\!\Omega \! \left[ \Lambda_{\omega} \! \! \left[ \FF_{\vec{y}_{k}}\!\!\left[  \!\hat{E}_{y_{k}} \!\dots \Omega \! \left[\Lambda_{\omega} \! \left[\FF_{\vec{y}_1}\!\!\left[\hat{E}_{y_{1}} \rho_{0}\hat{E}_{y_{1}}^{\dagger}\right]\right]\right]  \!\dots \hat{E}_{y_{k}}^{\dagger}\right]\right] \right]\!}\!. \nonumber
\end{align}

%%%%%%%%%%%%%%%%%%%%%%%%%%%%%%%%%%%%%%%%%%%%%%%%%%%%%%%%%%%%%%%%%%%%%%%%%%%%%%%%%%%%%%%%%%%%%%%%%%%%%%%%%%%%%%%%%%%%
\subsection{Convex decomposition of the likelihood}
Similarly to our previous work~\cite{Amoros-Binefa2021}, which dealt only with collective decoherence, our motivation is to find convex decomposition of the effective noisy $\omega$-encoding map, i.e.~$\Omega\left[\Lambda_{\omega}\left[ \; \cdot \; \right]\right]$ in \eqnref{eq:ap_ev_discr}, so that the discretised likelihood \eref{eq:likelihood_discr_2} can be decomposed as
\begin{align} \label{eq:likelihood_convex_decomp}
    p(\vec{y}_k|\omega) = \int \D\vec{\Zeta}_k \; q(\vec{\Zeta}_k|\omega) \; p(\vec{y}_k|\vec{\Zeta}_k)
\end{align}
where $\vec{\Zeta}_k = \{\vec{\zeta}_1,\vec{\zeta}_2,\dots,\vec{\zeta}_k\}$ is a sequence of sets, each containing $N$ auxiliary frequencylike random variables, e.g.~$\vec{\zeta}_\ell = \{\zeta^{(1)}_\ell,\zeta^{(2)}_\ell, \dots, \zeta^{(N)}_\ell\}$ indicates that within the $\ell$th step the first probe undergoes the Larmor precession for $\Dt$ with frequency $\zeta^{(1)}$, the second probe with $\zeta^{(2)}$, etc. 

While $q(\vec{\Zeta}_k|\omega)$ represents the mixing distribution that crucially contains all the $\omega$ dependence, $p(\vec{y}_k|\vec{\Zeta}_k)$ in \eqnref{eq:likelihood_convex_decomp} can be interpreted as a (fictitious) likelihood of obtaining the measurement record $\{\vec{y}_j\}_{j=1}^k$, while the discretised measurements are interspersed by CPTP maps within which each probe undergoes frequency encoding with frequencies specified by sequence $\vec{\Zeta}_k$, i.e.,
\begin{align} 
		\label{eq:likelihood_Zk}
    &p(\vec{y}_k|\vec{\Zeta}_k) =   \\
    &=\trace{\!\mathcal{U}_{\vec{\zeta}_k}\!\!\left[\FF_{\vec{y}_{k}}\!\left[\! \hat{E}_{y_{k}}\!\dots \mathcal{U}_{\vec{\zeta}_1}\!\!\left[\FF_{\vec{y}_1}\!\!\left[\!\hat{E}_{y_{1}}\rho_{0}\hat{E}_{y_{1}}^{\dagger}\right]\right]\!\dots\hat{E}_{y_{k}}^{\dagger}\right]\right]\!}\!. \nonumber
\end{align}

\subsubsection{The map $\Omega\circ\Lambda_{\omega}$ as a convex mixture of unitaries}
We express the overall map $\Omega\left[\Lambda_{\omega}\!\left[ \; \cdot \; \right]\right]$ as a mixture of unitaries by decomposing separately the collective map $\Omega[\cdot]$ that acts on all the probes, and $\Lambda_{\omega}[\cdot]$ that exhibits a tensor product structure with local maps acting independently on each probe. 

In the case of $\Omega$ representing the evolution of the atomic state under collective decoherence (dephasing along the magnetic field direction), we can simply use the results presented by us in~\citeref{Amoros-Binefa2021} and write
\begin{equation}
    \Omega[ \; \cdot \;] = \int d\xi \; p_\coll(\xi) \; \ee^{-\ii\xi \hat{J}_z \Dt} \; \cdot \; \ee^{\ii \xi \hat{J}_z \Dt},
    \label{eq:coll_overall_map}
\end{equation}
with a Gaussian distribution $p_\coll(\xi) = \mathcal{N}(0,V_\coll)$ of zero mean and variance:
\begin{equation}
    V_\coll \coloneqq \kcoll/\Dt. \label{eq:var_coll} 
\end{equation}

On the other hand, the overall map associated with the local decoherence, i.e.,
\begin{equation}
    \Lambda_\omega = \ee^{t \Lin},
    \label{eq:local_overall_map}
\end{equation}
can be described as the formal solution of the master equation
\begin{align}
    \frac{d\rho}{dt} &= \Lin \, \rho = -\ii [\hat{H},\rho(t)] + \sum_{i = 1}^{N} \D{[\hat{L}_i]}\rho(t)  \nonumber\\
    &= -\ii \omega [\hat{J}_z,\rho(t)] +  \frac{\kloc}{2} \sum_{i = 1}^{N} \D{[\sz^{(i)}]}\rho(t)  \nonumber\\
    &=  \left( -\ii \frac{\omega}{2} \sum_{i=1}^N  [ \sz^{(i)} , \; \cdot \; ] +  \frac{\kloc}{2} \sum_{i = 1}^{N} \D{[\sz^{(i)}]} \; \cdot \; \right) \rho(t)  \nonumber \\
    &= \left[ \bigoplus_{i=1}^N \Lin^{(i)} \right] \rho, \label{eq:local_SE}
\end{align}
where $\Jz$ is the collective angular momentum in the $z$ direction, $\Jz = \frac{1}{2} \sum_{i=1}^N \sz^{(i)}$, with the subscript $(i)$ denoting the position of $\sz$ in the tensor-product structure, and $\Lin^{(i)} = -\ii\omega/2[\sz^{(i)}, \; \cdot \; ] + \kloc/2 \D[\sz^{(i)}] \; \cdot $.

It then follows that the collective map $\Lambda_\omega$ can be written as a tensor product of the individual CPTP maps for each atom, $\Lambda_\omega^{(i)}$. Namely,
\begin{equation}
    \Lambda_\omega = \ee^{\oplus_{i=1}^{N} t \Lin^{(i)}} = \bigotimes_{i=1}^N \ee^{t \Lin^{(i)}} = \bigotimes_{i=1}^N \Lambda_\omega^{(i)}
\end{equation}
with the semigroup map $\Lambda_\omega^{(i)} = \ee^{t \Lin^{(i)}}$ defined by the Gorini–Kossakowski–Sudarshan–Lindblad generator $\Lin^{(i)}$ representing the unconditional evolution of the $i$th atom, i.e.,
\begin{align}
    \frac{d\rho_i(t)}{dt} = -i\omega [\jz^{(i)},\rho_i(t)] + 2\kloc \D[\jz^{(i)}]\rho_i(t) ,
\end{align}
where $\jz = \frac{1}{2} \sz$, and $\rho_i = \text{Tr}_{\forall \neq i}{(\rho)}$ is the state after tracing out all atoms except the $i$th one. Then, again using the results introduced in~\refcite{Amoros-Binefa2021}, the unconditional equation described above can be written as a convex combination of unitary evolutions
\begin{equation}
    \Lambda_{\omega}^{(i)}[ \; \cdot \; ] = \int d\upsilon^{(i)} \, p_{\loc}(\upsilon^{(i)}|\omega) \; \Unitary_{\upsilon^{(i)} \Dt}[\; \cdot \;],
\end{equation}
where $\upsilon^{(i)}$ is just a dummy random variable that follows $\upsilon^{(i)} \sim p_{\loc}(\upsilon^{(i)}|\omega) = \mathcal{N}(\omega,V_{\loc})$, a Gaussian with mean $\omega$ and variance 
\begin{equation}
    V_{\loc} \coloneqq 2 \kloc/\Dt. \label{eq:var_loc}
\end{equation}

The unitary channel $\Unitary_{\upsilon^{(i)} \Dt}[\, \cdot \,]$ is also parametrised w.r.t. the auxiliary variable  $\upsilon^{(i)}$, i.e., 
\begin{equation}
    \Unitary_{\upsilon^{(i)} \Dt} [\; \cdot \;] = \ee^{-\ii \, \upsilon^{(i)} \jz^{(i)} \Dt} \, \cdot \,  \ee^{\ii \, \upsilon^{(i)} \jz^{(i)} \Dt} .
\end{equation}
Hence, it follows that the overall map $\Lambda_\omega$ in \eqnref{eq:local_overall_map} is equivalent to a convex combination of tensor products of unitary maps:
\begin{equation}
      \Lambda_\omega [\, \cdot \,]  \! = \bigotimes_{i=1}^N  \! \Lambda_\omega^{(i)} [\, \cdot \,]  \! =  \! \! \int  \! \! \D \vec{\upsilon} \, \wp_\loc (\vec{\upsilon}|\omega) \bigotimes_{i = 1}^N \Unitary_{\upsilon^{(i)} \Dt}[\, \cdot \,]
\end{equation}
with $\vec{\upsilon} = (\upsilon^{(1)}, \dots , \upsilon^{(i)}, \dots ,\upsilon^{(N)})$, $\wp_\loc (\vec{\upsilon}|\omega) = \prod_{i=1}^N p_{\loc}(\upsilon^{(i)}|\omega)$ and $\D\vec{\upsilon} = \prod_{i=1}^N d\upsilon^{(i)}$. Note that, since $\exp{(A)}\otimes \exp{(B)} = \exp{(A \oplus B)}$, then
\begin{align}
    \Unitary_{\vec{\upsilon}}[ \, \cdot \, ] &\equiv \bigotimes_{i = 1}^N \Unitary_{\upsilon^{(i)} \Dt}[\, \cdot \,]  \nonumber \\
    &= \ee^{-\ii \Dt \sum_{i=1}^N \upsilon^{(i)} \IjzI^{(i)} } \; \cdot \;  \ee^{\ii \Dt \sum_{i=1}^N \upsilon^{(i)} \IjzI^{(i)} },
\end{align}
where $\IjzI^{(i)} = \underbrace{ \I \otimes \dots \otimes \I}_{i-1}  \otimes \, \jz^{(i)} \otimes \underbrace{ \I \otimes \dots \otimes \I}_{N-i}$, $\jz^{(i)} = \frac{1}{2} \sz^{(i)}$.
 
Finally, combining the maps \eref{eq:coll_overall_map} and \eref{eq:local_overall_map}, we get
\begin{equation}
    \Omega[\Lambda_\omega[\; \cdot \;]] = \int \!\!d\xi \, p_{\coll}(\xi) \int \D \vec{\upsilon} \, \wp_{\loc}(\vec{\upsilon}|\omega) \; \Unitary_{\xi,\vec{\upsilon}}[ \; \cdot \; ],
\end{equation}
where $\Unitary_{\xi,\vec{\upsilon}}[ \; \cdot \; ] =  \ee^{-i \Dt \sum_{i=1}^N (\xi + \upsilon^{(i)}) \IjzI^{(i)} } \; \cdot \;  \ee^{i \Dt \sum_{i=1}^N (\xi + \upsilon^{(i)}) \IjzI^{(i)} }$. Furthermore, for convenience, let us redefine $\upsilon^{(i)}$ as $\upsilon^{(i)} = \zeta^{(i)} - \xi$, so that the above decomposition becomes
\begin{align}
    &\Omega[\Lambda_\omega[\; \cdot \; ]] = \nonumber \\
    &\quad=\int \! \D \vec{\zeta} \! \left[  \! \int \! d\xi \, p_{\coll}(\xi) \prod_{i=1}^N  p_{\loc}(\zeta^{(i)} - \xi|\omega) \right] \! \Unitary_{\zeta}[ \; \cdot \; ]  \nonumber \\
    &\quad=\int \! \! \D\vec{\zeta} \! \left[ \! c_1 \! \! \int \! \!  d\xi\, \ee^{-\frac{\xi^2}{2V_\coll}} \ee^{-\!\sum_{i=1}^N \! \! \! \frac{(\zeta^{(i)} - \xi - \omega)^2}{2V_\loc}} \! \right] \! \Unitary_{\zeta}[ \; \cdot \; ]  \nonumber \\
    &\quad= \int \! \! \D\vec{\zeta} \, c_2 \, f(\vec{\zeta}) \, g(\overline{\vec{\zeta}}|\omega) \, \Unitary_{\vec{\zeta}}[ \; \cdot \; ],
    \label{eq:ap_collective_channel}
\end{align}
where vector $\vec{\zeta}$ represents a collection of each auxiliary frequency acting on each particle, i.e., $\vec{\zeta} = (\zeta^{(1)}, \dots , \zeta^{(i)}, \dots, \zeta^{(N)})$. The unitary map parametrised by the aforementioned auxiliary frequencies is denoted $\Unitary_{\vec{\zeta}}[ \; \cdot \; ] =  \ee^{-\ii \, \Dt \sum_{i=1}^N \zeta^{(i)} \IjzI^{(i)}} \, \cdot \,  \ee^{\ii \, \Dt \sum_{i=1}^N \zeta^{(i)} \IjzI^{(i)}}$, while we define the normalisation constant $c_2 \coloneqq c_1 \sqrt{2\pi V_\coll V_\loc /(N V_\coll + V_\loc)}$ for convenience, being proportional to $c_1 \coloneqq (2\pi V_\coll)^{-1/2} (2\pi V_{\loc})^{-N/2}$.

The final expression in \eqnref{eq:ap_collective_channel} is a consequence of the equality
\begin{equation}
     \int \!\!d\xi \;\ee^{-\frac{\xi^2}{2V_\coll}} \ee^{-\sum_{i=1}^N \frac{(\zeta^{(i)} - \xi - \omega)^2}{2V_\loc}} \!= \frac{c_2}{c_1} \, f(\vec{\zeta})\, g(\overline{\vec{\zeta}}|\omega),
     \label{eq:corollary_equation}
\end{equation}
where $\overline{\vec{\zeta}}$ is the average of the auxiliary frequencies experienced by the $N$ atoms, i.e., $\overline{\vec{\zeta}} \coloneqq \frac{1}{N} \sum_{i=1}^N \zeta^{(i)}$, whereas
\begin{equation}
    f(\vec{\zeta}) \coloneqq \exp\!\left\{- \frac{1}{2V_\loc} \left( \sum_{i} \left(\!\zeta^{(i)}\!\right)^2  - N \overline{\vec{\zeta}}^2 \right)\right\}
    \label{eq:ap_f_fun}
\end{equation}
and
\begin{equation}
    g(\overline{\vec{\zeta}}|\omega) \coloneqq \exp\!\left\{- \frac{(\overline{\vec{\zeta}} - \omega)^2}{2V_Q} \right\}, 
    \label{eq:ap_gauss_fun}
\end{equation}
are Gaussian-like functions---the latter exhibiting a new ``effective'' variance:
\begin{equation}
	V_Q \coloneqq V_\coll + \frac{V_\loc}{N} = \frac{\kcoll + 2 \kloc/N}{\Dt} = \frac{\kappa_Q}{\Dt}
	\label{eq:V_Q}
\end{equation}
with $\kappa_Q\coloneqq\kcoll + 2 \kloc/N$.

To prove \eqnref{eq:corollary_equation}, we first define a new variable $\nu^{(i)} = \zeta^{(i)} - \omega$ and expand the exponent
$(\zeta^{(i)} - \xi - \omega)^2 = (\xi - \nu^{(i)})^2 = \xi^2 -2\xi \nu^{(i)} + (\nu^{(i)})^2$. If we now take the sum  and rearrange terms, we obtain 
\begin{align}
    &\sum_i \frac{(\xi - \nu^{(i)})^2}{2 V_\loc} = \frac{\xi^2 -2\xi \frac{1}{N} \sum_i \nu^{(i)} + \frac{1}{N} \sum_i (\nu^{(i)})^2}{2 V_\loc / N}  \nonumber \\
    &= \frac{\xi^2 - 2\xi \, \overline{\vec{\nu}} + N (\overline{\vec{\nu}})^2}{2 V_\loc/N} - \frac{\frac{1}{N} \sum_{i\neq m} \nu^{(i)} \nu^{(m)}}{2 V_\loc/N}  \nonumber \\
    &= \frac{(\xi - \overline{\vec{\nu}})^2}{2 V_\loc/N} + \frac{(N - 1) \, \overline{\vec{\nu}}^2}{2 V_\loc/N} - \frac{\frac{1}{N} \sum_{i\neq m} \nu^{(i)} \nu^{(m)}}{2 V_\loc/N}, \label{eq:ap_exp1}
\end{align} 
where $\overline{\vec{\nu}} \coloneqq \frac{1}{N} \sum_i \nu^{(i)}$ and 
\begin{equation}
    \overline{\vec{\nu}}^2 = \frac{1}{N^2} \sum_{i=m} (\nu^{(i)})^2 + \frac{1}{N^2} \sum_{i\neq m} \nu^{(i)} \nu^{(m)}. \label{eq:ap_avg_sq}
\end{equation}
Crucially, the last two terms in \eqref{eq:ap_exp1} are independent of the frequency $\omega$. This will matter later on, but for now, let us show how they depend only on $\zeta$:
\begin{align}
    &(N-1) \, \overline{\vec{\nu}}^2 - \frac{1}{N} \sum_{i\neq m} \nu^{(i)} \nu^{(m)} = - \overline{\vec{\nu}}^2 +\frac{1}{N} \sum_{i} (\nu^{(i)})^2 \nonumber \\
    &= -\overline{\vec{\zeta}}^2 + 2 \overline{\vec{\zeta}} \omega - \omega^2 + \frac{1}{N} \sum_i \left( (\zeta^{(i)})^2 -2\omega \zeta^{(i)} + \omega^2 \right) \nonumber \\
    &= \frac{1}{N} \sum_i (\zeta^{(i)})^2 - \overline{\vec{\zeta}}^2.
\end{align}
Here we used \eqnref{eq:ap_avg_sq} and the fact that $\overline{\vec{\nu}} = \overline{\vec{\zeta}} - \omega$. Hence,
\begin{align}
    \!\!\exp{ \!\left\{\!\!-\!\!\sum_{i=1}^N \!\frac{(\zeta^{(i)} \!-\! \xi \!-\! \omega)^2}{2V_\loc} \!\right\} } \!=\! f(\vec{\zeta}) \exp{\!\left\{ \!-\frac{(\xi - \overline{\vec{\nu}})^2}{2V_\loc/N} \!\right\}},
\end{align}
which, upon substituting into the left-hand side of \eqnref{eq:corollary_equation}, allows us to directly evaluate the Gaussian integral over $\xi$, i.e.,
\begin{align}
    &f(\vec{\zeta}) \! \int \! d\xi \, \ee^{-\frac{\xi^2}{2V_\coll}} \ee^{ -\frac{(\xi - \overline{\vec{\nu}})^2}{2V_\loc/N} } = \nonumber \\
    &  = f(\vec{\zeta})\sqrt{2\pi \frac{V_\coll V_\loc}{N V_\coll + V_\loc}} \ee^{-\frac{(\overline{\vec{\zeta}}-\omega)^2}{2(V_\coll + V_\loc/N)}},
\end{align}
and arrive at the right-hand side of \eqnref{eq:corollary_equation}.

%%%%%%%%%%%%%%%%%%%%%%%%%%%%%%%%%%%%%%%%%%%%%%%%%%%%%%%%%%%%%%%%%%%%%%%%%%%%%%%%%%%%%%%%%%%%%%%%%%%%%%%%%%%%%%%%%%%%
\subsection{Upper bound on the Fisher Information}
Next, by substituting the convex combination of map $\Omega[\Lambda_\omega[\; \cdot \; ]]$ in terms of unitaries, that is, \eqnref{eq:ap_collective_channel}, into \eqnref{eq:likelihood_discr_2},
\begin{align} 
    &p(\vec{y}_k|\omega) = \nonumber \\
    &= \!\trace{\!\Omega \!\! \left[ \!\Lambda_{\omega} \!\! \left[ \!\FF_{\vec{y}_{k}}\!\!\left[ \!\measE_{y_k}\dots \Omega \! \!\left[\!\Lambda_{\omega} \!\! \left[\!\FF_{\vec{y}_1}\!\!\!\left[\! \hat{E}_{y_{1}} \rho_{0} \hat{E}_{y_{1}}^{\dagger}\!\right]\right]\right] \! \!\dots\! \hat{E}_{y_{k}}^{\dagger}\!\right]\right] \right] \!}\! \nonumber \\
    &= \! \! \int \!  \!\D\vec{\Zeta}_k \!  \! \left(  \!\prod_j^k  \! c_2  f(\vec{\zeta}_j)  g(\overline{\vec{\zeta}_j}|\omega)  \! \right) \! \! \mrm{Tr} \big[\mathcal{U}_{\vec{\zeta}_k}\!\big[\FF_{\vec{y}_{k}}\!\big[ \!\hat{E}_{y_{k}}\dots  \nonumber \\ 
    &\dots \mathcal{U}_{\vec{\zeta}_1}\!\big[\FF_{\vec{y}_1}\!\big[ \hat{E}_{y_{1}} \rho_{0} \hat{E}_{y_{1}}^\dagger \big]\big]\big]\!\dots \hat{E}_{y_{k}}^{\dagger}\big]\big]\! \nonumber \\
    & = \! \! \int \!  \!\D\vec{\Zeta}_k \!  \! \left(  \!\prod_j^k  \! c_2  f(\vec{\zeta}_j)  g(\overline{\vec{\zeta}_j}|\omega)  \! \right) \! p(\vec{y}_k|\vec{\Zeta}_k),
    \label{eq:comparison}
\end{align}
where we have used \eqref{eq:likelihood_Zk} in the last step. Note that, by comparing \eqref{eq:comparison} with \eqref{eq:likelihood_convex_decomp}, we can easily identify the auxiliary conditional probability $q(\vec{\Zeta}_{k}|\omega)$ as a product distribution 
\begin{equation}
     q(\vec{\Zeta}_k|\omega) \! = \prod_{j=1}^k q(\vec{\zeta}_j|\omega) =  c_2 \prod_{j=1}^k f(\vec{\zeta}_j) g(\overline{\vec{\zeta}}_j|\omega). \label{eq:ap_in_zeta_out_omega}
\end{equation}

As discussed at length in~\citeref{Amoros-Binefa2021}, the expression for $q(\vec{\Zeta}_k|\omega)$ allows us to directly construct an upper bound on the Fisher information evaluated with respect to the likelihood $p(\vec{y}_{k}|\omega)$ as 
\begin{align}
    \F[p(\vec{y}_{k}|\omega)] & = \F\left[\Smap \left(  q(\vec{\Zeta}_k|\omega) \right)\right] \leq \F \left[   q(\vec{\Zeta}_k|\omega) \right],
\end{align}
where $\Smap\!:  q(\vec{\Zeta}_k|\omega) \rightarrow p(\vec{y}_{k}|\omega)$ is a stochastic map $\Smap [ \; \cdot \; ] = \int d\vec{\Zeta}_k \;p(\vec{y}_k|\vec{\Zeta}_k) [\; \cdot \;]$, under which the Fisher information can only decrease due to its contractivity. 

We compute the Fisher information of $q(\vec{\Zeta}_k|\omega)$ with respect to the parameter $\omega$, i.e.,
\begin{align}
    \F\left[q(\vec{\Zeta}_k|\omega) \right] 
    & \coloneqq -\E{\frac{\partial^2}{\partial \omega^2} \ln q(\vec{\Zeta}_k|\omega)}{q(\vec{\Zeta}_k|\omega)} \label{eq:FI_def}\\
    & = \sum_{j=1}^k -\E{\frac{\partial^2}{\partial \omega^2} \ln g(\overline{\vec{\zeta}}_j|\omega)}{q(\vec{\Zeta}_k|\omega)},
    \label{eq:FI_g}
\end{align}
where \eqnref{eq:FI_g} follows from the fact that for each $j$th time step only the function $g(\overline{\vec{\zeta}}_j|\omega)$ depends on $\omega$. As a result, after further substituting the Gaussian form \eref{eq:ap_gauss_fun} of $g(\overline{\vec{\zeta}}_j|\omega)$, we obtain
\begin{equation}
    \F\left[q(\vec{\Zeta}_k|\omega) \right] = \sum_{j=1}^k \frac{1}{V_Q} = \frac{k}{V_Q} = \frac{k \Dt}{\kappa_Q},
\end{equation}
after also substituting for $V_Q$ according to \eqnref{eq:V_Q}.

%%%%%%%%%%%%%%%%%%%%%%%%%%%%%%%%%%%%%%%%%%%%%%%%%%%%%%%%%%%%%%%%%%%%%%%%%%%%%%%%%%%%%%%%%%%%%%%%%%%%%%%%%%%%%%%%%%%%
\subsubsection{Evaluating the $\Dt\to0$ limit}
It is then straightforward to see that when taking the limit of $\Dt \rightarrow 0$, the Fisher information of $q(\vec{\Zeta}_t|\omega)$ becomes $\F \left[   q(\vec{\Zeta}_t|\omega) \right] = t/\kappa_Q$, and, therefore, 
\begin{align}
    J_B &= J_P + J_M = \frac{1}{\sigma^2} + \int d\omega \; p(\omega) \, \F[p(\vec{y}_{\leq t}|\omega)]  \nonumber \\
    &\leq \frac{1}{\sigma^2}+ \int d\omega \; p(\omega) \, \F \left[ q(\vec{\Zeta}_t|\omega) \right]  = \frac{1}{\sigma^2} + \frac{t}{\kappa_Q}.
\end{align}

Hence, for a constant field, we obtain the following lower bound on the estimation AMSE,
\begin{align} \label{eq:ap_final_CSbound}
    \EE{ \Delta^2 \est{\omega}_t} \geq (J_B)^{-1} \geq V_t = \dfrac{1}{\dfrac{1}{\sigma_0^2} + \dfrac{1}{\dfrac{\kcoll}{t} + \dfrac{2\kloc}{t N}}}. \\
    \nonumber
\end{align}

\section{Retrieving the SME for homodyne measurement and field-compensating feedback}

Any conditional dynamics involving a \emph{continuous measurement}, such as SME \eref{eq:SME}, is generally derived as a continuous-time limit ($\Dt\to0$) of a discretised evolution consisting of a sequence of CPTP maps, $\Phi_\Dt$ of duration $\Dt$, that are interspersed by weak sequential measurements~\cite{Belavkin1989}. In fact, the continuous measurement record $\vec{y}_{\leq t}$ over time $t = k \Dt$ corresponds to the limiting case of a sequence of outcomes $\vec{y}_{k} = \{y_1, \dots , y_k\}$ after letting $k\to\infty$ as $\Dt \to 0$.

Additionally, by focusing on a single $\Dt$ step in \eqnref{eq:cm_discr}, we can show that consecutive conditional states are related as
\begin{equation}  \label{eq:deltat_map}
    \rho[k|\vec{y}_k]=\frac{\Phi_k[\hat{E}_{y_{k}}\rho[k\!\shortminus\!1|\vec{y}_{k\shortminus1}]\hat{E}_{y_{k}}^\dagger]}{\trace{\Phi_k[\hat{E}_{y_{k}}\rho[k\!\shortminus\!1|\vec{y}_{k\shortminus1}]\hat{E}_{y_{k}}^\dagger]}}.
\end{equation}
This iterative relationship can be further split to differentiate between the state before and after performing the measurement given by the POVM $\measE_{y_k}$. In other words, we label, analogously to the discrete Kalman filter,
\begin{align}
    \text{update : }&\rho[k\!\shortminus\!1|\vec{y}_{k}] \, = \, \text{state after measurement update $k$} \nonumber \\
    &\quad\quad\quad\quad\quad\;\;\;\text{but before evolution $k\!\shortminus\!1 \to k$} \nonumber \\
    \text{predict : }&\rho[k|\vec{y}_k] \, = \, \text{state after measurement update $k$} \nonumber \\
    &\quad\quad\quad\quad\;\;\,\text{and after evolution $k\!\shortminus\!1 \to k$} \nonumber,
\end{align}
such that the state at time $k$ is written as
\begin{equation} \label{eq:rho_after_evol}
    \rho[k|\vec{y}_k]=\frac{\Phi_k[\rho[k\!\shortminus\!1|\vec{y}_{k}]]}{\trace{\Phi_k[\rho[k\!\shortminus\!1|\vec{y}_{k}]]}},
\end{equation}
where $\rho[k\!\shortminus\!1|\vec{y}_{k}]$ is the updated state after we have performed the measurement but before evolving the system under the internal map $\Phi_k[\; \cdot \;]$, i.e., 
\begin{equation} \label{eq:rho_after_meas}
    \rho[k\!\shortminus\!1|\vec{y}_{k}] = \frac{\hat{E}_{y_{k}}\rho[k\!\shortminus\!1|\vec{y}_{k\shortminus1}]\hat{E}_{y_{k}}^\dagger}{\trace{\hat{E}_{y_{k}}\rho[k\!\shortminus\!1|\vec{y}_{k\shortminus1}]\hat{E}_{y_{k}}^\dagger}}.
\end{equation}
Both definitions readily emerge from \eqnref{eq:deltat_map} by dividing both the nominator and denominator by $\trace{\hat{E}_{y_{k}}\rho[k\!\shortminus\!1|\vec{y}_{k\shortminus1}]\hat{E}_{y_{k}}^\dagger}$. 

Let us now focus on deriving an expression for $\rho[k\!\shortminus\!1|\vec{y}_{k}]$ and its unnormalized version
\begin{equation}
    \rhoun[k\!\shortminus\!1|\vec{y}_{k}] \coloneqq \hat{E}_{y_{k}}\rho[k\!\shortminus\!1|\vec{y}_{k\shortminus1}]\hat{E}_{y_{k}}^\dagger.
\end{equation}

\subsection{Substituting the form of the continuous measurement}
Within the continuous measurement framework~\cite{Belavkin1989}, the Kraus operators of the $k$th measurement in \eqnref{eq:deltat_map}, $\{\hat{E}_{y_{k}}\}_{y_k}$, are generally associated with an interaction of the system with a bosonic mode, $\hat{B}_k$ satisfying $[\hat{B}_k,\hat{B}_{k'}^\dagger] = \delta_{kk'}$, that is subsequently measured, so that 
\begin{equation} 
	\hat{E}_{y_k} = \bra{y_k}  \UnitOp_{\Dt} \ket{0},
	\label{eq:Kraus_CM}
\end{equation} 
where the bosonic mode is initialised in the vacuum state $\ket{0}$ before being projected onto state $\ket{y_k}$ associated with a particular outcome $y_k$ of the quadrature operator
\begin{align} \label{eq:operator_y}
    \hat{y}_k \coloneqq \frac{1}{\sqrt{2}} (\Bcreatk + \Banihilk),
\end{align}
while the (weak) interaction is generated by the unitary operation~\cite{Belavkin1989,Wiseman2009,MarcoFrancescoNotes}:
\begin{equation}
	\UnitOp_{\Dt} = \exp\!\left\{ \sqrt{M \Dt} \, \left(\LinOp \otimes \hat{B}_k^\dagger - \LinOp^\dagger \otimes \hat{B}_k \right)\right\},
	\label{eq:unit_inter}
\end{equation}
with $M$ parametrising the strength of the continuous measurement and $\LinOp$ denoting \emph{any} system operator that is continuously probed.

Knowing that, we can expand the numerator of \eqnref{eq:rho_after_meas} as
\begin{align}
    &\rhoun[k\!\shortminus\!1|\vec{y}_{k}] \!=\! \braket{y_k|\UnitOp_\Dt|0}  \rho[k\!\shortminus\!1|\vec{y}_{k\shortminus1}]  \braket{0|\UnitOp_\Dt|y_k}  \\
    &\quad= \!\Tr_\mathcal{P} \{ \UnitOp_\Dt (\rho[k\!\shortminus\!1|\vec{y}_{k\shortminus1}] \!\otimes\! \projector{0}{0} ) \UnitOp_\Dt^\dagger (\I\! \otimes\! \projector{y_k}{y_k}) \}, \nonumber
\end{align}
where $\Tr_\mathcal{P}$ denotes the partial trace over the probe subspace. Note that the probe is modelled as a series of discretized uncorrelated modes interacting with the system in a conveyor-belt fashion \cite{MarcoFrancescoNotes}. In other words, before the system-probe interacting unitary is applied, the state of each probe mode is assumed to be in the vacuum state. Therefore, the joint state of the system and the probe before the interaction occurs is, at each step $\rho[k|\vec{y}_k]\!\otimes\!\projector{0}{0}$.  
To now derive a complete expression for the evolution of the reduced state of the atoms under the action of the Kraus operators, we have to first evolve the joint state under the action of the interaction Hamiltonian. Namely, 
\begin{align}
    &\UnitOp_\Dt (\rho[k\!\shortminus\!1|\vec{y}_{k\shortminus1}]  \!\otimes\! \projector{0}{0} ) \UnitOp_\Dt^\dagger = \rho[k\!\shortminus\!1|\vec{y}_{k\shortminus1}] \!\otimes\! \projector{0}{0} \nonumber \\
    &\; +\! M \LinOp \rho[k\!\shortminus\!1|\vec{y}_{k\shortminus1}]  \LinOp^\dagger  \!\otimes\!   \projector{1}{1}  \Dt  \nonumber \\
    &\; +  \!\left(   \LinOp\rho[k\!\shortminus\!1|\vec{y}_{k\shortminus1}]  \!\otimes\!  \projector{1}{0}  \!+\!  \rho[k\!\shortminus\!1|\vec{y}_{k\shortminus1}]  \LinOp^\dagger  \!\otimes\! \projector{0}{1} \right)  \!\! \sqrt{ \!M \Dt} \nonumber\\
    &\; + \!\frac{M}{2} \! \Big( \!\sqrt{2} \LinOp^2  \rho[k\!\shortminus\!1|\vec{y}_{k\shortminus1}]   \!\otimes\!  \projector{2}{0}  \!-\!  \LinOp^\dagger \! \LinOp \rho[k\!\shortminus\!1|\vec{y}_{k\shortminus1}]  \!\otimes\!  \projector{0}{0} \nonumber \\
    &\; + \!\sqrt{2} \rho[k\!\shortminus\!1|\vec{y}_{k\shortminus1}]  (\!\LinOp^\dagger)^2   \!\otimes\!  \projector{0}{2}  \!-\!  \rho[k\!\shortminus\!1|\vec{y}_{k\shortminus1}]  \LinOp^\dagger \!\LinOp \!\otimes\!\projector{0}{0} \!\Big) \Dt \nonumber \\
    &\; + \!\bigO(\Dt^{\sfrac{3}{2}}),
\end{align}
where we have used the expansion of the unitary operator up to order $\Dt$:
\begin{align}
    \UnitOp_{\Dt} 
    =& \I \!\otimes\! \I + \left(\!\LinOp \!\otimes\! \Bcreatk \!-\! \LinOp^\dagger \!\otimes\! \Banihilk \!\right) \!\sqrt{M \Dt} + \frac{1}{2} \Big(\!\LinOp^2 \!\otimes\! (\Bcreatk)^2 \nonumber \\
    &- \! \LinOp^\dagger \!\LinOp \!\otimes \!\Banihilk\Bcreatk \!-\! \LinOp \LinOp^\dagger \!\otimes \!\Bcreatk \Banihilk \! + \! (\LinOp^\dagger)^2 \!\otimes \!(\Banihilk)^2 \!\Big) M\Dt  \nonumber\\
    &+ \bigO(\Dt^{\sfrac{3}{2}}),
\end{align}
and the fact that $\Banihilk$ and $\Bcreatk$ are discretized annihilation and creation operators obeying the standard rules. 
Next, we simply project the state of the probe onto $\projector{y_k}{y_k}$ and trace out the bath (also referred to as the probe), yielding
\begin{align} 
		\label{eq:rhoc_unnorm_dt_y}
    & \rhoun[k\!\shortminus\!1|\vec{y}_{k}] = \braket{y_k|\UnitOp_\Dt|0}  \rho[k\!\shortminus\!1|\vec{y}_{k\shortminus1}]  \braket{0|\UnitOp_\Dt|y_k} \nonumber \\
    &=\abs{\braket{y_k|0}}^2\rho[k\!\shortminus\!1|\vec{y}_{k\shortminus1}]  \nonumber \\
    &+ \Big(\!\braket{y_k|1}\!\!\braket{0|y_k} \!\! \LinOp \rho[k\!\shortminus\!1|\vec{y}_{k\shortminus1}]  \nonumber \\
    &+ \braket{y_k|0}\!\!\braket{1|y_k}\!\rho[k\!\shortminus\!1|\vec{y}_{k\shortminus1}]  \!\LinOp^\dagger\!\Big)\!\sqrt{\!M\Dt} \nonumber \\
    &+ \frac{M}{2} \!\Big(\!\sqrt{2}\!\braket{y_k|2}\!\!\braket{0|y_k} \LinOp^2 \rho[k\!\shortminus\!1|\vec{y}_{k\shortminus1}]  \nonumber \\
    &- \abs{\braket{y_k|0}}^2 \! \LinOp^\dagger\!\LinOp \rho[k\!\shortminus\!1|\vec{y}_{k\shortminus1}]  \nonumber \\
    &+ \sqrt{2}\!\braket{y_k|0}\!\!\braket{2|y_k} \!\rho[k\!\shortminus\!1|\vec{y}_{k\shortminus1}] (\LinOp^\dagger)^2 \nonumber \\
    &- \abs{\braket{y_k|0}}^2\!\!\rho[k\!\shortminus\!1|\vec{y}_{k\shortminus1}] \LinOp^\dagger\!\LinOp \!\Big) \Dt \nonumber\\
    &+ M\! \abs{\braket{y_k|1}}^2 \LinOp\rho[k\!\shortminus\!1|\vec{y}_{k\shortminus1}] \LinOp^\dagger \,\Dt + \bigO(\Dt^{\sfrac{3}{2}}).
\end{align}
State $\rhoun[k\!\shortminus\!1|\vec{y}_{k}]$ represents the system's state after interacting with the probe for a time $\Dt$. Therefore, the trace of this unnormalized state gives the probability that the probe measurement yields $y_k$, \emph{after} its state has interacted with the system:
\begin{align}
     p_\Dt(y_k|\vec{y}_{k-1}) &= \trace{\rhoun[k\!\shortminus\!1|\vec{y}_{k}]}  \\
     &= \trace{\braket{y_k|\UnitOp_\Dt|0}  \rho[k\!\shortminus\!1|\vec{y}_{k\shortminus1}]  \braket{0|\UnitOp_\Dt|y_k}}. \nonumber
\end{align}
Here we have added a sub-index in the probability function to show that the interaction occurred during a time step $\Dt$. Additionally, it is useful to also derive the probability of the measurement of the probe yielding $y_k$ \emph{before} the system and probe interact, which should simply correspond to  $\abs{\braket{y_k|0}}^2$, since the probe is modelled as a conveyor belt of modes, each initialized in the vacuum state and interacting with the system sequentially \cite{MarcoFrancescoNotes}. Hence,
\begin{align}
    p_0(y_k|\vec{y}_{k-1}) &=\trace{\!\braket{y_k|\UnitOp_{\Dt=0}|0}  \rho[k\!\shortminus\!1|\vec{y}_{k\shortminus1}]  \braket{0|\UnitOp_{\Dt=0}|y_k}\!} \nonumber \\
    &=\trace{\bra{y_k}\left(\rho[k\!\shortminus\!1|\vec{y}_{k\shortminus1}]\otimes\proj{0}\right)\ket{y_k}} \nonumber\\
    &=\trace{\rho[k\!\shortminus\!1|\vec{y}_{k\shortminus1}]}\!\braket{y_k|0}\!\!\braket{0|y_k} \nonumber \\
    &=\abs{\braket{0|y_k}}^2\!=\!\frac{1}{\sqrt{\pi}} \ee^{-y_k^2} \eqqcolon p_0(y_k) ,
    \label{eq:py(t)}
\end{align}
where we have used the fact that the ground-state wave function of the 1D harmonic oscillator---with its energy eigenstates being the Fock states---in the position representation is a Gaussian function. 

With the definition of operator $\hat{y}_k$ in mind, we can further derive for a bosonic mode,
\begin{align}
  \braket{y_k|1} &= \braket{y_k|\Bcreat|0} = \braket{y_k|(\Bcreatk + \Banihilk)|0} = \sqrt{2} \braket{y_k|\hat{y}_k|0} \nonumber \\
  &=\sqrt{2}y_k\braket{y_k|0}, \label{eq:y_k|1} 
\end{align}
as well as
\begin{align} \label{eq:y_k|2}
    \sqrt{2}\braket{y_k|2}=(2y_k^2-1)\braket{y_k|0},
\end{align}
which can be derived from
\begin{align}
    &y_k^2 \braket{y_k|0}  
    = \braket{y_k|\hat{y}^2_k|0} = \Braket{y_k|\frac{\Bcreat + \Banihil}{\sqrt{2}}\frac{\Bcreat + \Banihil}{\sqrt{2}}|0} \nonumber \\
    &= \frac{1}{2} \! \braket{y_k|(\Bcreat)^2 \!\!+\!\Banihil \Bcreat|0} = \!\frac{1}{2}\!\left(\!\sqrt{2} \braket{y_k|2} \!+\! \braket{y_k|0}\!\right)\!.
\end{align}
Therefore, by recalling the expression for $p[k|\vec{y}_k]$ in \eqnref{eq:py(t)} and by substituting \eqnref{eq:y_k|1} and \eqnref{eq:y_k|2} into \eqnref{eq:rhoc_unnorm_dt_y}, we get
\begin{align}
    &\rhoun[k\!\shortminus\!1|\vec{y}_{k}] = \! p_0(y_k|\vec{y}_{k\shortminus1}) \bigg\{\!\rho[k\!\shortminus\!1|\vec{y}_{k\shortminus1}]  \nonumber \\
    &+ y_k\big(\LinOp\rho[k\!\shortminus\!1|\vec{y}_{k\shortminus1}]  \!+\! \rho[k\!\shortminus\!1|\vec{y}_{k\shortminus1}] \LinOp^\dagger\big)\sqrt{2M\Dt} \nonumber \\
    &+ 2 y_k^2 M \LinOp \rho[k\!\shortminus\!1|\vec{y}_{k\shortminus1}] \LinOp^\dagger \Dt \nonumber \\
    &+ \frac{M}{2}\!\Big[(2y_k^2 \!\shortminus\! 1)\!\left(\LinOp^2 \rho[k\!\shortminus\!1|\vec{y}_{k\shortminus1}] \!+\! \rho[k\!\shortminus\!1|\vec{y}_{k\shortminus1}] (\!\LinOp^\dagger)^2\right) \nonumber\\ 
    &-\{\LinOp^\dagger\!\LinOp,\rho[k\!\shortminus\!1|\vec{y}_{k\shortminus1}]\}\Big]\Dt + \bigO(\Dt^{\sfrac{3}{2}})\bigg\}.
    \label{eq:rhoc_unnorm_dt_py}
\end{align}

Furthermore, taking the trace of the above, we obtain
\begin{align}
    &p_\Dt(y_k|\vec{y}_{k\shortminus1}) = \trace{\rhoun[k\!\shortminus\!1|\vec{y}_{k}]} \nonumber \\
    &\quad=  p_0(y_k|\vec{y}_{k\shortminus1}) \! \left( \! 1 \!+\! y_k \langle \LinOp + \LinOp^\dagger \rangle \sqrt{2M\Dt} \!+\! \bigO(\Dt)\!\right) \label{p_x(t)_expansion}\\
    &\quad=  \frac{1}{\sqrt{\pi}} \ee^{-\left(y_k - \sqrt{M \Dt} \langle \LinOp + \LinOp^\dagger\rangle/2\right)^2} + \bigO(\Dt)
\end{align}
which thus constitutes a Gaussian distribution (up to the leading $\dt$ order) with mean $\sqrt{M\Dt} \langle \LinOp + \LinOp^\dagger \rangle /2 $ and variance $1/2$. As a result, we may introduce a new stochastic increment $\delta y_k$ that represents the above Gaussian fluctuations of the detected signal (up to order $\bigO(\Dt)$), i.e.,
\begin{equation} \label{eq:dy_increment}
    \delta y_k \coloneqq y_k \sqrt{2 \Dt} =  \sqrt{M} \langle \LinOp + \LinOp^\dagger \rangle \Dt + \delta W,
\end{equation}
where $\delta W\sim\mathcal{N}(0,\Dt)$ denotes the Wiener increment~\cite{Gardiner1985}. Physically, the derivative $I(t)\coloneqq\lim_{\delta t\to 0}\delta y/\Dt$ of the above corresponds to the stochastically fluctuating photocurrent being measured in real time in a homodyne setup~\cite{Wiseman2009}. Now, by noting that $y_k \sqrt{2 \Dt}=\delta y_k$ and $2y_k^2\Dt=\delta y_k^2=\Dt +\bigO(\Dt^{\sfrac{3}{2}})$, we can rewrite \eqnref{eq:rhoc_unnorm_dt_py} in terms of the increment $\delta y_k$ as

\begin{align}
     &\!\!\!\rhoun[k\!\shortminus\!1|\vec{y}_{k}] \!=\! p_0(y_k|\vec{y}_{k\shortminus1}) \bigg\{\!\rho[k\!\shortminus\!1|\vec{y}_{k\shortminus1}] \!+\! M \D[\LinOp] \rho[k\!\shortminus\!1|\vec{y}_{k\shortminus1}] \Dt  \nonumber \\
     & \!\!\!\!\!\!+\! \sqrt{\!M} \big(\!\LinOp\rho[k\!\shortminus\!1|\vec{y}_{k\shortminus1}]  \!+\! \rho[k\!\shortminus\!1|\vec{y}_{k\shortminus1}]\LinOp^\dagger  \big) \delta y_k \!+\! \bigO(\Dt^{\sfrac{3}{2}}) \!\bigg\}.
\end{align}
This enables us to write the conditional state at time $k$ after measurement as
\begin{align}
    &\rho[k\!\shortminus\!1|\vec{y}_{k}] = \frac{\rhoun[k\!\shortminus\!1|\vec{y}_{k}]}{\trace{\rhoun[k\!\shortminus\!1|\vec{y}_{k}]}} \nonumber\\ 
    &= \bigg(\rho[k\!\shortminus\!1|\vec{y}_{k\shortminus1}] + M \D[\LinOp]\rho[k\!\shortminus\!1|\vec{y}_{k\shortminus1}] \Dt \nonumber \\
    &\;\;+ \! \sqrt{M}\!\left(\LinOp\rho[k\!\shortminus\!1|\vec{y}_{k\shortminus1}] + \rho[k\!\shortminus\!1|\vec{y}_{k\shortminus1}] \LinOp^\dagger  \right)\!\delta y_k \bigg) \! \times \nonumber \\
    &\;\;\times \!\!
    \bigg( \!1 \!-\! \sqrt{\!M} \!\langle \LinOp \!+\! \LinOp^\dagger \rangle \delta y_k \!+\! M \!\langle \LinOp \!+\! \LinOp^\dagger \rangle^2 \Dt \bigg)  + \bigO(\Dt^{\sfrac{3}{2}}) = \nonumber \\
     &=\rho[k\!\shortminus\!1|\vec{y}_{k\shortminus1}] \!+\! M \D[\LinOp]\rho[k\!\shortminus\!1|\vec{y}_{k\shortminus1}] \Dt  \nonumber \\
    &\;\;+\!
    \sqrt{\!M} \!\left(\!\LinOp\rho[k\!\shortminus\!1|\vec{y}_{k\shortminus1}] \!+\! \rho[k\!\shortminus\!1|\vec{y}_{k\shortminus1}] \LinOp^\dagger \! \right)\! \delta y_k \nonumber \\
    &\;\;-\! 
    \sqrt{\!M} \langle \LinOp \!+\! \LinOp^\dagger \rangle \rho[k\!\shortminus\!1|\vec{y}_{k\shortminus1}] \delta y_k \nonumber \\
    &\;\;-\!
    M\!\!\left(\!\LinOp\rho[k\!\shortminus\!1|\vec{y}_{k\shortminus1}] \!+\! \rho[k\!\shortminus\!1|\vec{y}_{k\shortminus1}] \LinOp^\dagger  \!\right) \!\!\brkt{\LinOp \!+\! \LinOp^\dagger}\Dt \nonumber \\
    &\;\;+\! 
    M \brkt{\LinOp \!+\! \LinOp^2}^2\rho[k\!\shortminus\!1|\vec{y}_{k\shortminus1}]\Dt +\!\bigO(\Dt^{\sfrac{3}{2}}) \nonumber \\
    &=  \rho[k\!\shortminus\!1|\vec{y}_{k\shortminus1}]  \!+\! M \D[\LinOp] \rho[k\!\shortminus\!1|\vec{y}_{k\shortminus1}] \Dt  \nonumber \\
    &\;\;+\! 
    \sqrt{\!M} \H[\LinOp]\rho[k\!\shortminus\!1|\vec{y}_{k\shortminus1}] \delta W + \bigO(\Dt^{\sfrac{3}{2}})
    \label{eq:general_form_with_phi}
\end{align}
where the measurement-induced nonlinear superoperator is defined as~$\H[ \GenOp] \rho \coloneqq  \GenOp \rho + \rho  \GenOp^\dagger - \trace{(\GenOp+\GenOp^\dagger)\rho}\rho$ for any operator $\GenOp$ and state $\rho$, and we have employed the inverse of the normalisation constant, i.e.~of probability \eref{p_x(t)_expansion}, to the leading order in $\dt$:
\begin{align}
    &\frac{1}{p_\Dt(y_k|\vec{y}_{k\shortminus1})} = \nonumber \\
    &\!\!=\!\frac{1}{p_{0\!}(y_k|\vec{y}_{k\shortminus1}\!)\!} \!\left(\! 1 \!+\! y_k \langle \LinOp \!+\! \LinOp^\dagger \rangle \sqrt{2M\Dt} + \bigO(\Dt) \! \right)^{\!\!-1} \nonumber \\
    &\!\!=\! \frac{1}{p_{0\!}(y_k|\vec{y}_{k\shortminus1}\!)\!} \!\left( \!1 \!+\! \sqrt{\!M} \langle \LinOp \!+\! \LinOp^\dagger \rangle \delta y_k + \bigO(\Dt) \right)^{\!\!-1} \nonumber \\
    &\!\!=\!\frac{1}{p_{0\!}(y_k|\vec{y}_{k\shortminus1}\!)\!} \!\left( \!1 \!\shortminus\! \sqrt{\!M} \!\langle \LinOp \!+\! \LinOp^\dagger \rangle \delta y_k \!+\! M \!\langle \LinOp \!+\! \LinOp^\dagger \rangle^{\!2} \!\Dt \!+\! \bigO(\!\Dt^{\sfrac{3}{2}}) \!\right)\!. \nonumber
\end{align}
As a consequence, we may write \eqnref{eq:rho_after_evol} as 
\begin{align}
    \rho[k|\vec{y}_k]&=\frac{\Phi_k[\rho[k\!\shortminus\!1|\vec{y}_{k}]]}{\trace{\Phi_k[\rho[k\!\shortminus\!1|\vec{y}_{k}]]}} =\frac{\rhoun[k|\vec{y}_k]}{\trace{\rhoun[k|\vec{y}_k]}}
\end{align}
where
\begin{align} \label{eq:unnorm_cond_state_beforePhi}
    \rhoun[k|\vec{y}_k] &= \Phi_{k}\!\bigg[\rho[k\!\shortminus\!1|\vec{y}_{k\shortminus1}]  \!+\! M \D[\LinOp] \rho[k\!\shortminus\!1|\vec{y}_{k\shortminus1}] \Dt  \nonumber \\
    &+ \sqrt{\!M} \H[\LinOp]\rho[k\!\shortminus\!1|\vec{y}_{k\shortminus1}] \delta W + \bigO(\Dt^{\sfrac{3}{2}}) \bigg]
\end{align} 

\subsection{Recovering SME \eqref{eq:SME} with Bayesian feedback}
\label{subsec:recovering_the_sme}
As discussed in \appref{subsec:separability}, the internal map $\Phi_k$ can be decomposed into deterministic and stochastic components. One particular form of this decomposition arises when the evolution between measurements is described by
\begin{equation} \label{eq:phi_bayesian}
    \Phi_k = \ee^{\Lin_{\vec{y}_k} \Dt}.
\end{equation}
At first glance, this expression might appear purely deterministic. However, while the map itself does not explicitly include a stochastic increment (such as a Wiener process or a jump term), it remains stochastic due to the dependence of generator $\Lin_{\vec{y}_k}$ on the random measurement outcomes. Specifically, the generator is given by
\begin{align}
	\Lin_{\vec{y}_{k}}\rhoc
	&=-i (\omega + u(t)) [\Jz,\rhoc]\\  
	&+ \frac{\kloc}{2} \sum_{j=1}^N \D[\sz^{(j)}] \rhoc \, + \kcoll \D[\Jz]\rhoc,  \nonumber
\end{align}
where the feedback term $u(t) \equiv u(\vec{y}_k)$ introduces the stochasticity indirectly, as it varies with the random measurement record. However, for small time steps, we can still approximate \eqnref{eq:phi_bayesian} as
\begin{align}
    \Phi_{k} [\,\cdot\,] = [\,\cdot\,] + \Lin_{\vec{y}_k} [\,\cdot\,] \Dt + \bigO(\Dt^2).
\end{align}
Consequently, the unnormalized conditional state of \eqnref{eq:unnorm_cond_state_beforePhi} becomes:
\begin{align}
    \rhoun[k|\vec{y}_k] &= \rho[k\!\shortminus\!1|\vec{y}_{k\shortminus1}]  \!+\! M \D[\LinOp] \rho[k\!\shortminus\!1|\vec{y}_{k\shortminus1}] \Dt  \nonumber \\
    &+ \sqrt{\!M} \H[\LinOp]\rho[k\!\shortminus\!1|\vec{y}_{k\shortminus1}] \delta W \nonumber \\
    &+ \Lin_{\vec{y}_k} \rho[k\!\shortminus\!1|\vec{y}_{k\shortminus1}] \delta t + 
    \bigO(\Dt^{\sfrac{3}{2}}) \\
    &=\rho[k|\vec{y}_k]
\end{align}
which is simply the normalized conditional state since its trace is one. Therefore, by identifying the increment of the conditional density matrix as $\delta \rho[k|\vec{y}_k]=\rho[k|\vec{y}_k]-\rho[k\!\shortminus\!1|\vec{y}_{k\shortminus1}]$ and taking the continuous limit of $\Dt\to 0$, we arrive at SME \eref{eq:SME} central to our analysis in atomic magnetometry:
\begin{align} \label{eq:SME_gen_homodyne}
    \mrm{d}\rho(t|\vec{y}_{\leq t})&=\Lin_{\vec{y}_{\leq t}} \rho(t|\vec{y}_{\leq t}) \dt+M \D[\LinOp]\rho(t|\vec{y}_{\leq t}) \dt \nonumber \\
    &+\sqrt{M}\H[\LinOp]\rho(t|\vec{y}_{\leq t}) \dW,
\end{align}
where the Lindbladian $\Lin_{\vec{y}_{\leq t}}$ is
\begin{align}
	\Lin_{\vec{y}_{\leq t}}\rhoc
	&=-i (\omega + u(t)) [\Jz,\rhoc]\\  
	&+ \frac{\kloc}{2} \sum_{j=1}^N \D[\sz^{(j)}] \rhoc \, + \kcoll \D[\Jz]\rhoc,  \nonumber
\end{align}
and the probed system operator is $\LinOp=\Jy$.

\subsection{Markovian feedback of \citet{wiseman1994_Feedback}}
\label{subsec:Markov_feedback}
Consider the following general internal evolution given by a field Hamiltonian and a dissipator, plus a feedback Hamiltonian term:
\begin{align}
     &\Phi_{k} [\,\cdot\,] =  \ee^{-\ii\delta y_k \left[\Fop, \;\cdot\;\right] - \ii \left[\Ham, \;\cdot\;\right]\Dt + \D[\hat{T}]\;\cdot\;\Dt} [\,\cdot\,]
\end{align}
with $\delta y_k$ a stochastic increment. Then, as shown in \eqnref{eq:phik_det_and_stoch}, we can expand it to first order $\Dt$ as:
\begin{align} \label{eq:Wiseman_feedback_map}
     &\Phi_{k} [\,\cdot\,] = \nonumber \\
     &=  \!\bigg(\!\I \!-\!\ii\delta y_k \!\left[\!\Fop, \;\cdot\;\right] \!-\! \ii \!\left[\!\Ham, \;\cdot\;\right]\!\Dt \!+\! \D[\hat{T}]\,\cdot\,\Dt \!+\! \D[\Fop] \, \cdot \, \Dt \nonumber \\
     &+\! \bigO(\Dt^{\sfrac{3}{2}})\bigg)\! [\,\cdot\,] = \ee^{ - \ii \left[\Ham, \;\cdot\;\right]\Dt + \D[\hat{T}]\;\cdot\;\Dt}\left[\ee^{-\ii \delta y_k \left[\Fop, \,\cdot\,\right]} [\,\cdot\,]\right] \nonumber \\
     &= \Xi_\omega \left[\FF_{y_k}\left[\,\cdot\,\right]\right]
\end{align}
Furthermore, if we now apply the map of \eqnref{eq:Wiseman_feedback_map} to \eqnref{eq:unnorm_cond_state_beforePhi} then we can retrieve the Markovian feedback SME equation of \refcite{wiseman1994_Feedback}. Namely, 
\begin{align} \label{eq:unnorm_cond_state_afterWiseman}
    \rhoun[k|\vec{y}_k] &= \rho[k\!\shortminus\!1|\vec{y}_{k\shortminus1}]  - \ii \left[\Ham,\rho[k\!\shortminus\!1|\vec{y}_{k\shortminus1}]\right]\Dt \nonumber \\
    &+ \D[\hat{T}]\rho[k\!\shortminus\!1|\vec{y}_{k\shortminus1}]\Dt \!+\! M \D[\LinOp] \rho[k\!\shortminus\!1|\vec{y}_{k\shortminus1}] \Dt  \nonumber \\
    &+ \sqrt{\!M} \H[\LinOp]\rho[k\!\shortminus\!1|\vec{y}_{k\shortminus1}] \delta W + \D[\Fop] \rho[k\!\shortminus\!1|\vec{y}_{k\shortminus1}] \Dt \nonumber \\
    &-i\sqrt{\!M} \!\left[\Fop,\LinOp\rho[k\!\shortminus\!1|\vec{y}_{k\shortminus1}]\!+\!\rho[k\!\shortminus\!1|\vec{y}_{k\shortminus1}]\LinOp^\dagger\right] \!\Dt \nonumber \\
    &-\ii\!\left[\Fop,\rho[k\!\shortminus\!1|\vec{y}_{k\shortminus1}]\right]\!\delta W \!+\! \bigO(\Dt^{\sfrac{3}{2}}), 
\end{align}
where we have used
\begin{align}
    &\ii\left[\sigma \!+\! \sqrt{M} \H[\LinOp] \sigma \, \delta W , \Fop \right]\! \delta y_k \nonumber \\ 
    &= \ii\left[\sigma \!+\! \sqrt{M} \H[\LinOp] \sigma \, \delta W , \Fop \right]\!\! \left(\!\!\sqrt{\!M} \!\braket{\LinOp\!+\!\LinOp^\dagger}\! \delta t \!+\! \delta W \!\right) \nonumber \\
    &=\ii\left[\sigma \sqrt{\!M} \!\braket{\LinOp\!+\!\LinOp^\dagger}\! \delta t \!+\! \sqrt{M} \H[\LinOp] \sigma \, \delta t \!+\! \sigma \delta W,\Fop\right] \nonumber \\
    &=\ii\left[\!\sqrt{\!M}\!\left(\!\LinOp \sigma \!+\! \sigma \LinOp^\dagger\!\right)\!\Dt\!+\! \sigma \, \delta W,\Fop\right] \nonumber \\
    &=-\ii\sqrt{\!M} \!\left[\Fop,\LinOp\sigma\!+\!\sigma\LinOp^\dagger\right] \!\Dt -\ii\!\left[\Fop,\sigma\right]\!\delta W.
\end{align}
Additionally, note that the trace of \eqnref{eq:unnorm_cond_state_afterWiseman} is one. Therefore, 
\begin{align}
    \rhoun[k|\vec{y}_k] = \rho[k|\vec{y}_k], 
\end{align}
which leads us to rearrange \eqnref{eq:unnorm_cond_state_afterWiseman} using the definition $\delta \rho[k|\vec{y}_k]=\rho[k|\vec{y}_k]-\rho[k\!\shortminus\!1|\vec{y}_{k\shortminus1}]$, and then take the limit of $\Dt\to0$ in order to get
\begin{align}
    \dd \rho(t|\vec{y}_{\leq t}) &= -\ii \left[\Ham,\rho(t|\vec{y}_{\leq t})\right] \! \dt \nonumber \\
    &-\ii \sqrt{M} \left[\frac{\LinOp^\dagger \Fop +\Fop \LinOp}{2} ,\rho(t|\vec{y}_{\leq t}) \right] \! \dt \nonumber \\
    &+\D[\hat{T}] \rho(t|\vec{y}_{\leq t}) \dt \nonumber \\
    &+ \D[\sqrt{M}\LinOp -\ii\Fop] \rho(t|\vec{y}_{\leq t}) \dt \nonumber \\
    &+ \H[\sqrt{M}\LinOp -\ii\Fop ]\rho(t|\vec{y}_{\leq t}) \dW,
\end{align}
where we have rearranged terms by noting that
\begin{align}
     &-\ii \sqrt{\!M} \!\left[\!\frac{\LinOp^\dagger \Fop +\Fop \LinOp}{2} ,\sigma \!\right] \!+\! \D[\sqrt{\!M}\LinOp \!-\!\ii\Fop] \sigma  = \nonumber \\
     &\quad= -\ii\sqrt{\!M} \!\left[\Fop,\LinOp\sigma\!+\!\sigma\LinOp^\dagger\right] + M\D[\LinOp]\sigma + \D[\Fop]\sigma ,
\end{align}
and
\begin{align}
    \H[\sqrt{M}\LinOp -\ii\Fop ] = \sqrt{M}\H[\LinOp]\sigma - \ii[\Fop,\rho],
\end{align}
since $\H[\GenOp][\;\cdot\;]$ is linear with respect to $\GenOp$ and $\H[-\ii \Fop]\sigma = -\ii[\Fop,\sigma]$. Note that the feedback creates new effective measurement and Hamiltonian operators:
\begin{align}
    &\Ham_{\text{eff}} = \Ham + \sqrt{\!M} \frac{\LinOp^\dagger \Fop + \Fop \LinOp}{2} , \\
    &\LinOp_{\text{eff}} = \sqrt{\!M}\LinOp \!-\!\ii\Fop.
\end{align}

%%%%%%%%%%%%%%%%%%%%%%%%%%%%%%%%%%%%%%%%%%%%%%%%%%%%%%%%%%%%%%%%%%%%%%%%%%%%%%%%%%%%%%%%%%%%%%%%%%%%%%%%%%%%%%%%%%%%
%%%%%%%%%%%%%%%%%%%%%%%%%%%%%%%%%%%%%%%%%%%%%%%%%%%%%%%%%%%%%%%%%%%%%%%%%%%%%%%%%%%%%%%%%%%%%%%%%%%%%%%%%%%%%%%%%%%%
%%%%%%%%%%%%%%%%%%%%%%%%%%%%%%%%%%%%%%%%%%%%%%%%%%%%%%%%%%%%%%%%%%%%%%%%%%%%%%%%%%%%%%%%%%%%%%%%%%%%%%%%%%%%%%%%%%%%
\section{Extended Kalman Filter construction} \label{ap:EKF}
A crucial step in the construction of the Extended Kalman Filter (EKF) is to analytically compute the Jacobian matrices (Jacobians) for the nonlinear model studied, here the one of Eqs.~\eref{eq:dynamical_model}. As defined in \secref{sec:EKF}, there are always three Jacobians to be found:~$F$ -- for the deterministic part of the model, $G$ -- for the stochastic part of the model; and $H$ -- for the measurement dynamics. For the model \eref{eq:dynamical_model}, we obtain the following Jacobians:
\vspace{-10pt}
\begin{widetext}
\begin{align}
    &F = \nabla_{\vec{x}} \vec{f} |_{(\vec{\Tilde{x}},u,0)}  = \nonumber \\
    &\left(
     \scalemath{0.7}{
    \begin{array}{ccccccc}
     -(\kcoll+2\kloc + M)/2 & -(\omega + u) & 0 & 0 & 0 & 0 & -\brktc{\estJy} \\ 
    (\omega + u) & -(\kcoll+2\kloc)/2 & 0 & 0 & 0 & 0 & \brktc{\estJx} \\ 
    0 & 2\kcoll \brktc{\estJy} & -(\kcoll+2\kloc+M) & \kcoll & M & -2(\omega + u) - 8\eta M \estCxy^{\cc} & -2\estCxy^{\cc}\\ 
    2\kcoll\brktc{\estJx} & 0 & \kcoll & -\kcoll - 2\kloc - 8\eta M \estVy^{\cc} & 0 & 2(\omega + u) & 2 \estCxy^{\cc} \\
    2M\brktc{\estJx} & 0 & M & 0 & -M & 0 & 0 \\
    -\kcoll \brktc{\estJy} & -\kcoll \brktc{\estJx} & \omega + u & -(\omega + u) -4\eta M \estCxy^{\cc} & 0 & -\left(2\kcoll + 2\kloc + \frac{M}{2} \right) - 4\eta M \estVy^{\cc} & \estVx^{\cc} - \estVy^{\cc}\\
    0 & 0 & 0 & 0 & 0 & 0 & -\chi 
    \end{array}
    }
  \right) 
\end{align}

\begin{align}
    G = \nabla_{\vec{\xi}} \vec{f}|_{\vec{\Tilde{x}}} = 
    \begin{pmatrix} 
    2\sqrt{\eta M} \estCxy^{\cc} & 0 \;\; \\ 
    2\sqrt{\eta M} \estVy^{\cc} & 0 \;\; \\ 
    0 & 0 \;\; \\ 
    0 & 0 \;\; \\ 
    0 & 0 \;\; \\ 
    0 & 0 \;\; \\ 
    0 & 1 \;\;
    \end{pmatrix}, 
\end{align}
\begin{align}
    H = \nabla_{\vec{x}} \vec{h} &= \nabla_{\vec{x}} \left(2\eta \sqrt{M} \, \brktc{\Jy} + \sqrt{\eta} \, \xi \right) = 2\eta \sqrt{M} \begin{pmatrix} 0 & 1 & 0 & 0 & 0 & 0 & 0 \end{pmatrix},
\end{align}
\end{widetext}
of which $F$ and $H$ act on the vector of dynamical parameters $\vec{x}(t) = ( \brktc{\Jx},\brktc{\Jy},\Vx^{\cc},\Vy^{\cc},\Vz^{\cc},\Cxy^{\cc},\omega )^\TT$, whereas $G$ on the vector of stochastic increments $\vec{\xi} = (\xi, \xi_\omega)^\TT$. However, as we restrict to non-fluctuating fields in this work, see also~\refcite{Amoros-Binefa2025}, we set $\chi,\xi_\omega=0$ above.

%%%%%%%%%%%%%%%%%%%%%%%%%%%%%%%%%%%%%%%%%%%%%%%%%%%%%%%%%%%%%%%%%%%%%%%%%%%%%%%%%%%%%%%%%%%%%%%%%%%%%%%%%%%%%%%%%%%%
%%%%%%%%%%%%%%%%%%%%%%%%%%%%%%%%%%%%%%%%%%%%%%%%%%%%%%%%%%%%%%%%%%%%%%%%%%%%%%%%%%%%%%%%%%%%%%%%%%%%%%%%%%%%%%%%%%%%
\section{Benchmarking against a classical strategy with a strong measurement} 
\label{app:classical_strategy}
As demonstrated already by the numerical considerations in \secref{sec:numerical_results}, the optimal EKF+LQR strategy not only attains the lowest AMSE in estimating the Larmor frequency $\omega$, but also steers the ensemble into a (conditionally) spin squeezed and, hence, entangled state. However, as the continuous probing also introduces extra decoherence into the dynamics---note the term $\propto~\!\!M\D[\Jy]$ in \eqnref{eq:SME}---one may question whether such an entanglement-enhanced approach is actually beneficial. As an alternative, we consider a \emph{classical strategy} within which the atomic ensemble evolves without being disturbed until a given time $t$, at which point a \emph{strong (destructive) measurement} is performed that can in principle provide much more information. As we now show, such an approach is deemed useless, even when considering the most general strong measurements, as it is crucial for the single-shot (Bayesian) estimation scenario for the continuous measurement to constantly gather more and more information about $\omega$ over time. 

\subsubsection{Quantum Bayesian Cram\'{e}r-Rao bound} 
Considering such a classical strategy with atoms just undergoing precession and local ($\kloc$) or collective ($\kcoll$) decoherence after being initialised in a CSS state, we can lower bound the minimal AMSE---as defined in \eqnref{eq:AMSE} but with $\vec{y}_t$ now representing the outcomes of a \emph{single} strong measurement performed at time $t$---by resorting to a straightforward generalisation of BCRB \eqref{eq:BCRB}, i.e.~the \emph{quantum Bayesian Cram\'{e}r-Rao Bound} (QBCRB):
\begin{equation}
    \EE{\Delta^2 \est{\omega}} \geq \frac{1}{\F[p(\omega)] + \F_Q[\rho_t,\Jz]},
    \label{eq:QBCRB}
\end{equation}
which now applies to any possible quantum measurement performed at time $t$, with the Fisher information $\F[p(\vec{y}_t|\omega)]$ appearing in \eqnref{eq:BCRB} being replaced by the QFI~\cite{Toth2014,Demkowicz2015}
\begin{equation}
 	\F_Q[\rho_t,\Jz] \coloneqq t^2\,\trace{\rho_t L^2}=t^2\,\trace{\ii[\rho_t,\hat{J}_z] L}
 	\label{eq:QFI}
\end{equation}
that generally satisfies $\F[p(\vec{y}_t|\omega)]\le\F_Q[\rho_t,\Jz]$~\cite{Helstrom1976}, but here it is also $\omega$ independent. The $L$ operator in \eqnref{eq:QFI} is the solution to $\ii[\rho_t,\hat{J}_z]=\frac{1}{2}\{\rho_t,L\}$~\cite{Helstrom1976}.

\subsubsection{Local decoherence} 
When only local decoherence ($\kloc >0$, $\kcoll = 0$) is present, since the initial CSS state is a product of $N$ single-spin states, i.e., $\ket{\mathrm{CSS}_x} = \ket{+}^{\otimes N}$, we can write the state of the atoms at time $t$ as
\begin{equation}
    \rho_t = \Lambda_{\loc}^{\otimes N} \left[ (|+ \rangle \langle +|)^{\otimes N} \right] = \varrho_t ^{\otimes N},
\end{equation}
where $\varrho_t=\Lambda_{\loc}\left[|+ \rangle \langle +| \right]=\frac{1}{2} (1, \ee^{-\kloc t} ; \ee^{-\kloc t}, 1)$.
Now, as QFI \eref{eq:QFI} is additive on product states~\cite{Toth2014,Demkowicz2015}, we can explicitly evaluate it for local decoherence as:
\begin{equation}
    \F_Q[\rho_t,\Jz] = N \F_Q\!\left[\rho_t,\tfrac{1}{2}\hat{\sigma}_z\right] = N t^2 \ee^{-\kloc t}.
\end{equation}
Hence, assuming the prior distribution of $\omega$ to be Gaussian with $p(\omega)=\mathcal{N}(\mu_0,\sigma_0)$, from QBCRB \eref{eq:QBCRB} we obtain the following benchmark constraining the best classical strategy in the presence of local decoherence:
\begin{equation}
    \EE{\Delta^2 \est{\omega}} \geq \frac{1}{1/\sigma_0^2 + N t^2 \ee^{-\kloc t}}.
    \label{eq:BRCB_loc}
\end{equation}

\begin{figure}[t!]
    \centering
    \includegraphics[width=\columnwidth]{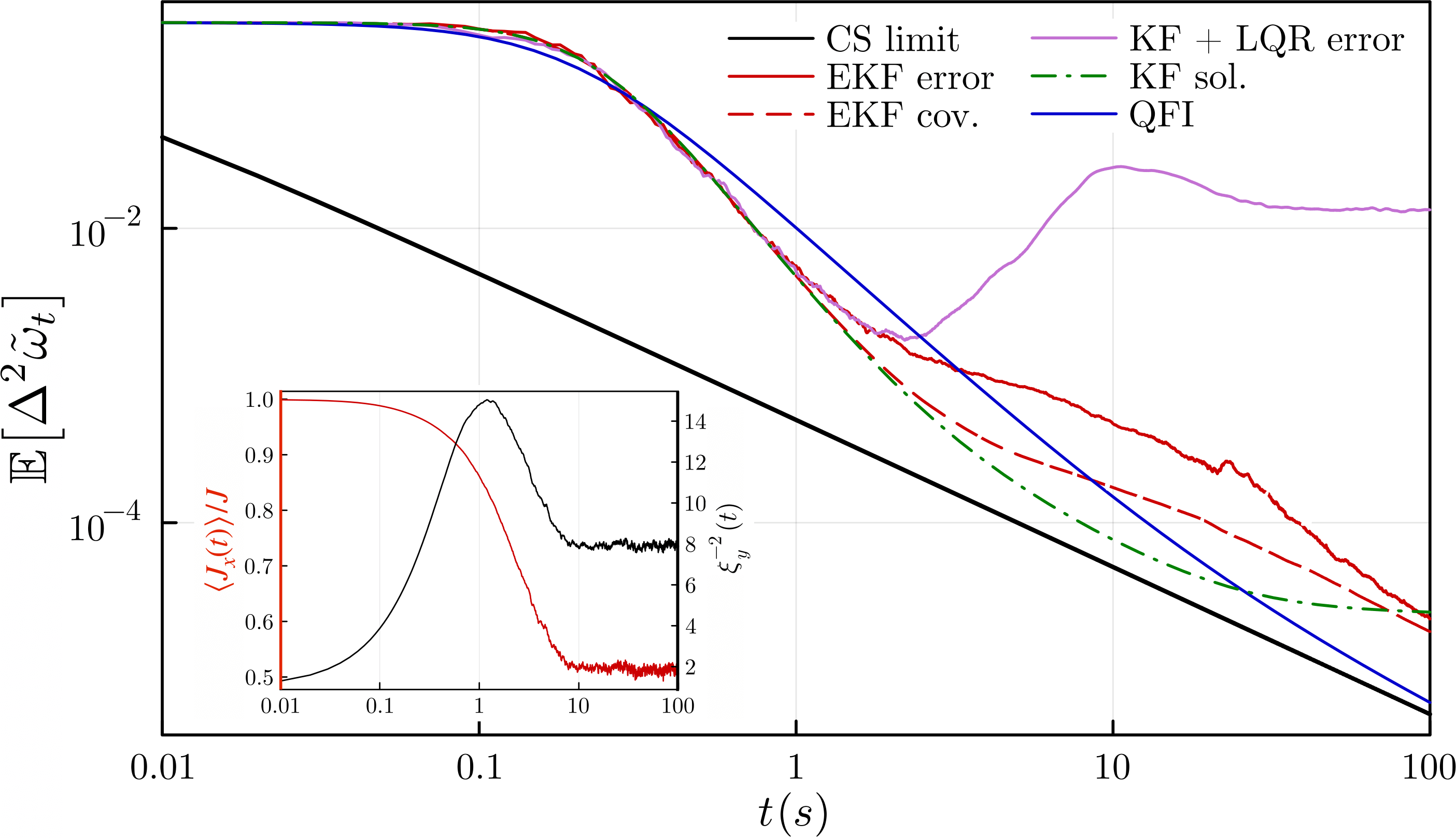}
    \caption{\textbf{Benchmarking against the classical strategy with weak collective decoherence} (${\kcoll\ll M}$,~${\kloc=0}$). The evolution of the AMSE in estimating $\omega$ for various estimation+control schemes is plotted against QCRB \eref{eq:QBCRB} applicable to classical strategies with a strong measurement (\emph{blue solid line}). The covariances provided by the estimators (EKF, \emph{red dashed line};~KF, \emph{green dash-dot line}) correctly predict the true error only within the LG regime (${t \lesssim (M + \kcoll)^{-1}\!\!\approx\!3\mrm{s}}$), while the CS limit \eref{eq:CSlimit} (\emph{black solid line}) is not attained. The inset depicts the evolution of the spin-squeezing parameter \eref{eq:spin_squeez_par} (\emph{black line}) and the ensemble polarisation (\emph{red line}) for the EKF+LQR scheme, for which the atoms maintain spin squeezing. Parameters chosen:~$N = 100$, $\kcoll = 0.0005$, $M = 0.3$, $\omega=1$ and $\eta = 1$. The data is averaged over $\nu=1000$ trajectories and the EKF is initialised as in \figref{fig:err}.
    }
    \label{fig:weak_dec}
\end{figure}

\subsubsection{Collective decoherence}
In the case of collective decoherence, we resort to computing the QFI \eref{eq:QFI} numerically, which we demonstrate to be possible efficiently in the angular momentum basis. In particular, by rewriting the initial CSS state in the $\{\hat{J}^2,\hat{J}_z\}$-basis, i.e.~$\ket{\mathrm{CSS}_x}=\sum_{m=-J}^{J}b_{J,m}\ket{J,m}$ with $b_{J,k} = \frac{1}{2^{J}}\binom{2 J}{J + k}^{1/2}$ and $J = N/2$, we obtain the expression for the state of the atomic ensemble at time $t$ as
\begin{align} \label{eq:totally_sim_rho}
    &\rho_t = \Lambda_\coll\!\left[\ket{\mrm{CSS}_x}\!\bra{\mrm{CSS}_x}\right] = \nonumber \\
    &=\!\!\!\!\! \sum_{n,m = -J}^J \!\!\! b_{J,n} b_{J,m} \, \ee^{-\kcoll \, t \, (m-n)^2 / 2 } |J,n\rangle \langle J,m|, 
\end{align}
whose dimension scales linearly with $N$. Hence, we can numerically perform its eigendecomposition, $\rho_t |k\rangle = \lambda_k |k \rangle$, for all $N\lesssim300$ of relevance. As a result, we may compute the QFI using the expression~\cite{Toth2014,Demkowicz2015}
\begin{equation}
		\F[\rho_t,\Jz] = 2t^2\, \sum_{k,l} \frac{(\lambda_k - \lambda_l)^2}{(\lambda_k + \lambda_l)} |\brkt{k|\Jz|l}|^2,
		\label{eq:QFI_coll}
\end{equation}
with every eigenvector $\ket{k}$ known in the angular momentum basis $\ket{J,m}$, and $\Jz=\sum_{m=-J}^Jm\ket{J,m}\!\bra{J,m}$. Substituting QFI \eref{eq:QFI_coll} into QBCRB \eref{eq:QBCRB}, we obtain an expression similar to \eqnref{eq:BRCB_loc}, which we evaluate numerically to obtain a universal lower bound on the AMSE for the classical strategy with collective decoherence.

\begin{figure}[t!]
    \includegraphics[width=\columnwidth]{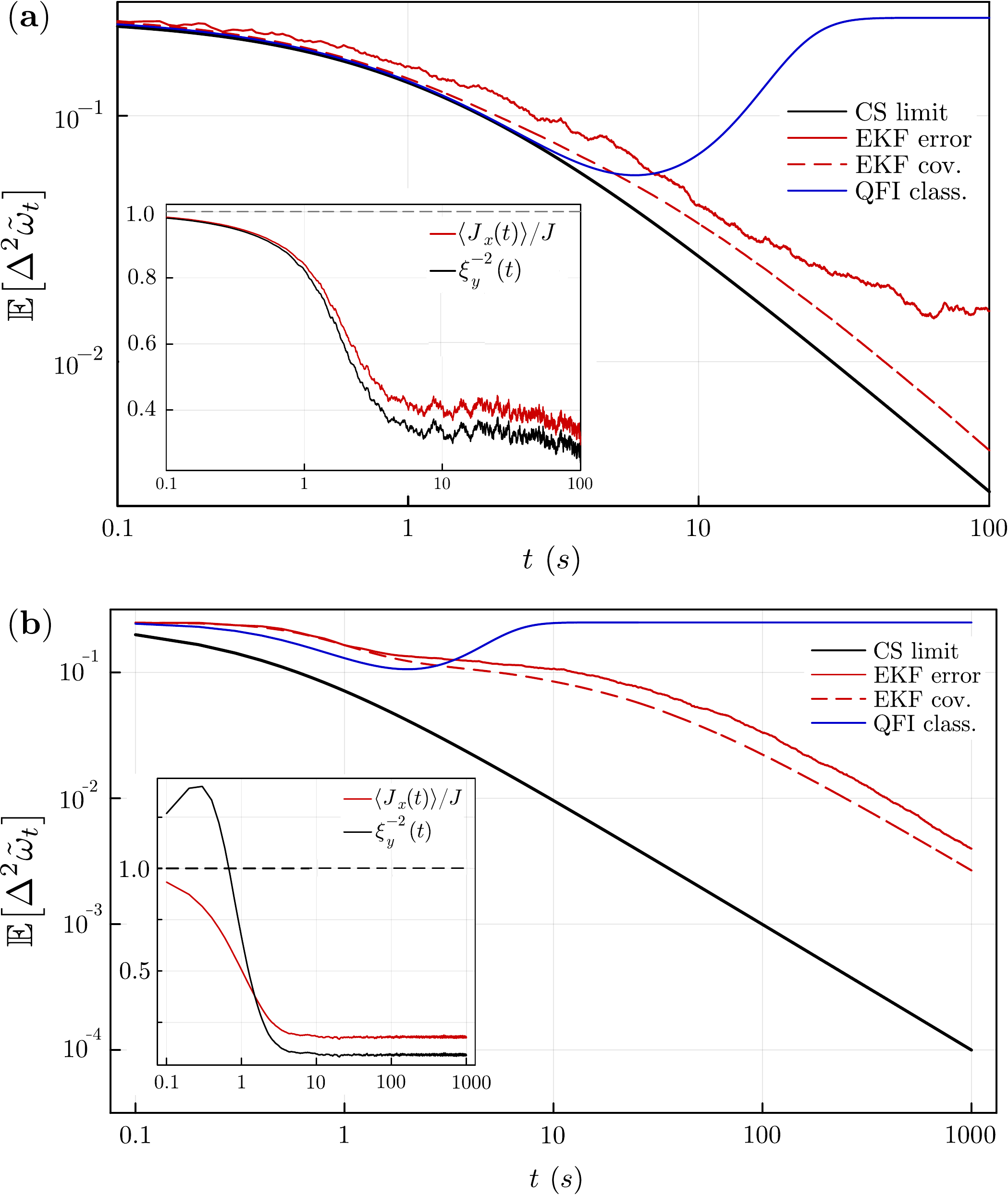}
    \caption{\textbf{Benchmarking against the classical strategy with strong decoherence.} The AMSE attained by the EKF+LQR scheme (\emph{red line}) is shown as a function of time against QCRB \eref{eq:QBCRB} applicable to classical strategies with a strong measurement (\emph{blue line}) for (\textbf{a}) \emph{collective} decoherence (${\kcoll\approx M}$, ${\kloc = 0}$) and (\textbf{b}) \emph{local} decoherence (${\kloc\approx M}$, ${\kcoll = 0}$). Covariances of the EKF are shown as \emph{red dashed lines} along with the CS limit \eref{eq:CSlimit} (\emph{black line}). The insets depict evolution of the spin-squeezing parameter \eref{eq:spin_squeez_par} (\emph{black line}) and the ensemble polarisation (\emph{red line}), but observing spin squeezing ($\EE{\xi^{-2}_{y}}>1$) is impossible for $\kcoll\ge M$~\cite{Amoros-Binefa2021}. In (\textbf{a}) $N=100$ and $\kcoll= M = 0.3$, whereas in (\textbf{b}) $N=10$ and $\kloc= M = 0.5$ ($\omega=1$, $\eta=1$ in both). The AMSEs are averaged over $\nu=1000$ trajectories and the EKF is initialised as in \figref{fig:err}.
    }
    \label{fig:strong_dec}
\end{figure}

\subsubsection{Benchmarking the estimation+control schemes}
In \figref{fig:weak_dec}, we consider \emph{weak collective decoherence} ($\kloc=0$, $\kcoll\ll M$) and compare the AMSEs attained by different schemes involving continuous probing against QBCRB \eref{eq:QBCRB} applicable for the classical strategy (\emph{blue line}). As expected, out of all the estimation+control schemes the EKF+LQR strategy (\emph{red solid line}) yields the best results, and clearly surpasses the limit imposed on classical strategies. However, this occurs as short timescales within the LG regime, $t \lesssim (M + \kcoll)^{-1}$, at which the KF (\emph{pink line}) could also be used, despite quickly becoming unreliable at longer times. Moreover, due to the low atomic number, $N=100$, the covariance $\sigma_{\omega\omega}$ of the EKF (\emph{red dashed line}) and KF (\emph{green dash-dot line}; derived by us in \refcite{Amoros-Binefa2021}) correctly predicts the true errors only at very short times for which the CoG model \eref{eq:dynamical_model} can be trusted. Although the QBCRB \eref{eq:QBCRB} in \figref{fig:weak_dec} indicates that at longer times the classical strategy involving a strong measurement may attain lower AMSEs, this is only a consequence of choosing very weak decoherence and, hence, a very long coherence time, $\kcoll^{-1}\approx10^4$s, beyond which the strong measurement provides no information about $\omega$. 

In order to emphasise this behaviour and see the clear benefits of using the EKF+LQR scheme, we significantly increase the strength of decoherence, either collective or local in \fref{fig:strong_dec}(\textbf{a}) or (\textbf{b}), respectively, so that it is comparable with the strength of the QND-measurement used for continuous probing. From \fref{fig:strong_dec}, it is clear that in both cases, while the EKF+LQR schemes keeps the atomic state polarised (see the \emph{red lines} within the insets) and, hence, the AMSE decreasing, the classical strategy becomes useless beyond the coherence time ($1/\kcoll$ or $1/\kloc$) of the atomic ensemble.

%%%%%%%%%%%%%%%%%%%%%%%%%%%%%%%%%%%%%%%%%%%%%%%%%%%%%%%%%%%%%%%%%%%%%%%%%%%%%%%%%%%%%%%%%%%%%%%%%%%%%%%%%%%%%%%%%%%%
%%%%%%%%%%%%%%%%%%%%%%%%%%%%%%%%%%%%%%%%%%%%%%%%%%%%%%%%%%%%%%%%%%%%%%%%%%%%%%%%%%%%%%%%%%%%%%%%%%%%%%%%%%%%%%%%%%%%
%%%%%%%%%%%%%%%%%%%%%%%%%%%%%%%%%%%%%%%%%%%%%%%%%%%%%%%%%%%%%%%%%%%%%%%%%%%%%%%%%%%%%%%%%%%%%%%%%%%%%%%%%%%%%%%%%%%%
\begin{figure}[b!]
    \centering \includegraphics[width = 0.77\columnwidth]{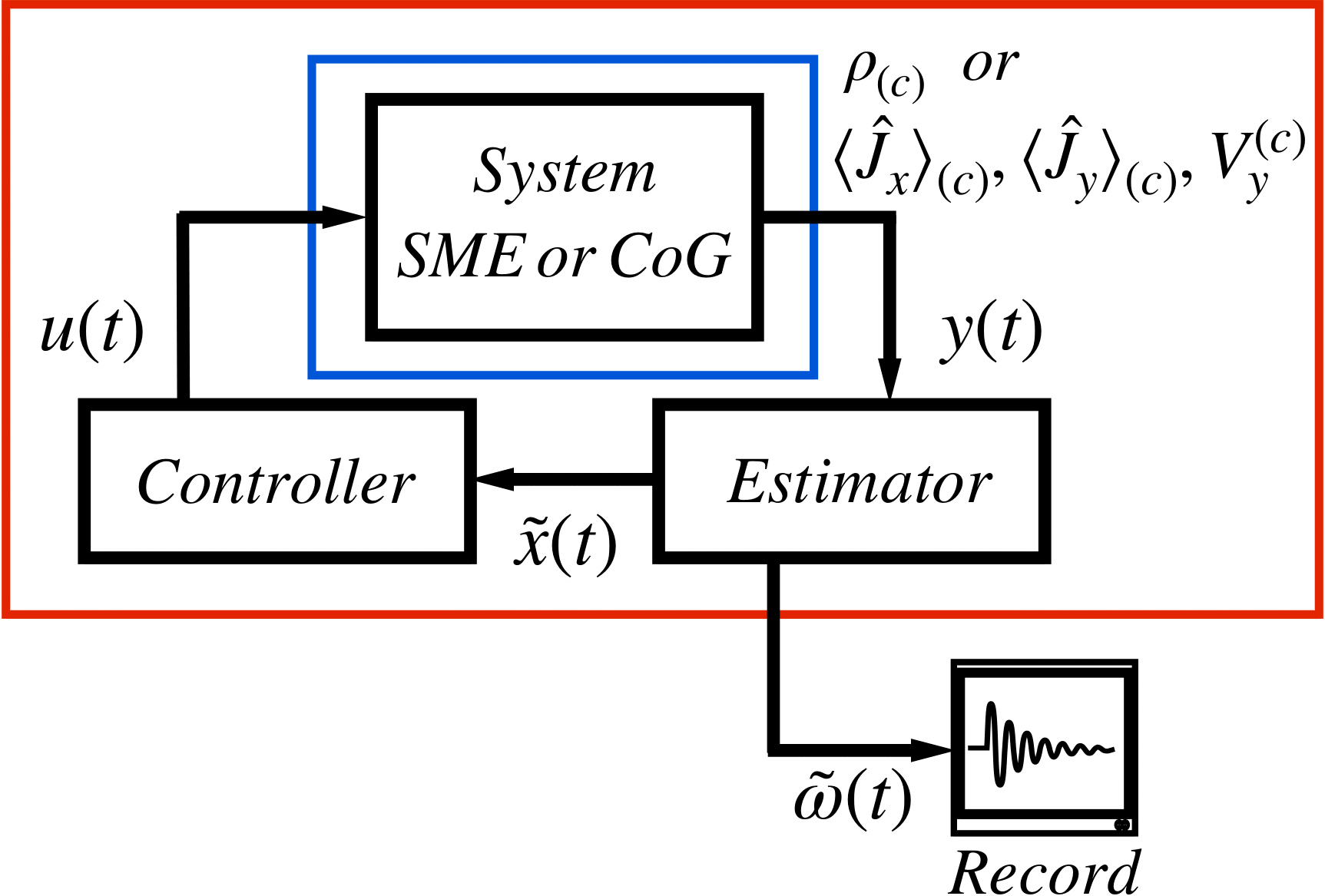}
    \caption{\textbf{Information flow within the setup of \figref{fig:setup}}. Basic building blocks: \emph{system}, \emph{estimator} and \emph{controller}, all connected in a closed feedback loop. The outputs of the system can either be the conditional state $\rho_{\cc}(t)$ or the state-vector $\vec{x}(t)$ containing first and second moments (e.g.~$\brktc{\Jx}$, $\brktc{\Jy}$, $\Vy^{\cc}$), depending on how we simulate the system, either with SME \eqref{eq:SME} or CoG \eqref{eq:dynamical_model}, respectively. From the system output we construct the photocurrent $y(t)$ according to \eqnref{eq:photocurrent}, which is fed into a filter that produces real-time estimates $\est{\vec{x}}(t)$, which are then used by a controller to devise a control signal $u(t)$ to steer the state of the system. Assessing the fidelity of the CoG approximation in replicating the evolution of the system can be performed at two different stages:~by analysing the system dynamics itself (output of the blue box) or focussing only on the $\omega$-estimation task (output of the red box).}
    \label{fig:scheme}
\end{figure}
\section{Verification of the CoG approximation}
\label{ap:verify_CoG}
\figref{fig:scheme} presents the architecture of the feedback loop exploited within our atomic magnetometry scheme. The detection data $\vec{y}(t)$ in each round is generated by simulating the `\emph{System}' either exactly---its full conditional density matrix $\rhoc$ with the help of SME \eref{eq:SME}---or approximately---tracking only the dynamics of its relevant first and second moments, $\brktc{\Jx}$, $\brktc{\Jy}$, and $\Vy^{\cc}$, assuming them to be governed by the CoG model \eref{eq:dynamical_model}. The so-simulated measurement record is interpreted by the `\emph{Estimator}' (i.e.,~the EKF) providing in real time not only the estimate of the Larmor frequency $\est{\omega}(t)$, but also dynamical parameters $\est{\vec{x}}(t)$ that are in turn used by the `\emph{Controller}' (i.e., the LQR) to modify ``on the fly'' the system dynamics being simulated by changing $u(t)$.

To assess the accuracy of the CoG approximation \eref{eq:dynamical_model} in simulating the system dynamics---in contrast to the estimator construction, in which case the EKF is always built based on the CoG model---we benchmark it against the exact SME \eref{eq:SME} solution for a moderate atomic number, for which the latter can still be efficiently performed. In what follows, we perform the comparison at two levels. Firstly, we focus only on the estimation task and, in particular, compute the average discrepancy only when requiring the scheme to accurately provide (on average) $\est{\omega}(t)$ in real time (red box in \figref{fig:scheme}). Secondly, we are more restrictive and further require the relevant moments, $\brktc{\Jx}$, $\brktc{\Jy}$ and $\Vy^{\cc}$, to be accurately reproduced by comparing their average error when compared to their exact values computed with help of $\rhoc$ (blue box in \figref{fig:scheme})---so that they may be directly used, e.g.~to estimate the spin squeezing as in \figref{fig:attaining_bounds}. 

\setcounter{figure}{\value{figure}+1}
\begin{figure*}[t]
    \centering 
    \includegraphics[width = 0.89\textwidth]{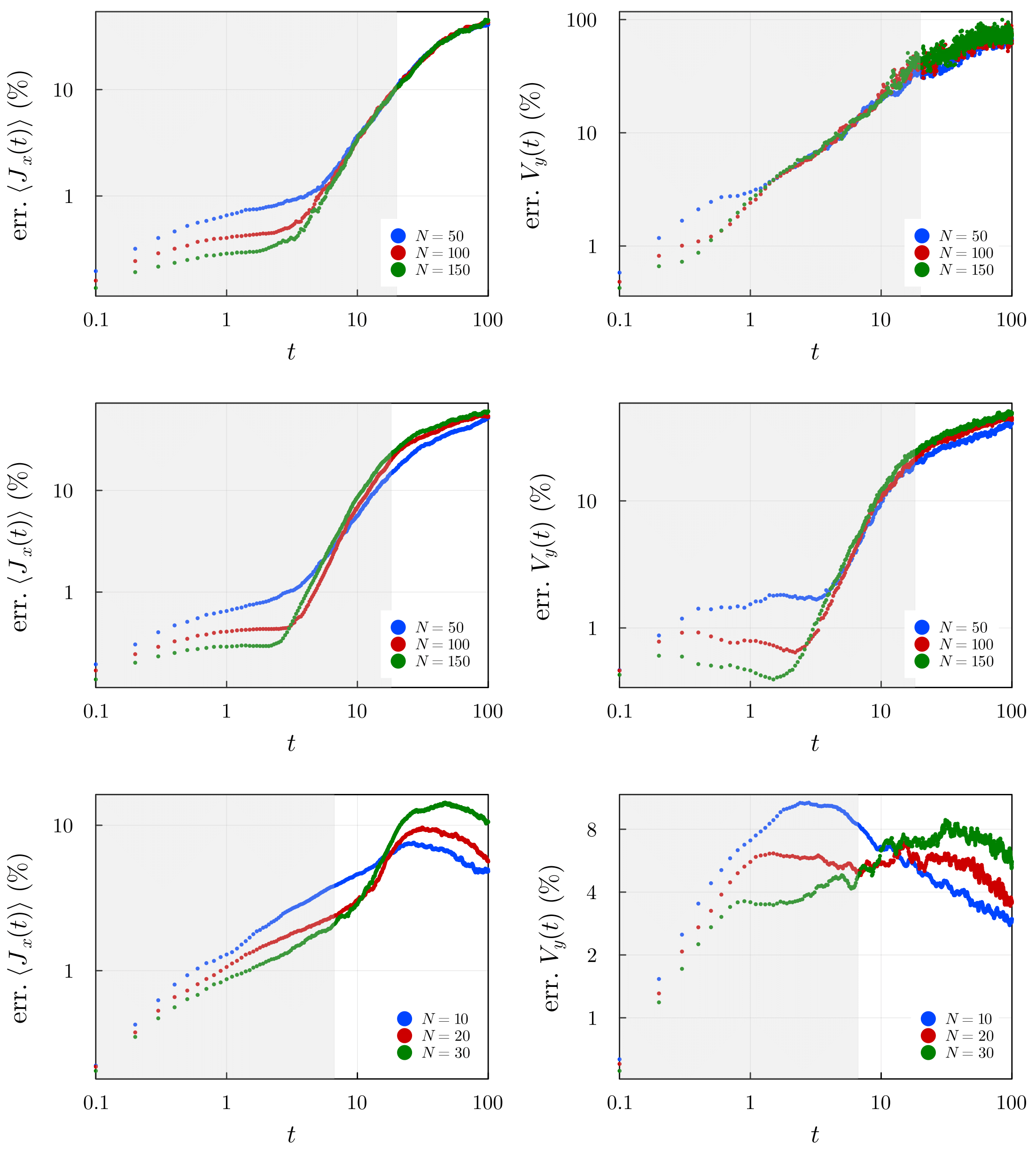}
    \caption{\textbf{Comparative error analysis of moments $\brkt{\Jx}$ and $\Vy$ between the exact and the CoG model.} Each plot presents the relative error (in \%) in simulating the first and second moments, $\mrm{x}\in\{\brkt{\Jx},\Vy\}$, with the approximate CoG model \eref{eq:dynamical_model} as versus solving the exact SME \eref{eq:SME}. The formula for the presented error, $\EE{\delta_{\mrm{x}}}$, can be found in the text below. The comparative analysis is done for three different types of decoherence:~induced by the continuous measurement ($M = 0.05$, $\kcoll=\kloc=0$;~\emph{top row}), measurement-induced and collective ($M = 0.05$, $\kcoll = 0.005$, $\kloc=0$;~\emph{middle row}), measurement-induced and local ($M = 0.05$, $\kcoll = 0$, $\kloc=0.05$;~\emph{bottom row}). In each graph we plot the error for systems of increasing size (either $N = 50, 100, 150$ or $N = 10, 20, 30$ for local noise) to show how, for short times, we expect the CoG approximation to become more accurate as $N$ increases. All errors above are obtained upon averaging over $\nu = 1000$ measurement trajectories.}
    \label{fig:model_vs_exact_moments}
\end{figure*}

In \figref{fig:model_vs_exact_Larmor}, we present in percentages the \emph{average relative error} (ARE) between the real-time estimate $\est{\omega}(t)$ of $\omega$ obtained using the exact model (full SME solution) and the approximate model (CoG), i.e.~$\EE{\delta_{\est{\omega}}} \coloneqq 100\% \times \EE{|(\est{\omega}_{SME} - \est{\omega}_{CoG})/\est{\omega}_{SME}|} = 100\% \times \int\!d\omega\,p(\omega) \int\!\mathcal{D}\vec{y}_{\leq t}\,p(\vec{y}_{\leq t}|\omega)|(\est{\omega}_{SME} - \est{\omega}_{CoG})/\est{\omega}_{SME}|$. \figref{fig:model_vs_exact_Larmor} showcases three plots, each corresponding to a different noise scenario: measurement decoherence, combined measurement and collective decoherence, and combined measurement and local decoherence, arranged from top to bottom. Each plot shows the averaged relative error for increasing system sizes---specifically, $N = 50,100,150$ for the first two scenarios, and $N = 10,20,30$ for the local case. The plots demonstrate%\linebreak
\setcounter{figure}{\value{figure}-2}
\begin{figure}[H]
    \centering \includegraphics[width = 0.9\columnwidth]{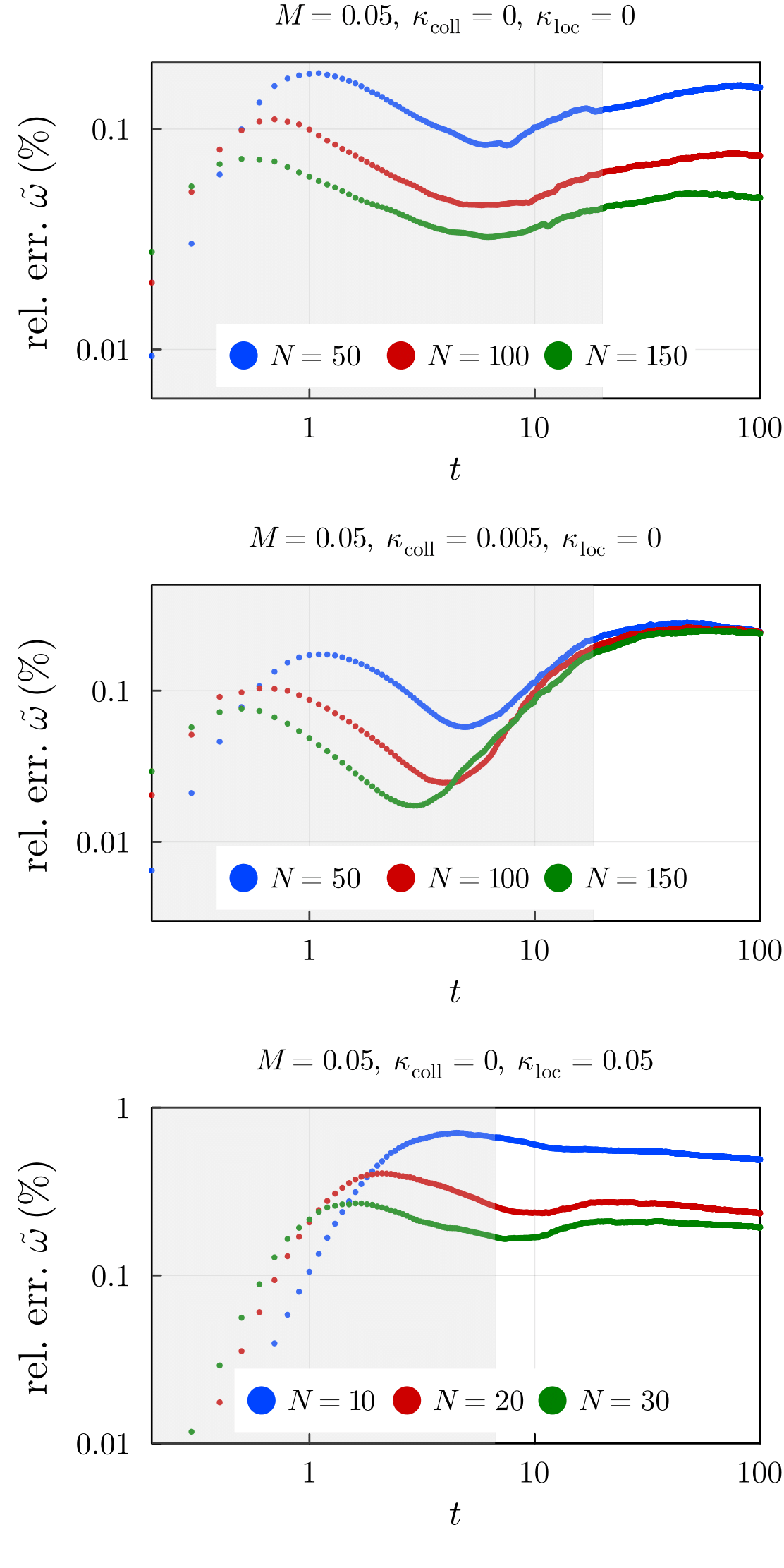}
    \caption{\textbf{Evolution of the relative error in estimating the Larmor frequency when comparing the exact (SME \eqref{eq:SME}) and the approximate (COG \eqref{eq:dynamical_model}) models.} Each graph shows the relative error in percentages for three different noisy scenarios:~with the decoherence induced solely by the continuous measurement, with both measurement-induced and collective decoherence, with all measurement-induced, collective and local decoherence (in order from \emph{top to bottom}). In each plot, various system sizes are considered:~$N = 50,100,150$ (\emph{blue, red and green}, respectively) for the first two cases $N = 10,20,30$ (also \emph{blue, red and green}, respectively) also in the presence of the local decoherence. Each plot demonstrates that the CoG approximation can be used to simulate the dynamics of the system and still get accurate estimations of the Larmor frequency compared to using the exact model (errors always below 1\%). Additionally, all plots show a decrease in the relative error when increasing the system size within the LG regime (shaded pink area), and for the case of only measurement decoherence and local decoherence, also outside of it. All errors above are obtained upon averaging over $\nu = 1000$ measurement trajectories.}
    \label{fig:model_vs_exact_Larmor}
\end{figure}
\noindent that, as the system size increases, the Larmor frequency estimate derived using the CoG approximation either aligns more closely with the estimate given by solving the SME exactly, or maintains an error margin below 1\%. This trend leads us to conclude that for large ensembles with $N \sim 10^5 - 10^{14}$, the CoG approximation proves to be sufficiently accurate to generate the measurement data needed to estimate the Larmor frequency. 

However, to evaluate the reliability of the CoG approximation \eqref{eq:dynamical_model} itself, we verify its ability to reproduce the key system dynamical properties, in particular, the moments necessary to compute the spin-squeezing parameter \eqref{eq:spin_squeez_par}, $\brktc{\Jx}$ and $\Vy^{\cc}$, as compared to their true values obtained from $\rhoc$ provided by the exact solution of SME \eqref{eq:SME} (compare the outputs of the blue box in \figref{fig:scheme}). For a quantitative assessment, as detailed in \figref{fig:model_vs_exact_moments}, we employ the error metric $\EE{\delta_{\mrm{x}}} = 100 \times \EE{|\mrm{x}_\mrm{SME} - \mrm{x}_\mrm{CoG}|} / \EE{|\mrm{x}_\mrm{SME}|}$, where $\mrm{x}$ in our case can be either $\brktc{\Jx}$ or $\Vy^{\cc}$. In \figref{fig:model_vs_exact_moments}, each subplot illustrates that, for short times, the simulation error decreases as the system size increases. However, as time progresses, this trend is halted, with the error in simulating first and second moments escalating significantly. Nonetheless, for times shorter than $t \lesssim 1/(M+\kcoll+2 \kloc)$ , the CoG simulation error for these moments is kept below about 10\%. This finding supports the use of the CoG approximation to accurately predict the squeezing parameter of large atomic ensembles, e.g.~with $N$ as large as $10^5 - 10^{14}$, at short times below $t \lesssim 1/(M+\kcoll+2 \kloc)$.

%%%%%%%%%%%%%%%%%%%%%%%%%%%%%%%%%%%%%%%%%%%%%%%%%%%%%%%%%%%%%%%%%%%%%%%%
\bibliographystyle{myapsrev4-2}
\bibliography{EKFfeed}
%%%%%%%%%%%%%%%%%%%%%%%%%%%%%%%%%%%%%%%%%%%%%%%%%%%%%%%%%%%%%%%%%%%%%%%%
\end{document}